\documentclass[sn-basic,iicol]{sn-jnl}

\usepackage{graphicx}%
\usepackage{multirow}%
\usepackage{amsmath,amssymb,amsfonts}%
\usepackage{amsthm}%
\usepackage{mathrsfs}%
\usepackage[title]{appendix}%
\usepackage{xcolor}%
\usepackage{textcomp}%
\usepackage{manyfoot}%
\usepackage{booktabs}%
\usepackage{algorithm}%
\usepackage{algorithmicx}%
\usepackage{algpseudocode}%
\usepackage{listings}%

\usepackage{subcaption} 
\usepackage{siunitx} 
\usepackage{pdfpages} 

\graphicspath{{pictures/}}

\renewcommand{\orcid}[1]{\href{https://orcid.org/#1}{\textcolor[HTML]{A6CE39}{\includegraphics{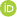}}}}

\begin{document}

\title{A Comprehensive Numerical Approach to Coil Placement in Cerebral Aneurysms: Mathematical Modeling and In Silico Occlusion Classification}
\date{\today}


\author*[1]{\fnm{Fabian} \sur{Holzberger}  \orcid{0000-0003-3577-5466}}\email{holf@cit.tum.de}

\author[1]{\fnm{Markus} \sur{Muhr}}

\author[1]{\fnm{Barbara} \sur{Wohlmuth}}

\affil[1]{\orgdiv{Department of Mathematics}, \orgname{Technical University of Munich}, \orgaddress{\street{Boltzmannstr. 3/III}, \city{Garching b. M\"unchen}, \postcode{85748}, \country{Germany}}}


\abstract{Endovascular coil embolization is one of the primary treatment techniques for cerebral aneurysms. Although it is a well established and minimally invasive method, it bears the risk of sub-optimal coil placement which can lead to incomplete occlusion of the aneurysm possibly causing recurrence. One of the key features of coils is that they have an imprinted natural shape supporting the fixation within the aneurysm. For the spatial discretization our mathematical coil model  is based on the Discrete Elastic Rod model which results in a dimension-reduced 1D system of differential equations. We include bending and twisting responses to account for the coils natural curvature. Collisions between coil segments and the aneurysm-wall are handled by an efficient contact algorithm that relies on an octree based collision detection. The numerical solution of the model is obtained by a symplectic semi-implicit Euler time stepping method. Our model can be easily incorporated into blood flow simulations of embolized aneurysms.

In order to differentiate optimal from sub-optimal placements, we employ a suitable in silico Raymond–Roy type occlusion classification and measure the local packing density in the aneurysm at its neck, wall-region and core. We investigate the impact of uncertainties in the coil parameters and embolization procedure. To this end, we vary the position and the angle of insertion of the microcatheter, and approximate the local packing density distributions by evaluating sample statistics.}




\maketitle


\section{Introduction}
\label{sec:introduction}

Besides from surgical clipping and the use of flow diverters\,/\,stents \citep{briganti2015endovascular, pierot2011flow, sindeev2019evaluation} and Woven EndoBridge (WEB)-Devices \citep{goyal2020web, pierot2015web}, endovascular coiling \citep{pierot2013endovascular, guglielmi2007history, hui2014history, eddleman2013endovascular} is one of the most commonly used methods \citep{zhao2018current} in clinical treatment of cerebral aneurysms. It is a volumetric occlusion technique, where the sack of an aneurysm is filled with a thin metal wire, usually platinum, that coils up therein. This causes a stagnation of the blood flow in the aneurysm which together with the intrinsic thrombogenicity of the metal wire leads to embolization \citep{ngoepe2018thrombosis, byrne1997nature}.

\begin{figure*}[!htb]
    \centering
    \begin{subfigure}[b]{0.4\textwidth}
        \includegraphics[width=1\textwidth]{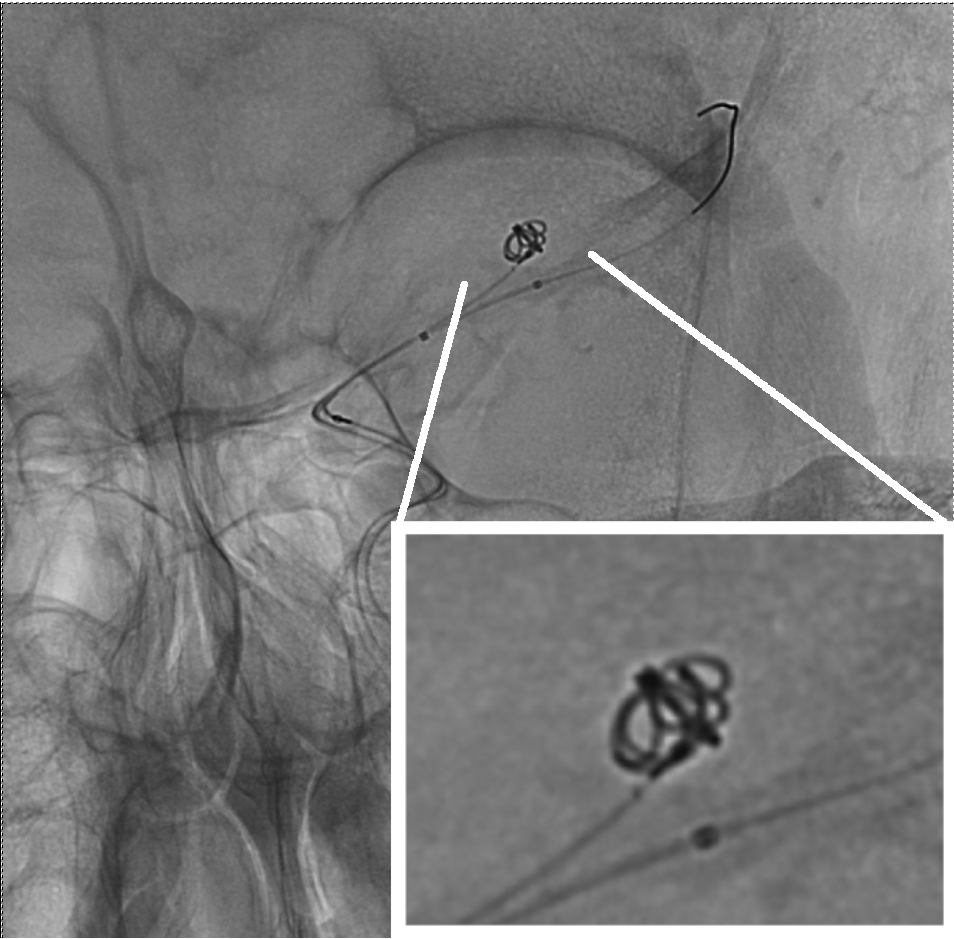}
        \caption{}
        \label{fig:coiling_photo1}
    \end{subfigure}
    \hfill
    \begin{subfigure}[b]{0.4\textwidth}
        \includegraphics[width=1\textwidth]{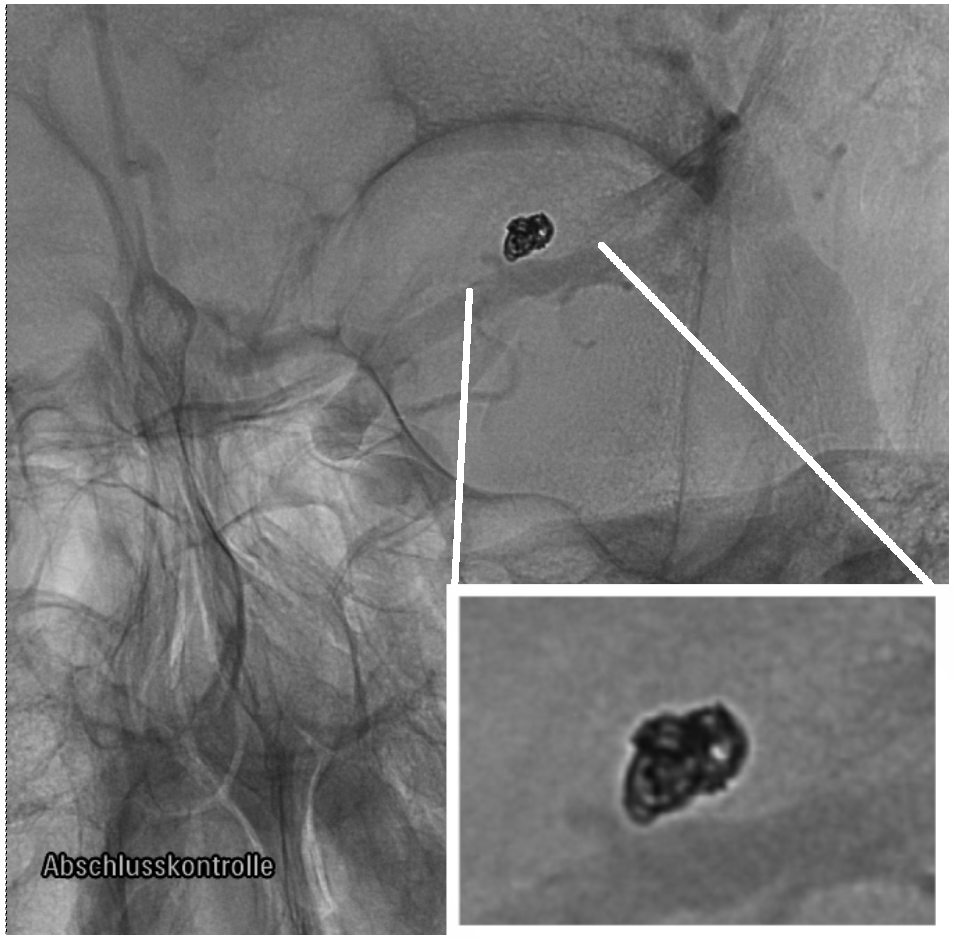}
        \caption{}
        \label{fig:coiling_photo2}
    \end{subfigure}
    \caption{Coiling of an aneurysm at different packing densities. (\subref{fig:coiling_photo1}) Incomplete occlusion (low packing density), with only framing coil inserted. (\subref{fig:coiling_photo2}) Completed (high volumetric occlusion) coiling procedure with additional filling coil in place.}
    \label{fig:main}
\end{figure*}

The coiling procedure is conducted under continuous  imaging supervision in digital subtraction angiography, a fluoroscopy technique, where an interventionist inserts a catheter through a femoral or radial access to reach for the brain along the upstream  of blood, carefully moving to the part where the aneurysm is located. Then, via a micro-catheter, the wire is protruded into the aneurysm dome. Coils do exist in various shapes and sizes as well as with various material properties \citep{ito2018experimental, kanenaka2016comparative, white2008coils}. A patient-specific choice, e.g., with respect to the length or the imprinted natural shape of the inserted coil(s), is made based on preceding aneurysm measurements as well as the surgeons experience with respect to optimal placement \citep{neki2018optimal}. Possible choices are, e.g., a stiffer framing coil followed by one or several softer filling and finishing coils depending on the shape of the aneurysm, see Fig. \ref{fig:main}. 
An angiography in combination with the injection of a tracer fluid allows to evaluate the actual occlusion quality and serves as a decision-making tool whether more coiling wires have to be inserted. Typically, a \textit{packing density} $\psi$, defined as the volume of the inserted coil $V_{\textup{coil}}$ relative to the volume of the aneurysm sack $V_{\textup{aneu}}$, of 20-25\,\% is desired \citep{sluzewski2004relation}, with the coil sitting tightly within the aneurysm, not protruding into the parent vessel. As soon as the aneurysm is sufficiently packed, the catheter is retracted from the aneurysm and clot-formation is about to begin. The presence of a sufficient amount of coil within the aneurysm reduces the perfusion of the aneurysm sack. This effect is even further enhanced by a growing thrombus, eventually 
completely shutting of the aneurysm from the bloodstream.
On one hand, a too high packing density may result either in an occlusion of the parent vessel or partial damage to the wall of the aneurysm sack. 
On the other hand, insufficient packing of the aneurysm  bears the risk of either a prolapse within the adjacent vessel or the occurrence of continued blood flow into the aneurysm in a region with poor occlusion. The latter, especially when close to the aneurysm wall, might trigger an aneurysm-regrowth \citep{mascitelli2015update, kim2021recurrence}.\\

Computational simulations serve as a tool to provide deep insight into the complicated procedures taking place during aneurysm occlusion procedures and subsequent healing mechanisms. In \citep{Horn2018}, a system  of 28 advection–diffusion–reaction partial differential equations is used to model the clotting induced by bare metal coils in cerebral aneurysms. They show that a more comprehensive model of the clotting cascade results in more realistic thrombus growth when compared to simplified models. The influence of stent induced vessel deformation on the hemodynamics of intracranial aneurysms is investigated in \citep{Sabernaeemi2023}. Here the computational fluid dynamics methods provide new insight into the flow patterns triggered by stent assisted coiling and show that the deformation of the aneurysm considerably decreases the wall shear stress on the aneurysm wall due to limited entrance of blood into the aneurysm. The occlusion capabilities of memory polyurethane foams in cerebral aneurysms have been simulates in \citep{Jarrah2021}. In their study they develop a model that selects the best fitting memory polyurethane foam for a patient specific aneurysm by considering thermo-mechanical responses on the foam.

In order to test different coiling configurations, insertion strategies as well as material properties already \textit{before} the clinical surgery, several mathematical models for coiling (and similar, wire-like) structures have been applied. These models can be grouped roughly into mechanically motivated and phenomenologically motivated ones.

Within the category of phenomenologically motivated models, dynamic path planning algorithms are employed to generate coil distributions in aneurysms. Their major advantage is that they have a relatively low computational complexity. However, they do mostly neglect structure mechanical effects. The method presented in \citep{morales2012virtual} applies such algorithms to achieve high packing densities of coils, reaching up to $\SI{30}{\percent}$ in aneurysms. In \citep{Patel_2021}, the Authors noted that such methods produce artifacts such as very small loops and kinks that are nonphysical. They improved them by developing the \textit{pre-shape path planning algorithm} which considers the natural shape of the coil by including its tendency to minimize the strain energy. This yields more realistic coil placements.

Most of the mechanically motivated models are based on the theory of elastic beams. One of the first models was proposed in \citep{dequidt2008interactive} and used Timoshenko beam finite elements. This model allows taking curved rest shapes into account and was a first step towards realistic coil simulations. In
\citep{Babiker_2013}, the model is used to generate coil-deployments which are then incorporated into CFD simulations. This allowed to analyze the post-treatment hemodynamics and occlusion properties. Since then several studies have been performed to improve upon the modelling and statistical analysis. The works \citep{otani2015computational, otani2020modeling} focus on external forces and internal friction during the coil deployment. Further investigations of post treatment hemodynamics were made in \citep{fujimura2016hemodynamic} to analyze the reduction of the flow velocity triggered by different coil distributions. In \citep{fujimura2016hemodynamic, otani2017computational} the small-scale coil geometry was replaced by a volumetric porosity field in combination with a porous media model reducing the computational cost of a fully resolved free flow simulation. To reduce the complexity of a beam element model, a spring mass system was introduced in \citep{otani2017computational}. However, it is not straightforward to include the imprinted natural shape of the coil within this approach.

Our approach combines the advantages of both, the beam element and the spring mass, approaches. We adapt the ``\textit{Discrete Elastic Rods}'' (DER) approach to the problem of generating realistic coil deployments. Contacts are modelled in an explicit manner and are efficiently detected. Our model enables us to simulate a large number of coil deployments in parallel, which will then be used in a subsequent statistical analysis. To perform this analysis, we introduce an in silico Raymond Roy \citep{greve2021initial, mascitelli2015update} type classification for the deployments with respect to their occlusion quality. We account for material and geometric uncertainties by introducing small variations in the coils material parameters as well as in the initial placement location.\\

The rest of the paper is structured as follows: In Section \ref{sec:methodology}, we derive the mathematical model used for our coiling simulations including the data and the numerical methods that are employed. Section \ref{sec:sensitivity-study} then describes our approach towards a statistical occlusion analysis motivated by the Raymond Roy Occlusion Classification. Numerical results are discussed in Section \ref{sec:SimulationOfCoilPlacements} including both, specific coil deployment simulations validating our model against actual deployments as well as results of the statistical analysis. In Section \ref{sec:conclusion}, we then give a conclusion of our findings.


\section{Mathematical modelling}
\label{sec:methodology}
This section starts by introducing the data basis of the three aneurysm shapes that we are using in this study. Following that, we delve into an in depth discussion of the mechanical model used for our coil deployment simulations including its numerical realization and algorithmic aspects.

\subsection{Aneurysm geometry}
\begin{figure*}[!htb]
    \centering
    \begin{subfigure}[t]{0.32\textwidth}
    \centering
        \includegraphics[height=4cm]{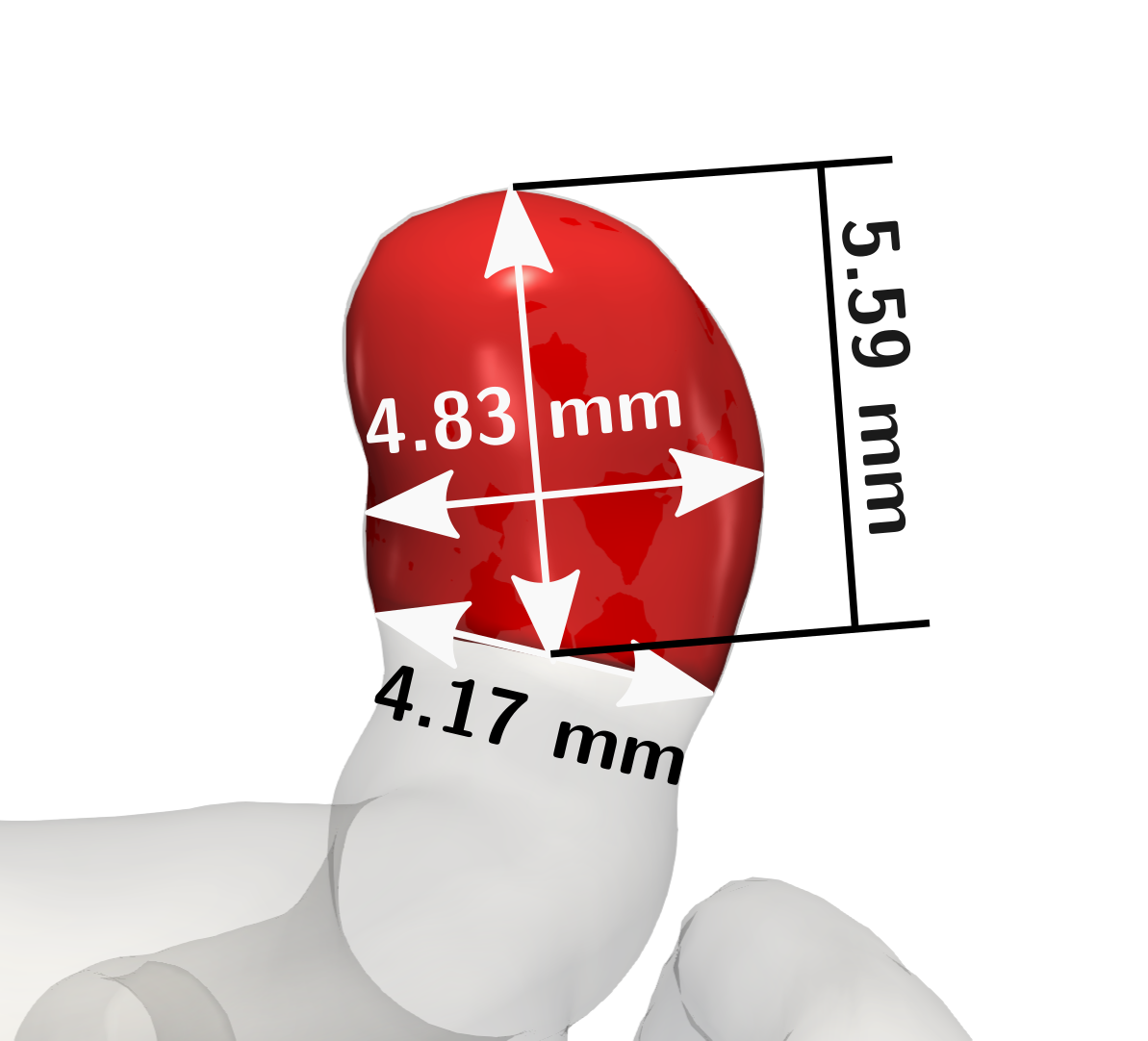}
        \subcaption{}
        \label{fig:Geom_Small_1}
    \end{subfigure}
    \begin{subfigure}[t]{0.32\textwidth}
    \centering
        \includegraphics[height=4cm]{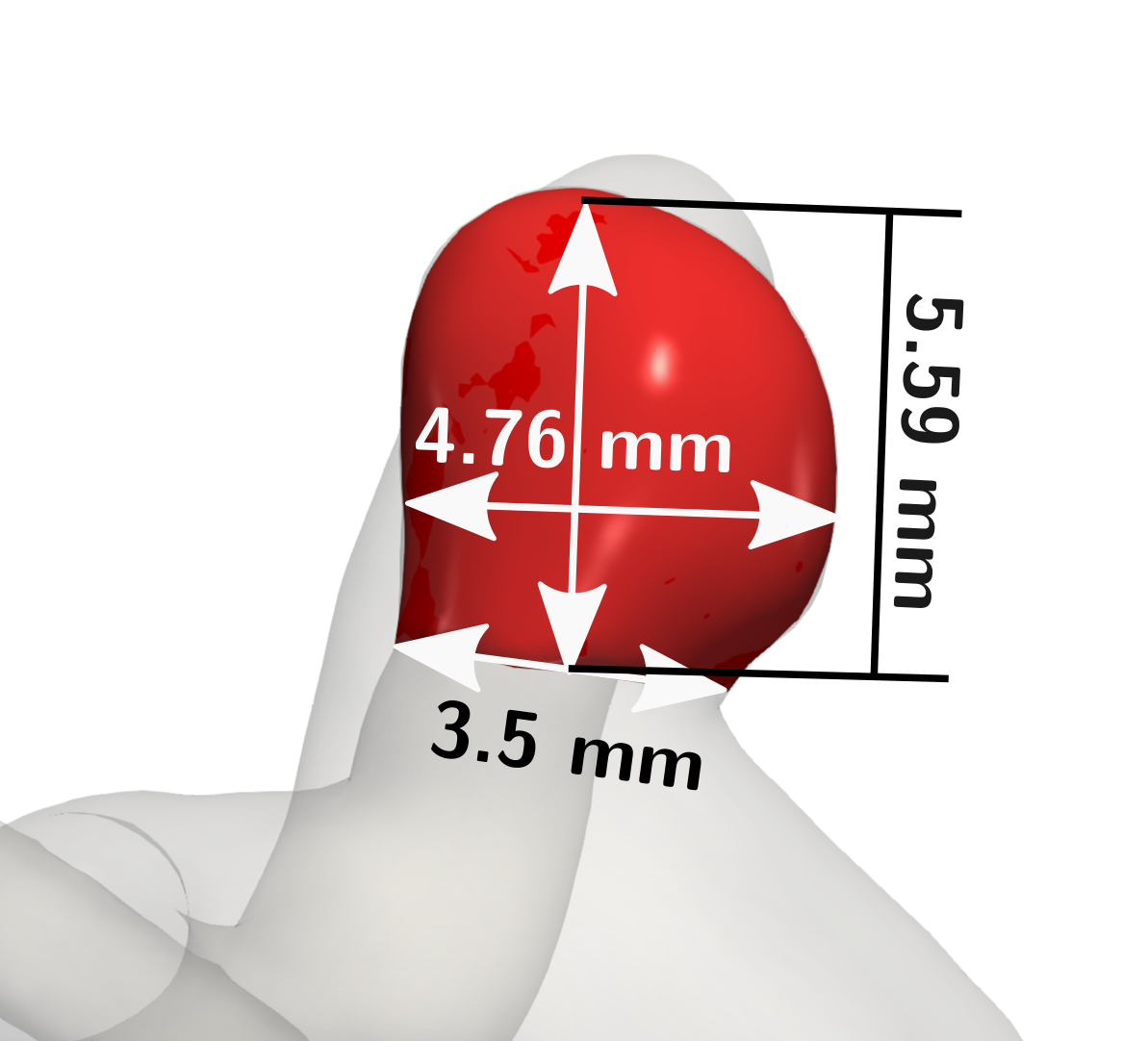}
        \subcaption{}
        \label{fig:Geom_Small_2}
    \end{subfigure}
    \begin{subfigure}[t]{0.32\textwidth}
    \centering
        \includegraphics[height=4cm]{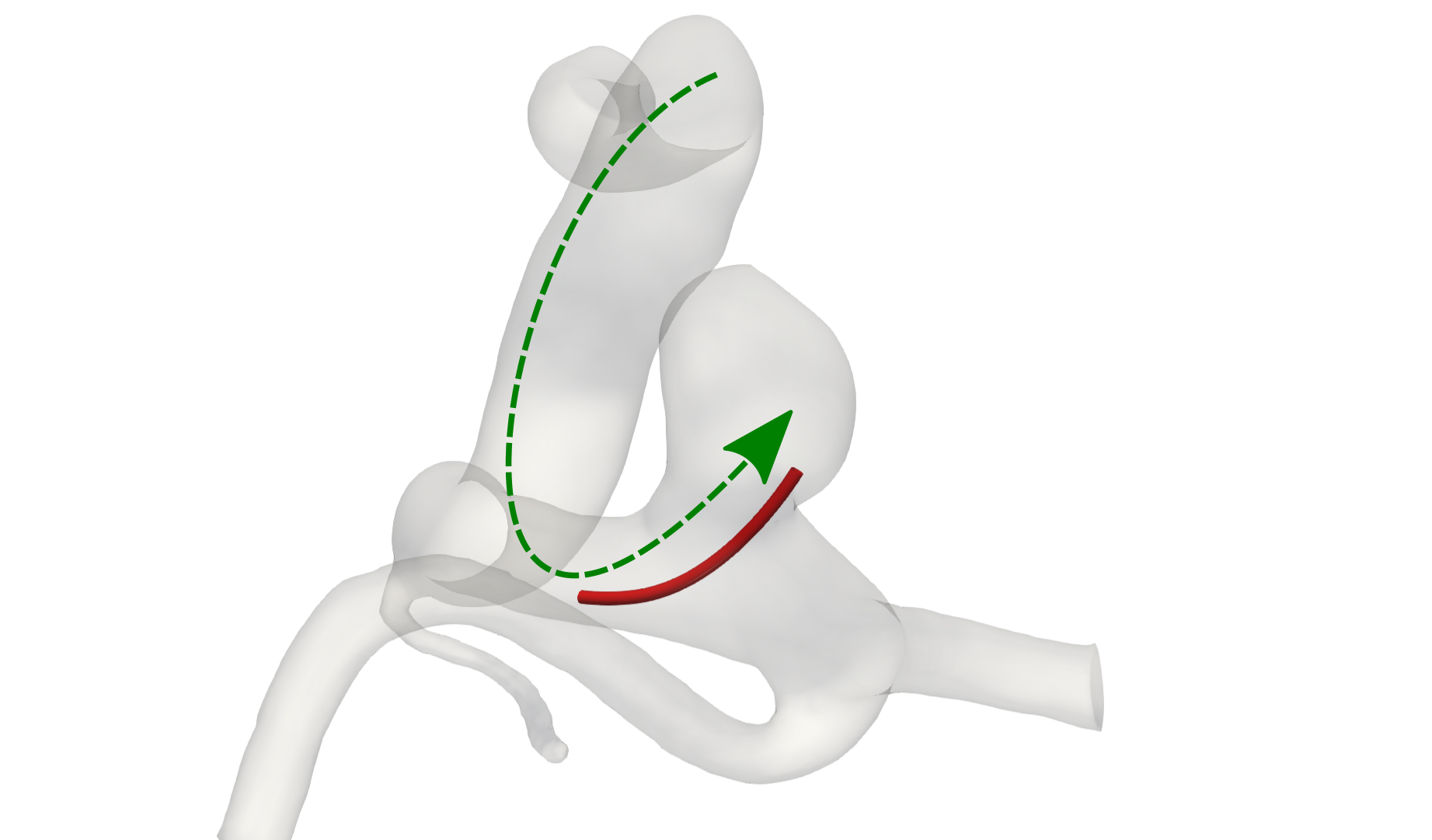}
        \subcaption{}
        \label{fig:Geom_Small_3}
    \end{subfigure}\\
    \begin{subfigure}[t]{0.32\textwidth}
    \centering
        \includegraphics[width=4cm]{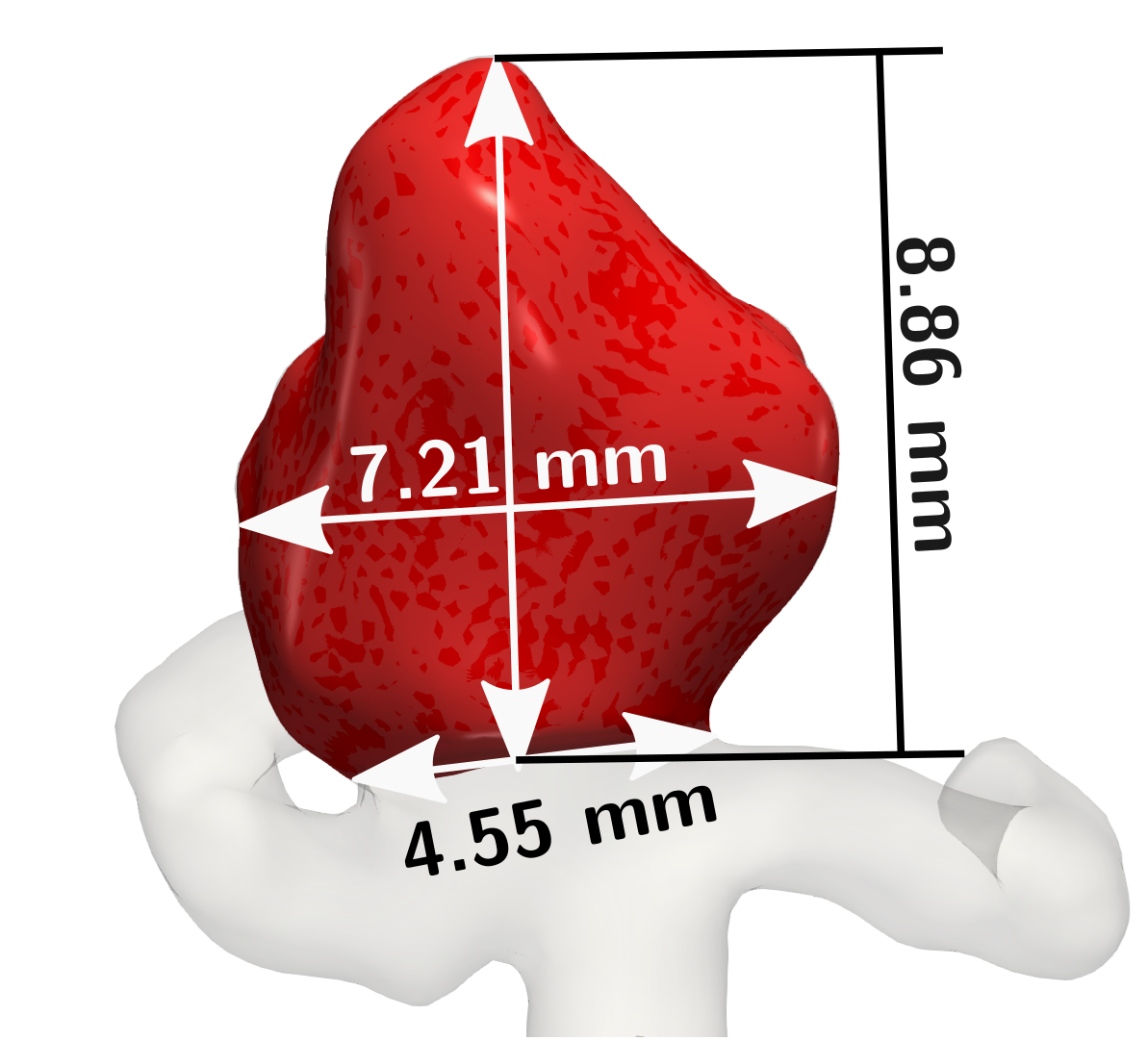}
        \subcaption{}
        \label{fig:Geom_Bir_1}
    \end{subfigure}
    \begin{subfigure}[t]{0.32\textwidth}
    \centering
        \includegraphics[width=4cm]{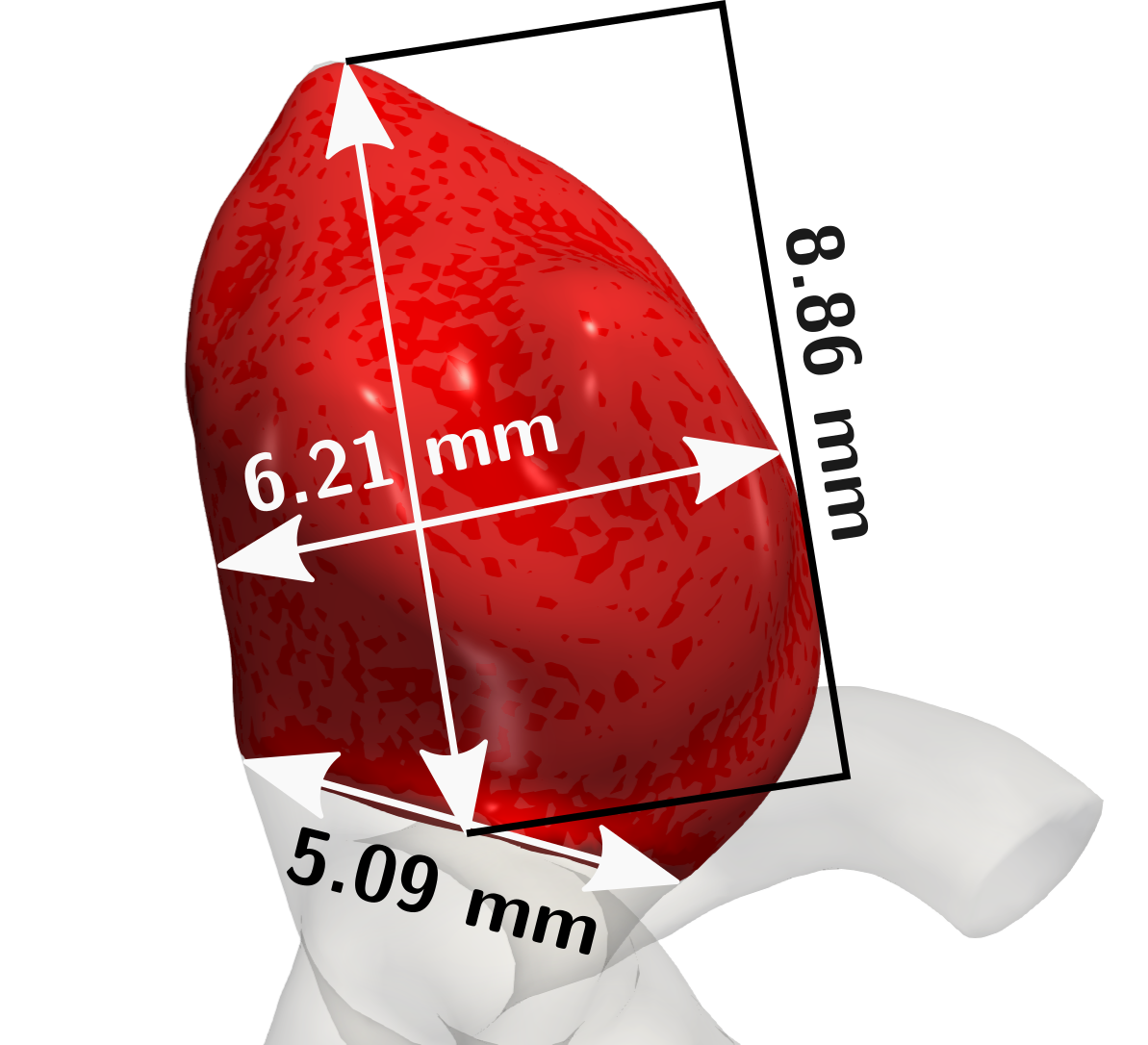}
        \subcaption{}
        \label{fig:Geom_Bir_2}
    \end{subfigure}
    \begin{subfigure}[t]{0.32\textwidth}
    \centering
        \includegraphics[width=4cm]{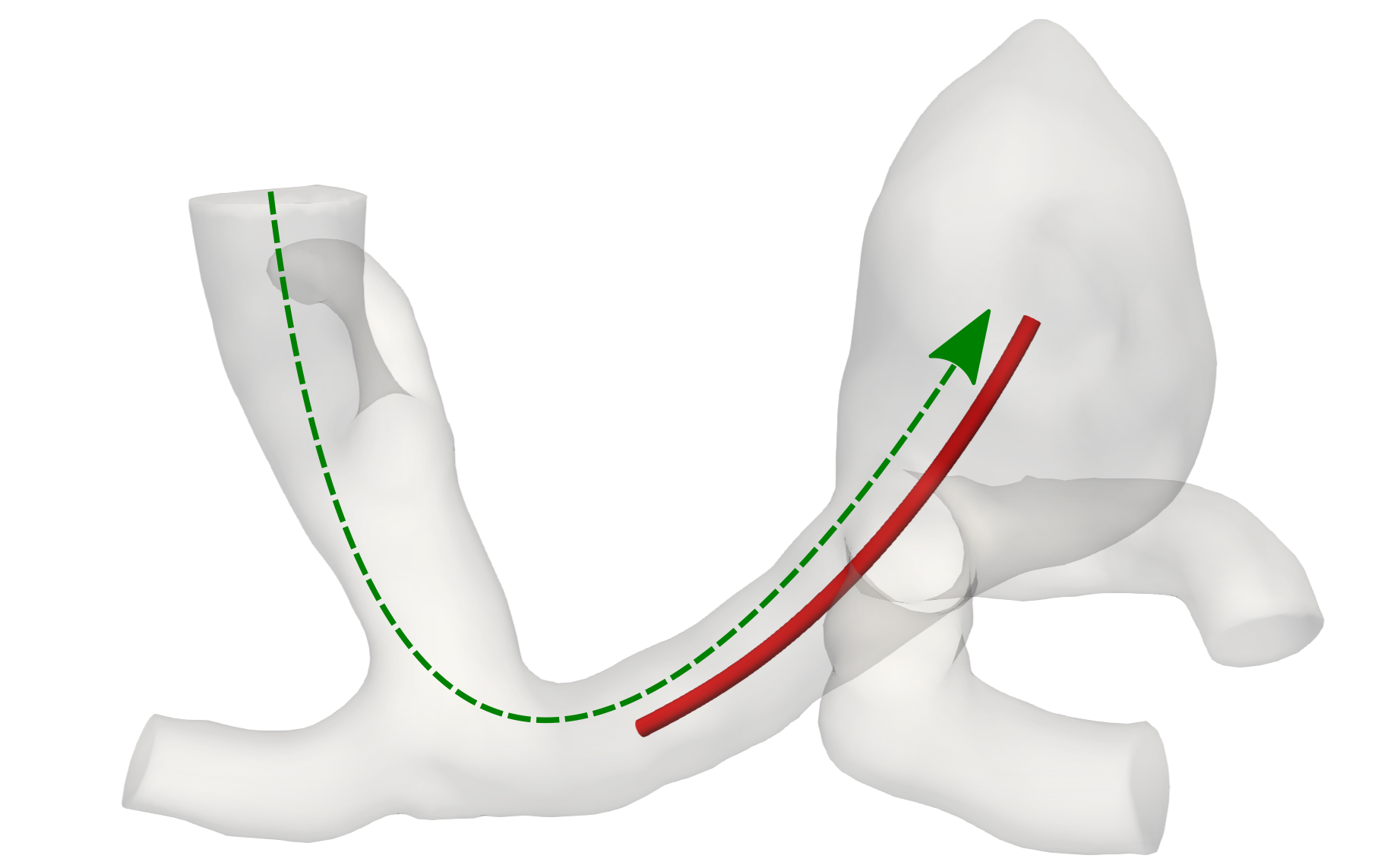}
        \subcaption{}
        \label{fig:Geom_Bir_3}
    \end{subfigure}\\
    \begin{subfigure}[t]{0.32\textwidth}
    \centering
        \includegraphics[width=4cm]{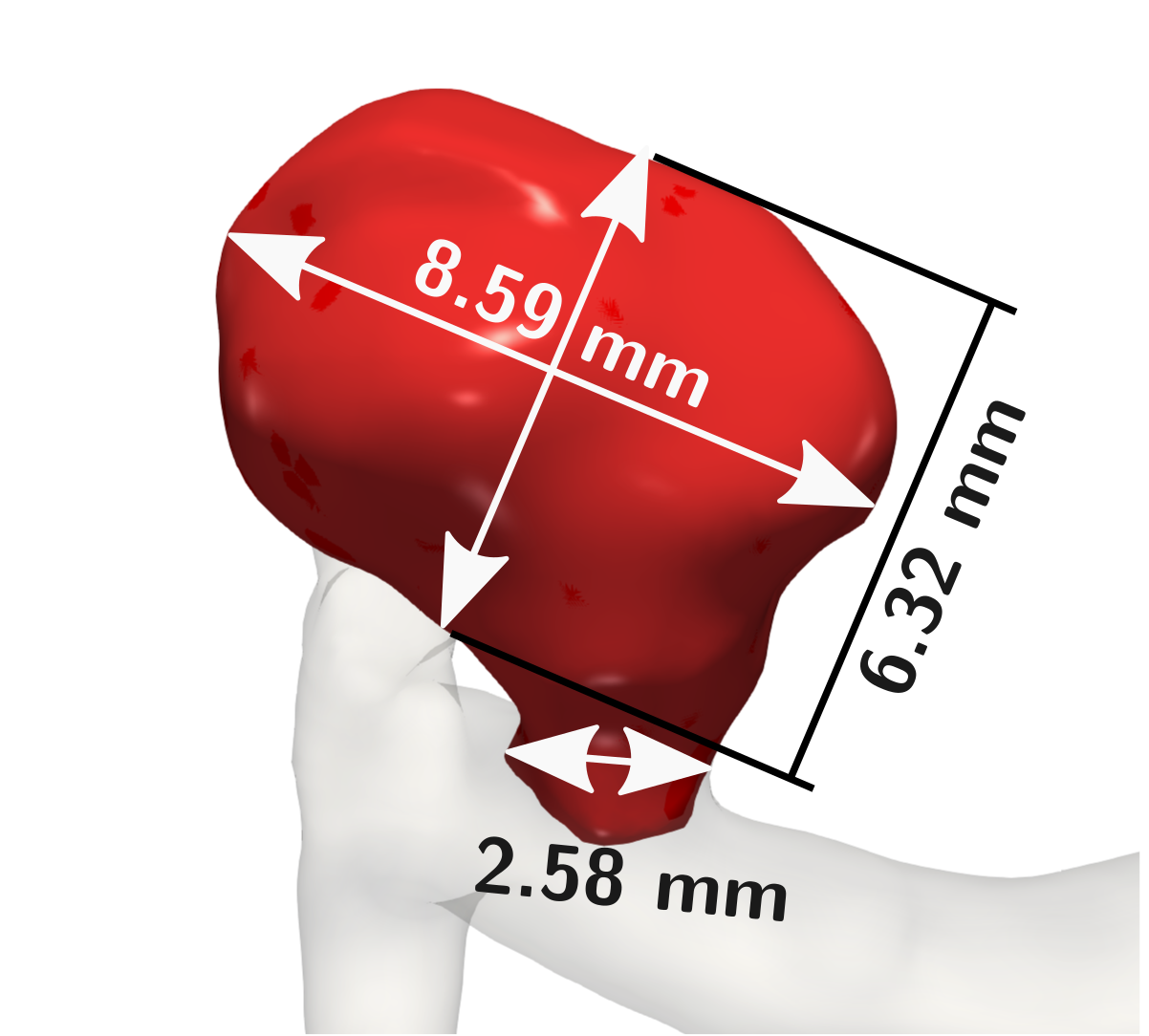}
        \subcaption{}
        \label{fig:Geom_Gamm_1}
    \end{subfigure}
    \begin{subfigure}[t]{0.32\textwidth}
    \centering
        \includegraphics[width=4cm]{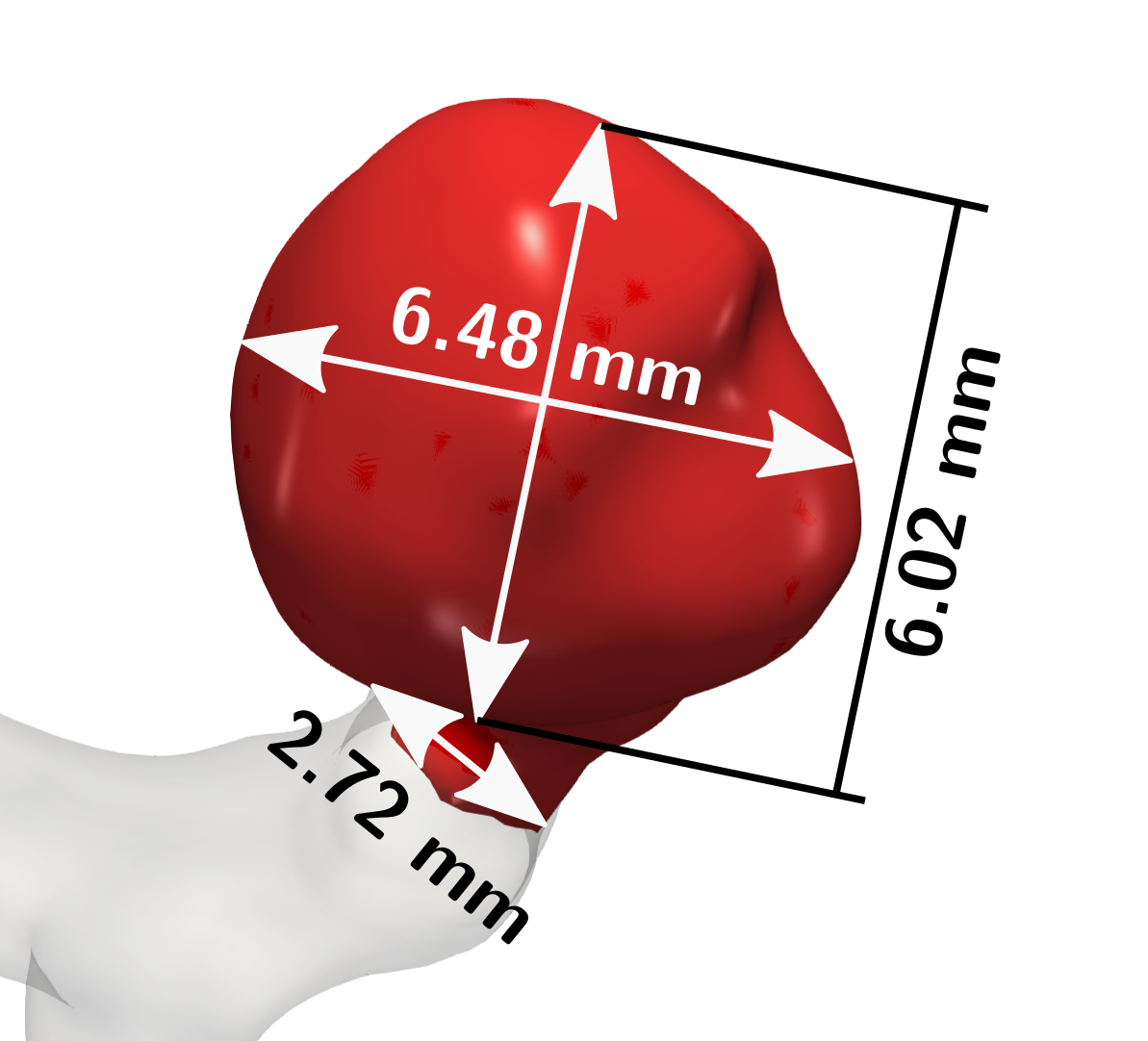}
        \subcaption{}
        \label{fig:Geom_Gamm_2}
    \end{subfigure}
    \begin{subfigure}[t]{0.32\textwidth}
    \centering
        \includegraphics[width=4cm]{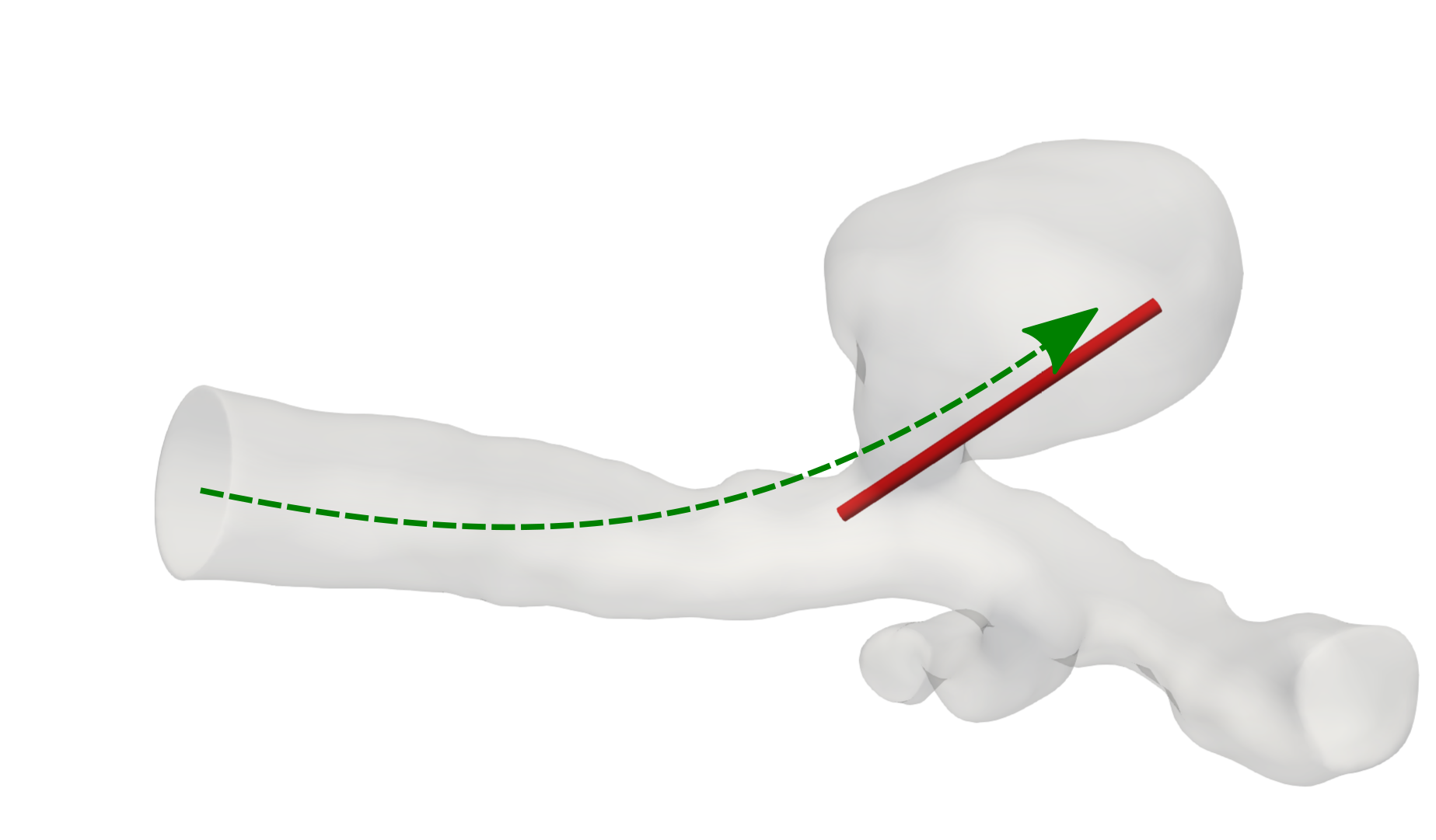}
        \subcaption{}
        \label{fig:Geom_Gamm_3}
    \end{subfigure}\\
    \caption{Morphological features of the three aneurysms considered in this study. (\subref{fig:Geom_Small_1})--(\subref{fig:Geom_Small_2}) is denoted as \textit{small} aneurysm, (\subref{fig:Geom_Bir_1})--(\subref{fig:Geom_Bir_2}) is a \textit{bifurcation aneurysm} and (\subref{fig:Geom_Gamm_1})--(\subref{fig:Geom_Bir_2}) a \textit{narrow neck aneurysm}. From the left to the middle column the perspective is changed by a rotation of 90 degrees. The neck width, dome width and height are given. In the right column, at (\subref{fig:Geom_Small_3}), (\subref{fig:Geom_Gamm_3}), (\subref{fig:Geom_Bir_3}), the insertion direction of the catheter is sketched in green and the position of the catheter is shown in red.}
    \label{fig:Aneurisms}
\end{figure*}

In this study, we consider the three aneurysms as given in Fig. \ref{fig:Aneurisms} which where obtained from \citep{pozosoler2017}. For convenience of the reader, we have annotated each of the aneurysms by its neck width, dome diameter as well as its height, as it is usually done before an actual clinical intervention. We can classify each of them by their size, see, e.g., \citep{wiebers2003unruptured}, with \SI{2}{\milli \meter}--\SI{7}{\milli \meter} as \textit{very small}, \SI{7}{\milli \meter}--\SI{12}{\milli \meter} as \textit{small} and \SI{12}{\milli\meter}--\SI{30}{\milli\meter} as \textit{large}.

The top case in Fig. \ref{fig:Aneurisms} is an example of a very small aneurysm, the middle case is a bifurcation aneurysm and an intermediate case between a very small and a small one. The bottom case stands out since it has a relatively small neck and is in the size range of a small aneurysm. Throughout the paper, we refer to these cases as follows: ``\textit{small aneurysm}'' for the first one, ``\textit{bifurcation aneurysm}'' for the second one and ``\textit{narrow neck aneurysm}'' for the last one mentioned. The geometries where chosen in such a way that they represent sufficiently different but relevant cases to showcase that our coiling model is applicable to a wide range of cases. This means that the dome sizes are chosen with regard to the occurrence frequency when considering treatment studies \citep{wermer2007risk}, where the category ``\textit{very small}'' constitutes the vast majority, followed by ``\textit{small}'' ones and ``\textit{large}'' ones being much less frequent.

The raw geometries from the database \citep{pozosoler2017} are associated by default with a triangular surface mesh. For the embolization of the coil, we initially cut off the aneurysm from the parent artery at the approximate position of the neck (see red section in Fig. \ref{fig:Aneurisms}). The coil is then inserted into the separated aneurysm. Finally, after the coil is fully inserted, we enable the full geometry to allow a movement into the direction of the parent artery. For the modification of the mesh, we use blender~\footnote{\url{https://www.blender.org/}} and for the re-meshing meshlab~\footnote{\url{https://www.meshlab.net/}}. The micro-catheter, through which the coil is pushed into the aneurysm, is considered as an obstacle in our simulation. We model it as a B-spline consisting of three nodes. The position of the micro-catheter is fixed as shown in Fig. \ref{fig:Aneurisms} (right column) if not stated otherwise.

\subsection{Strain energy in elastic rods} \label{sec:coil_model}
\begin{figure}[!htb]
    \centering
    \includegraphics[width=0.4\textwidth]{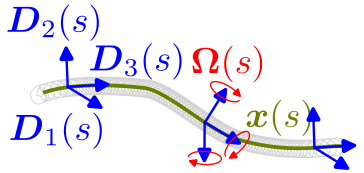}
    \caption{A space curve $\boldsymbol{x} $ parameterized by its arc-length $s$ together with the material directors $\boldsymbol{D}_{1}(s), \boldsymbol{D}_{2}(s), \boldsymbol{D}_{3}(s)$ and their corresponding Darboux-vector $\boldsymbol{\Omega}(s)$.}
    \label{fig:CoilContiCurve}
\end{figure}

We assume that the coil can be modelled by Kirchhoffs theory of elastic rods as presented in \citep{audoly2000elasticity}. Herein a three dimensional space curve $\boldsymbol{x}:[0,L]\rightarrow\mathbb{R}^3,~s\mapsto\boldsymbol{x}(s)$ with arc-length parameter $s\in [0,L]$ is used to trace the center-line of a rod (see Fig. \ref{fig:CoilContiCurve}). 
For each position on the space curve, we attach a right-handed frame of orthonormal directors $\boldsymbol{D}_{1}(s), \boldsymbol{D}_{2}(s) \textup{ and } \boldsymbol{D}_{3}(s)$, each $\in\mathbb{R}^3$ used as columns to form the matrix $\big\{\boldsymbol{D}_{1}(s), \boldsymbol{D}_{2}(s), \boldsymbol{D}_{3}(s)\big\}\in \operatorname{SO}(3)$. They represent the orientation of the rods cross-section and are material to the curve, hence they are called material directors.  As shown in \citep{eddleman2012endovascular}, an extension of the coil while surgery may lead to serious complications and has to be suppressed by the design of the coil. Thus, we assume that our rod is inextensible, with $\|\boldsymbol{x}'(s)\|=1$. 
Further the material frame is assumed to be adapted, i.e., $\boldsymbol{D}_{3}(s) = \boldsymbol{x}'(s)$ at each position $s$ (the arc-length derivative is denoted by $\partial(\cdot)/\partial s=(\cdot)'$). This expresses that no shear occurs and implies that bending and twisting are the most significant modes of deformation.
A standard concept from differential geometry is the Darboux-vector $\boldsymbol{\Omega(s)}$ defined by
\begin{align}
    \boldsymbol{D}_i'(s) = \boldsymbol{\Omega}(s) \times  \boldsymbol{D}_i(s) \quad \forall i \in\{1,2,3\}
    \label{EQ:Darboux}
\end{align}
which can be represented as $\boldsymbol{\Omega}(s) = \kappa_1(s)  \boldsymbol{D}_1(s) + \kappa_2(s)  \boldsymbol{D}_2(s) + \tau(s)  \boldsymbol{D}_3(s)$; we refer to \citep{audoly2000elasticity}. We interpret $\boldsymbol{\Omega}(s)$ as the rotation velocity around the directors. By inserting $\boldsymbol{\Omega}(s)$ into the Frenet-Serret equations \citep{audoly2000elasticity}, the elastic strains for the curvature $\boldsymbol{\kappa}=(\kappa_1,\kappa_2)^T$ and the torsion / twist $\tau$ can be obtained from (\ref{EQ:Darboux}) as
\begin{align}
    \boldsymbol{D}'_3 \cdot \boldsymbol{D}_1 = - \kappa_2,
\boldsymbol{D}'_3 \cdot \boldsymbol{D}_2 =  \kappa_1,
\boldsymbol{D}'_2 \cdot \boldsymbol{D}_1 =  \tau.
    \label{EQ:StrainsDarboux}
\end{align}
Here we applied the identity $\boldsymbol{a}\cdot (\boldsymbol{b}\times \boldsymbol{c}) = \boldsymbol{b}\cdot (\boldsymbol{c}\times \boldsymbol{a})$.
Having expressions for the elastic strains, we are in the position to formulate the elastic potential of the rod as
\begin{align}
    E(\boldsymbol{x}) = \frac{1}{2}\int\limits_{0}^L \boldsymbol{\kappa}^T\boldsymbol{B}\boldsymbol{\kappa} + \beta \tau^2 \operatorname{ds},
    \label{EQ:StrainEnergyConti}
\end{align}
where $\boldsymbol{B}\in \mathbb{R}^{2\times 2}$ and $\beta\in\mathbb{R}$ are dependent on the material parameters and the geometry of the cross-section.

\begin{figure}[!htb]
    \centering
    \includegraphics[width=0.4\textwidth]{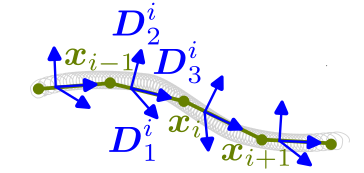}
    \caption{A discrete space curve with material points $\boldsymbol{x}_i$ for $i \in \{1,...,N\}$ and edges $\boldsymbol{e}^j=\boldsymbol{x}_{j+1}-\boldsymbol{x}_{j}$ for $j \in \{1,...,N-1\}$ on which the material frames $\boldsymbol{D}_{1}^j, \boldsymbol{D}_{2}^j, \boldsymbol{D}_{3}^j$ are attached.}
    \label{fig:CoilDiscontiCurve}
\end{figure}

From this point on, it is standard to obtain the continuous Kichhoff equations for the equilibrium of an elastic rod by a variational argument, see, e.g., \citep{audoly2000elasticity}. We follow the approach of \citep{bergou2008discrete, bergou2010discrete} called "Discrete Elastic Rods" (DER) which discretizes the strain energy directly. We ommit many details in our short review of the DER but refer the interested reader to a recent primer on the topic \citep{jawed2018primer}. We start by discretizing the rod as a finite set of $N$ material points $\boldsymbol{x}_i := \boldsymbol{x}(s_i)$ for $i\in\{0,1,...,N-1\}$, where two subsequent points are connected by an edge $\boldsymbol{e}^i=\boldsymbol{x}_{i+1}-\boldsymbol{x}_i$ (see Fig.\ref{fig:CoilDiscontiCurve}), following the notation of \citep{bergou2008discrete, bergou2010discrete}, we denote node quantities by subscript-indices and edge quantities by superscript-indices. 
For each edge, we define one material frame $\{ \boldsymbol{D}_1^i, \boldsymbol{D}_2^i, \boldsymbol{D}_3^i \} \in \operatorname{SO}(3)$ for $i\in\{0,1,...,N-1\}$, where the constraint $\boldsymbol{D}_3(s)=x'(s)$ is explicitly fulfilled by letting the third director be the tangent $\boldsymbol{t}^i:=\boldsymbol{e}^i/\| \boldsymbol{e}^i \| = \boldsymbol{D}_3^i$ for all $i\in\{0,1,...,N-1\}$. The inextensibility constrained can be satisfied by adding an axial strain component in the stretching energy, see (\ref{EQ:StrainEnergyDisc}). The strain energy in (\ref{EQ:StrainEnergyConti}) considers only rods that are pre-shaped straight when no external forces act. For medical coils, it is important to drop this assumption since they are naturally curved \citep{eddleman2013endovascular, wallace2001, ito2018experimental, white2008coils}. 
For our upcoming discussion, barred quantities are used to refer to the curve in its natural position $\overline{\boldsymbol{x}}$ while quantities that are not barred refer to the curve in deformed configurations. We denote by $\overline{\boldsymbol{e}}^j=\overline{\boldsymbol{x}}_{j+1}-\overline{\boldsymbol{x}}_{j}$ the edge vector in the natural position and by $\overline{l}_i=1/2(\|\overline{\boldsymbol{e}}^{i-1}\| + \|\overline{\boldsymbol{e}}^{i}\|)$ the Voronoi length created by two consecutive edges in the natural position.
A curve $\boldsymbol{x}$, that is not straight in its natural curved position $\overline{\boldsymbol{x}}$, can have a curvature $\overline{\boldsymbol{\kappa}}$ and twist $\overline{\tau}$ different from the current curvature and twist $\boldsymbol{\kappa}, \tau$. This allows to write the enriched discrete strain energy in the following way
\begin{align}
E_{tot} &= 
\sum\limits_{j=0}^{n-2} \frac{1}{2}\alpha \big(\frac{\|\boldsymbol{e}^j\|}{\|\overline{\boldsymbol{e}}^j\|} - 1\big)^2 \|\overline{\boldsymbol{e}}^j\| \notag\\
&+\sum\limits_{i=1}^{n-2} \frac{1}{2 \overline{l}_i}(\boldsymbol{\kappa}_i - \overline{\boldsymbol{\kappa}}_i) \boldsymbol{B} (\boldsymbol{\kappa}_i - \overline{\boldsymbol{\kappa}}_i)^T \notag\\
&+ \sum\limits_{j=1}^{n-2}  \frac{1}{2 \overline{l}_j} \beta(\tau^j-\overline{\tau}^j)^2,
    \label{EQ:StrainEnergyDisc}
\end{align}
where the first term corresponds to the axial strain energy ensuring with $\alpha>0$ chosen sufficiently large the inextensibility of the rod (see Appendix \ref{app:ax_energy}). To find expressions for $\boldsymbol{\kappa}_i$ and $\tau^i$, we note that since $\boldsymbol{D}_3^i$ is adapted to the curve, the directors $\boldsymbol{D}_1^i, \boldsymbol{D}_2^i$ form a plane to which the curvature binormal $(\kappa \boldsymbol{b})$ is co-planar. By straight forward calculations as, demonstrated in \citep{jawed2018primer}, one can find an expression for $(\kappa \boldsymbol{b})$ at the node $i$ by: 
\begin{align}
(\kappa \boldsymbol{b})_i = \frac{2 \boldsymbol{t}^{i-1}\times \boldsymbol{t}^{i}}{1+\boldsymbol{t}^{i-1}\cdot \boldsymbol{t}^{i}}.
\end{align}
This expression is referred to as nodal discrete integrated curvature binormal.
With help of $(\kappa \boldsymbol{b})_i$, one derives the discrete nodal curvatures
\begin{align}
    &\kappa_{i1} = \frac{1}{2} \big(\boldsymbol{D}_2^{i-1} + \boldsymbol{D}_2^{i}\big) \cdot (\kappa \boldsymbol{b})_i, \notag\\
    &\kappa_{i2} = -\frac{1}{2} \big(\boldsymbol{D}_1^{i-1} + \boldsymbol{D}_1^{i}\big) \cdot (\kappa \boldsymbol{b})_i.
    \label{EQ:DiscNatCurv}
\end{align}
We note that this is the finite difference formulation for the curvatures in (\ref{EQ:StrainsDarboux}) since 
\begin{align}
\lim\limits_{\overline{\ell}\rightarrow 0} \frac{ \boldsymbol{t}(s-\overline{\ell}/2)\times \boldsymbol{t}(s+\overline{\ell}/2))}{\overline{\ell}} =\boldsymbol{t}'=\boldsymbol{D}'_3.
\end{align}

\begin{figure*}[!htb]
    \centering
    \begin{subfigure}{0.32\textwidth}
        \includegraphics[width=\linewidth]{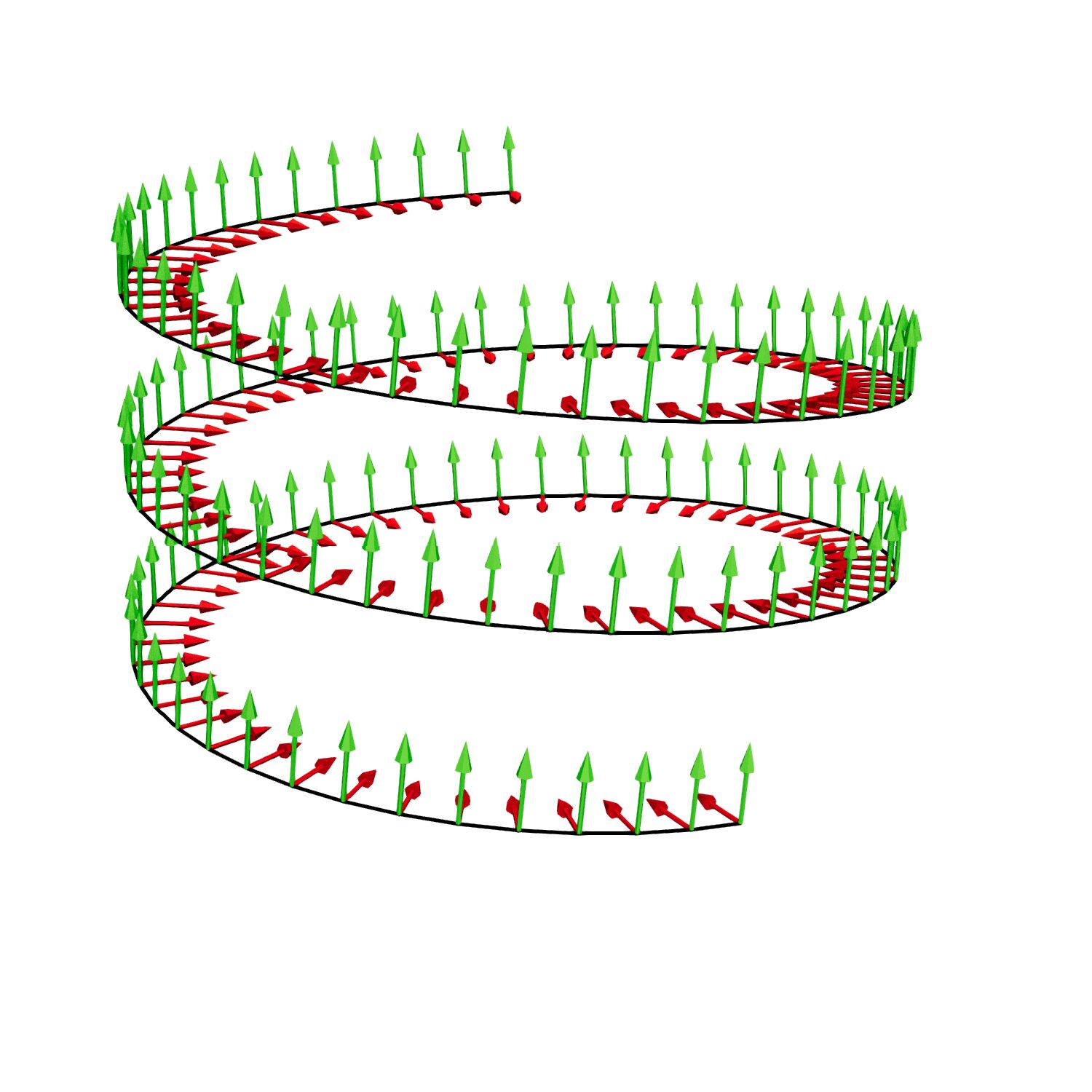}
        \caption{}
        \label{fig:Frenet}
    \end{subfigure}
    \begin{subfigure}{0.32\textwidth}
        \includegraphics[width=\linewidth]{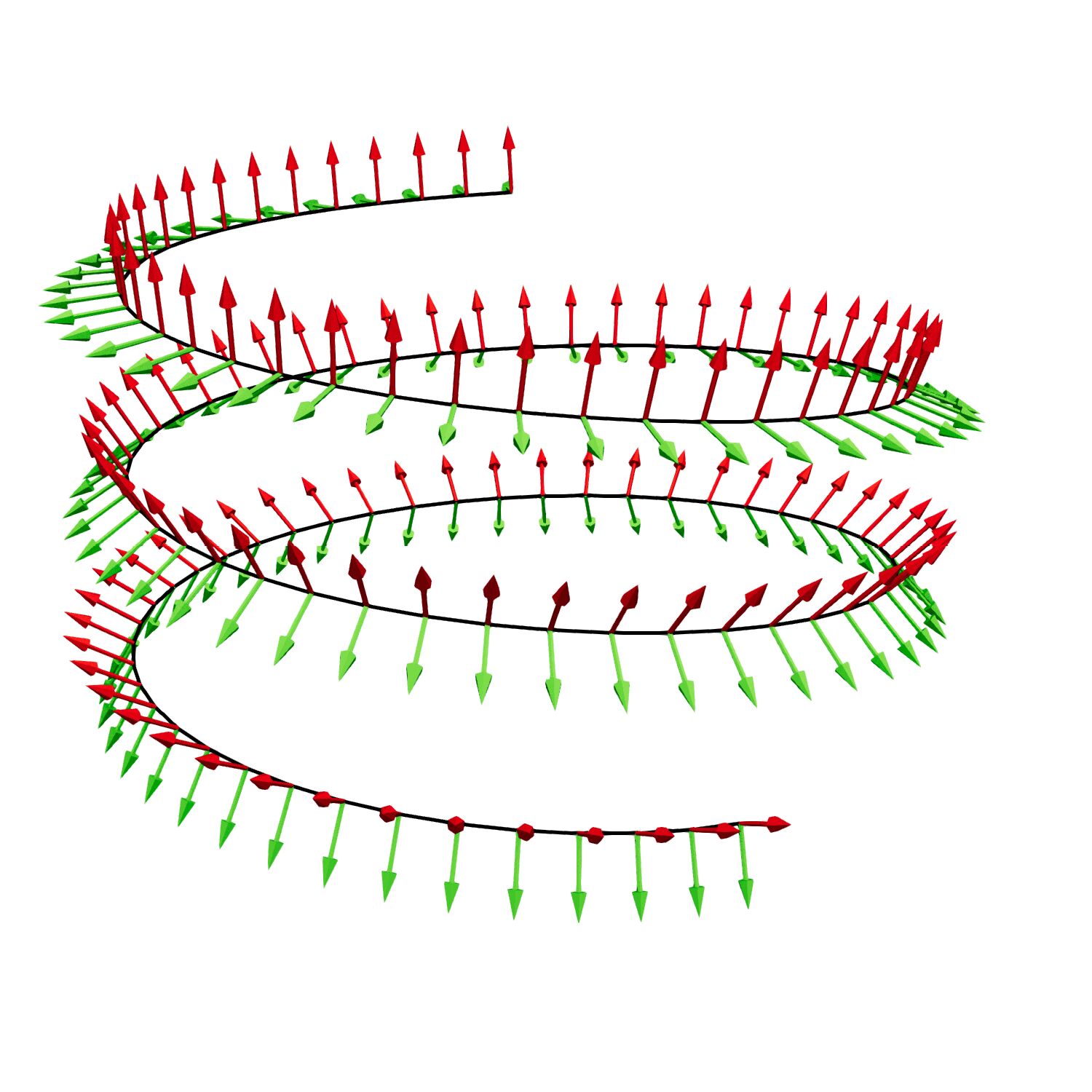}
        \caption{}
        \label{fig:Bishop}
    \end{subfigure}
    \begin{subfigure}{0.32\textwidth}
        \includegraphics[width=\linewidth]{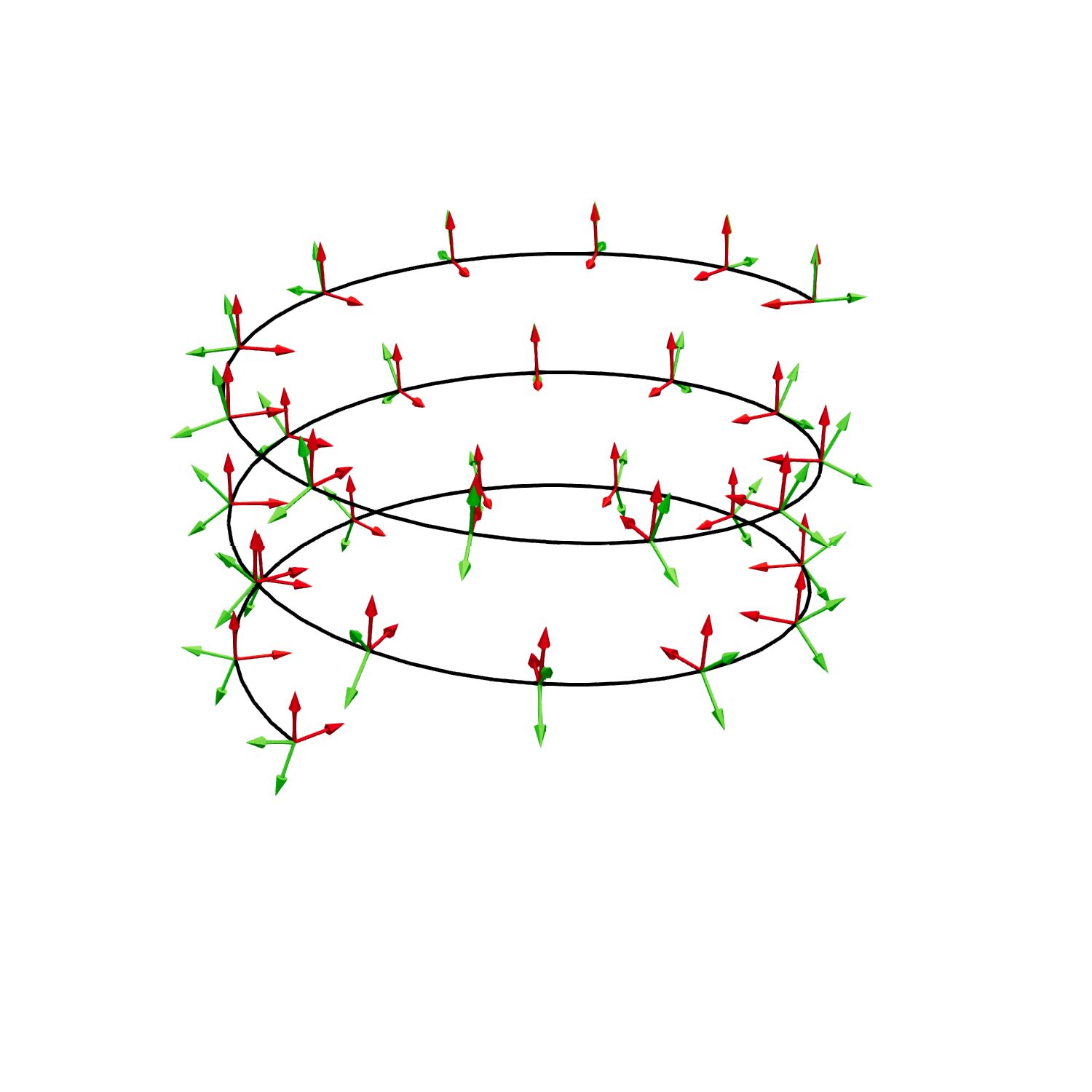}
        \caption{}
        \label{fig:Combined_Frames}
    \end{subfigure}
    \caption{Framing of a Helix where $\boldsymbol{D}_1$ is represented in green and $\boldsymbol{D}_2$ in red (\subref{fig:Frenet}) An example for a material Frame with the director $\boldsymbol{D}_1$ in green and $\boldsymbol{D}_2$ in red (\subref{fig:Bishop}) Bishop Frame with directors $\boldsymbol{U}_1$ in green and $\boldsymbol{V}_2$ in red. (\subref{fig:Combined_Frames}) In red, we show a material Frame of directors $\boldsymbol{D}_1,\boldsymbol{D}_2$ and in green a Bishop frame for the curve $\boldsymbol{U}_1,\boldsymbol{V}_2$. Since the Bishop frame is twist free we can rotate it on each edge into the material frame to get the absolute reference angle with respect to the Bishop frame. On two consecutive edges $(j,j+1)$, the difference in this angle is the twist of the material frame $\tau^j$.}
    \label{fig:Frames}
\end{figure*} 

For the discrete torsion, we first introduce the parallel transport given by the rotation matrix $\boldsymbol{R}_{\boldsymbol{t}^{j-1}}^{\boldsymbol{t}^{j}}$ that rotates the tangent of edge $j-1$ into the tangent of edge $j$. An explicit representation of the matrix follows from the Rodrigues formula \citep{dai2015euler}:
\begin{align}
    &\boldsymbol{R}_{\boldsymbol{t}^{j-1}}^{\boldsymbol{t}^{j}} = \boldsymbol{I} + \boldsymbol{K} + \boldsymbol{K}^2\frac{1}{1-(\boldsymbol{t}^{j-1}\cdot{\boldsymbol{t}^{j})}} \notag\\ 
    &\text{ with } \boldsymbol{K}\boldsymbol{v}=(\boldsymbol{t}^{j-1}\times{\boldsymbol{t}^{j})}\times \boldsymbol{v} \text{ for } \boldsymbol{v}\in\mathbb{R}^3.
    \label{EQ:Bishop_prop}
\end{align}
In \citep{bergou2008discrete}, the twist degree of freedom is parametrized with respect to the Bishop frame \citep{bishop1975}$\{\boldsymbol{U}^j,\boldsymbol{V}^j,\boldsymbol{D}_3^j\}$. It is for any continuous space-curve uniquely defined provided $\boldsymbol{U}^0,\boldsymbol{V}^0$ is fixed, see Fig. \ref{fig:Frames} which shows the frames on a helix. Applying the parallel transport operator, we find
\begin{align}
    \boldsymbol{U}^{j} = \boldsymbol{R}_{\boldsymbol{t}^{j-1}}^{\boldsymbol{t}^{j}}\boldsymbol{U}^{j-1},\quad 
    \boldsymbol{V}^{j} = \boldsymbol{R}_{\boldsymbol{t}^{j-1}}^{\boldsymbol{t}^{j}}\boldsymbol{V}^{j-1}.
\end{align}
In Fig. \ref{fig:Frames}, the difference between a frame that has a non zero twist and a torsion free Bishop frame is shown. In case of the  frame that has non zero twist, the first director points into the helix center for any point on the helix, which is only possible when then frame is rotated by a constant angle increment while traversing it. For the Bishop frame, we choose the first director at the lower end of the helix and then propagate it by (\ref{EQ:Bishop_prop}). This does not introduce any twist but causes the Bishop frame to turn upwards while propagating it along the curve.
One can show that the Bishop frame has always zero twist and can be used as a reference frame to measure the twist in the material directors $\{\boldsymbol{D}_1^j,\boldsymbol{D}_2^j,\boldsymbol{D}_3^j\}$. Therefore, we first measure the signed angle between either $\boldsymbol{D}_1^j$ and $\boldsymbol{U}^j$ or $\boldsymbol{D}_2^j$ and $\boldsymbol{V}^j$ which we call $\phi^j$.  In a second step, we calculate the discrete integrated twist as the increment $\tau_j=\phi^j-\phi^{j-1}$. This increment can also be directly expressed as the angle-increment to align $\boldsymbol{D}_i^{j}$ with $\boldsymbol{R}_{\boldsymbol{t}^{j-1}}^{\boldsymbol{t}^{j}}\boldsymbol{D}_i^{j-1}$ around $\boldsymbol{D}_3^{j}$ for $i\in\{1,2\}$ such that we have 

\begin{align}
    \boldsymbol{D}_1^{j} = \cos(\tau^j)\boldsymbol{R}_{\boldsymbol{t}^{j-1}}^{\boldsymbol{t}^{j}}\boldsymbol{D}_1^{j-1}
		       +\sin(\tau^j)\boldsymbol{R}_{\boldsymbol{t}^{j-1}}^{\boldsymbol{t}^{j}}\boldsymbol{D}_2^{j-1},\notag\\
\boldsymbol{D}_2^{j} = \cos(\tau^j)\boldsymbol{R}_{\boldsymbol{t}^{j-1}}^{\boldsymbol{t}^{j}}\boldsymbol{D}_2^{j-1}
		       -\sin(\tau^j)\boldsymbol{R}_{\boldsymbol{t}^{j-1}}^{\boldsymbol{t}^{j}}\boldsymbol{D}_1^{j-1}.
\end{align}
To close the discretization of the elastic energy, it remains to formulate expressions for the natural curvature and twist $\overline{\boldsymbol{\kappa}_i}$ and $\overline{\tau}_i$ of the rod, respectively. To this end, we employ again the Bishop frame as the material directors of the discrete-curve $\boldsymbol{x}$ in its natural shape $\overline{\boldsymbol{x}}_i$. In that way $\{\overline{\boldsymbol{D}}_1^{j},\overline{\boldsymbol{D}}_2^{j},\overline{\boldsymbol{D}}_3^{j}\}$ coincide with the Bishop frame of $\overline{\boldsymbol{x}}_i$. This allows us to measure the natural curvatures by (\ref{EQ:DiscNatCurv}), and moreover we note that the twist for this configuration is zero. We point out that the Bishop frame is not the only possible framing for the curve in its natural position. In fact any frame will lead to the same equilibrium position of the rod when minimizing the energy potential in (\ref{EQ:StrainEnergyDisc}) as shown in \citep{bertails2018inverse}.

\subsection{The coil dynamics}
To model the dynamics of the rod, we apply Newtons second law which yields a system of ordinary differential equations
\begin{align}
\begin{bmatrix}
 \boldsymbol{M}\ddot{\boldsymbol{X}} \\
  \boldsymbol{0}
\end{bmatrix}
+
\begin{bmatrix}
 \boldsymbol{D}_{\boldsymbol{X}} \dot{\boldsymbol{X}}\\
 \boldsymbol{D}_{\boldsymbol{\Phi}} \dot{\boldsymbol{\Phi}}
\end{bmatrix}
 = \underbrace{
-\begin{bmatrix}
 \nabla_{\boldsymbol{X}} E_{tot} \\
 \nabla_{\boldsymbol{\Phi}} E_{tot}
\end{bmatrix}
}_{\boldsymbol{F}_{int}} + 
\begin{bmatrix}
 \boldsymbol{F}_{ext} \\
 \boldsymbol{0}
\end{bmatrix},
\end{align}
with the vector  $ \boldsymbol{X}=\big(\boldsymbol{x}_0,\boldsymbol{x}_1,...,\boldsymbol{x}_{n-2},\boldsymbol{x}_{n-1} \big)$ containing the translational degrees of freedom (DOFs), and  $\boldsymbol{\Phi}=\big(\boldsymbol{\phi}_0,\boldsymbol{\phi}_1,...,\boldsymbol{\phi}_{n-2} \big)$ containing the rotational DOFs. $\boldsymbol{F}_{ext}$ are the external forces and moments due to wire-wire and wire-aneurysm-surface collisions. Expressions for $\nabla_{\boldsymbol{X}} E_{tot}$ are derived in \citep{jawed2018primer}, and $\nabla_{\boldsymbol{\Phi}} E_{tot}$ can be found in Appendix \ref{app:twist_mom}. $\boldsymbol{M}$ and $\boldsymbol{D}_{\boldsymbol{X}}, \boldsymbol{D}_{\boldsymbol{\Phi}} $ are the mass and damping matrices for translational and rotational DOFs, respectively. We assume that they are diagonal and set $\boldsymbol{M}=m\boldsymbol{I}$, $\boldsymbol{D}_{\boldsymbol{X}}=\eta_{\boldsymbol{X}}\boldsymbol{I}$, $\boldsymbol{D}_{\boldsymbol{\Phi}}=\eta_{\boldsymbol{\Phi}}\boldsymbol{I}$ with $\boldsymbol{I}$ denoting the identify matrix and $m$, $\eta_{\boldsymbol{X}},\eta_{\boldsymbol{\Phi}}$ denoting the masses that are lumped into the vertices and scalar damping coefficients. Note that we have set the mass matrices for the rotational DOFs to zero. To motivate this simplification, we assume that the node mass of the elastic rod is $m$ then the rotational moment of inertia is proportional to $m D_2^2 $ making it excessively small due to the small diameter $D_2$, hence we neglect it.  
In \citep{bergou2008discrete}, a similar simplification is applied with the difference that there the rotational moment and the rotational damping are set to zero resulting in $\nabla_{\boldsymbol{\Phi}} E_{tot}=0$. To ensure this constraint, one needs to optimize in each time-step of the numerical method the energy with respect to the rotational DOFs which is costly. In our implementation, we add the diagonal damping matrix $\boldsymbol{D}_{\boldsymbol{\Phi}}$ which when chosen sufficiently small approximates the condition $\nabla_{\boldsymbol{\Phi}} E_{tot}=0$. Further this enables us to circumvent the minimization of $E_{tot}=0$ with respect to $\boldsymbol{\Phi}$.
To solve the ODE system, we apply the symplectic Euler method \citep{hairer2006geometric} as time stepping scheme
\begin{subequations}
\begin{align}
    \boldsymbol{M}\dot{\boldsymbol{X}}(t_{n+1}) &= \Delta t \big( - \nabla_{\boldsymbol{X}} E_{tot}(t_n)+ \notag\\ 
    &\boldsymbol{F}_{ext}-\boldsymbol{D}_{\boldsymbol{X}}(t) \dot{\boldsymbol{X}}(t_n) \big) +  \boldsymbol{M}\dot{\boldsymbol{X}}(t_n),\label{EQ:SympEuler0}\\
\begin{bmatrix}
 \boldsymbol{X}(t_{n+1})\\
 \boldsymbol{\Phi}(t_{n+1})
\end{bmatrix}
 & = 
 \Delta t
\begin{bmatrix}
 \dot{\boldsymbol{X}}(t_{n+1}) \\
 - \boldsymbol{D}_{\boldsymbol{\Phi}}^{-1}\nabla_{\boldsymbol{\Phi}} E_{tot}(t_n)
\end{bmatrix}
+
\begin{bmatrix}
 \boldsymbol{X}(t_n)\\
 \boldsymbol{\Phi}(t_n)
\end{bmatrix}.
\label{EQ:SympEuler1}
\end{align}
\end{subequations}
Where $\Delta t:=t_{n+1}-t_n$ refers to the time step size of the discrete time steps $t_n$. Note that for the symplectic Euler method, we first solve (\ref{EQ:SympEuler0}) yielding $\dot{\boldsymbol{X}}(t_{n+1})$ that is then used in (\ref{EQ:SympEuler1}).

\subsection{Boundary / initial conditions for endovascular coiling}
Here we  formulate appropriate boundary conditions that approximate the real endovascular coiling procedure. We assume that the coil is pushed out of the micro-catheter at a constant speed $\boldsymbol{v}_{push}$. Let $I_{micro}$ be the set of node indices such that the corresponding nodes are still inside the micro-catheter at time $t_n$. Then for any node with index $i\in I_{micro}$, we set the velocity as $\dot{\boldsymbol{x}}_i=\boldsymbol{v}_{push}$ and for any node with index $i\not\in I_{micro}$, the velocity can be calculated from (\ref{EQ:SympEuler0})--(\ref{EQ:SympEuler1}). In the rotational degrees of freedom, we assume that the coil can freely rotate due to the lubrication therein. Note that this cannot lead to a blowup in the angular velocity since we have applied the damping $\boldsymbol{D}_{\Phi}$. We assume do nothing boundary conditions for the free ends of the coil. To do so, we apply different forces and moments on the first and last node and edge, respectively. The forces are stated in \citep{jawed2018primer} and the moments can be found in Appendix \ref{app:twist_mom}. 
As initial condition, we assume that the coil is placed in a straight configuration in front of the micro-catheter inlet pointing into the direction of the insertion velocity $\boldsymbol{v}_{push}/\|\boldsymbol{v}_{push}\|$. Additionally, the natural curve for the centerline of the coil is parametrized by a Bishop frame. Further we assume that the initial orientation of the directors is given as a twist free Bishop frame.

\subsection{Elastic spring constants of coils} 
\begin{figure}[!htb]
    \centering
    \includegraphics[width=0.5\textwidth]{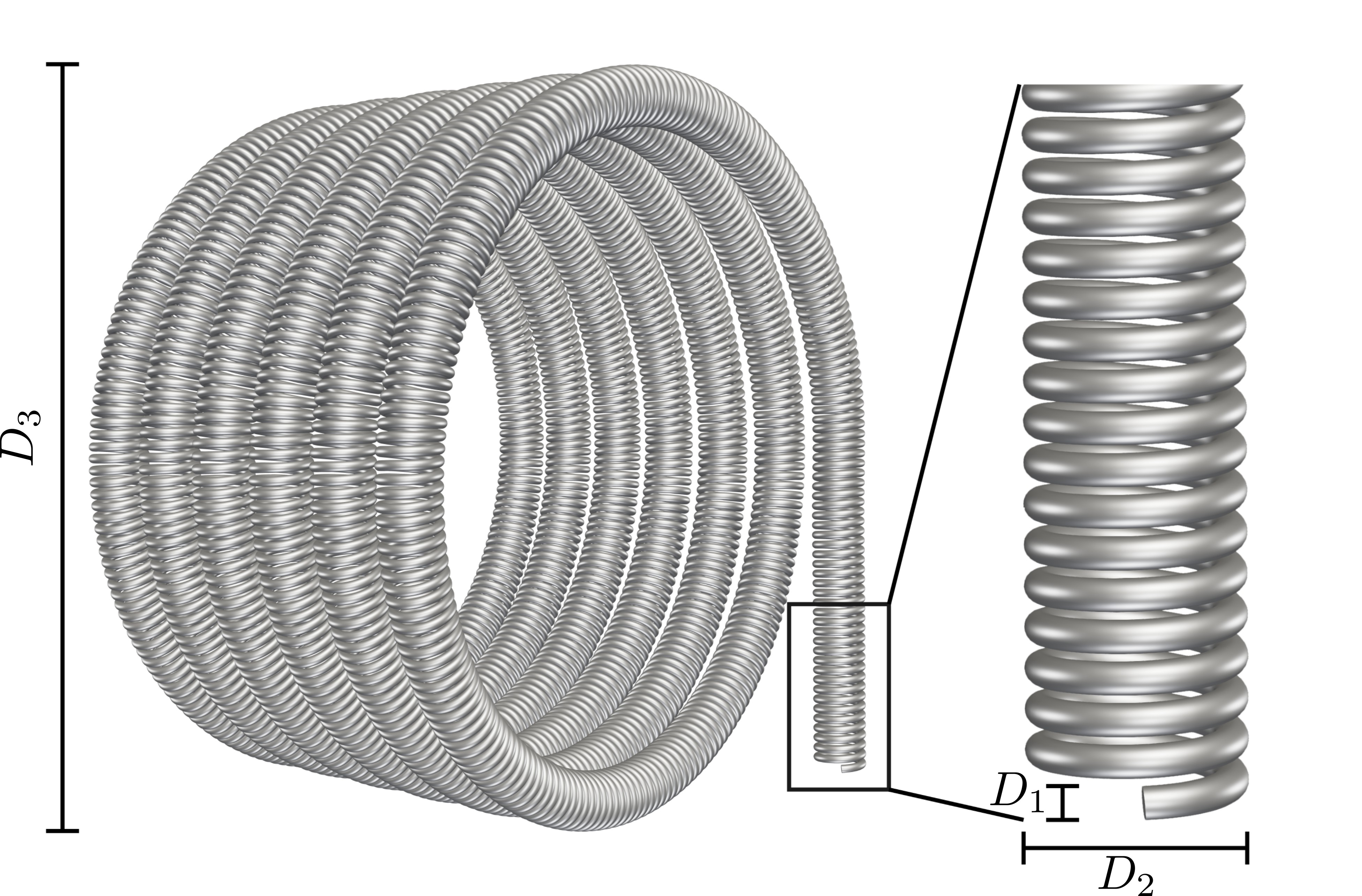}
    \caption{Characteristic lengths $D_1, D_2$ and $D_3$ of the coil with a natural shape corresponding to a helix. }
    \label{fig:CoilDims}
\end{figure}

The material parameters $\boldsymbol{B}=b\boldsymbol{I}$ and $\beta$ in the discrete energy potentials (\ref{EQ:StrainEnergyDisc}) can be obtained by integration over the rod cross section as described in \citep{jawed2018primer}. Considering the coil structure in detail, one identifies three characteristic length scales as shown in Fig. \ref{fig:CoilDims}. The first one called $D_1$ is the diameter of the coil stock wire which is then coiled up into a helical shape with radius $D_2$. The helical shape essentially makes up the coil. The natural curvature is then imprinted into the $D_3$ diameter. In our coiling model, we do not resolve the details of the stock wire. We rather consider the helical structure with $D_2$ as a rod and express the shape of the stock wire via the constants $b,\beta$.  This can be done by calculating the bending and twisting stiffness for a torsion string as done in \citep[Chapter 9,17]{wahl1964mechanical}, \citep{Otani_2020} which yields the following expressions
\begin{flalign}
&&b = \frac{E_w D_1^4p_c}{32(2+\mu_w) D_2}, && \beta = \frac{E_w D_1^4p_c}{64 D_2}.&&
\label{EQ:spring_constants}
\end{flalign}
In these $G_w$, $E_w$ and $\mu_w$ denote the stock wires Shear modulus, Young's modulus and Poisson ratio, respectively. The quantity $p_c$ is the distance between the center-line of each turn in the $D_2$ structure and defined by $p_c=D_1 p$ where $p>1$ is often refereed to as the pitch \citep{white2008coils}. In the remainder of this work we set $p=1.1$.

\subsection{Contact algorithm: collision detection}
The contacts in our simulation can be split up into contacts of the coil with the aneurysm wall and contacts of the coil with itself. This section describes how contacts are identified and collisions are modelled.

 Naively one can check for a coil with $N$ edges and a wall triangulated by $M$ facets if a coil self-collision occurs in $\mathcal{O}(N^2)$ and if a coil-wall collision occurs in $\mathcal{O}(NM)$. Since $N$ and $M$ are usually relatively large (at least $10^3$), a naive approach increases the computational complexity. Therefore, more advanced methods have been proposed \citep{eberly20063d, jimenez20013d} that only consider contacts in close vicinity by dividing the simulation domain into subsections. In this study, we use the octree method \citep{behley2015} that belongs to this class of methods. The simulation domain $\Omega$ is partitioned into a finite set of intervals; each of it roughly containing  the same number of contact objects. Crucial about the method is that the hierarchy of intervals can be represented in a tree structure. Therefore, the search for a contact partner of a certain object can be restricted to the partners in a specific interval which is found in the tree structure in logarithmic time complexity. In our case, the time complexity for coil-wall collision detection is reduced to $\mathcal{O}(N\log(M))$ and respectively for coil self collisions detection to $\mathcal{O}(N\log(N))$.

\begin{figure}[!htb]
    \centering
    \begin{subfigure}{0.4\textwidth}
        \includegraphics[width=\linewidth]{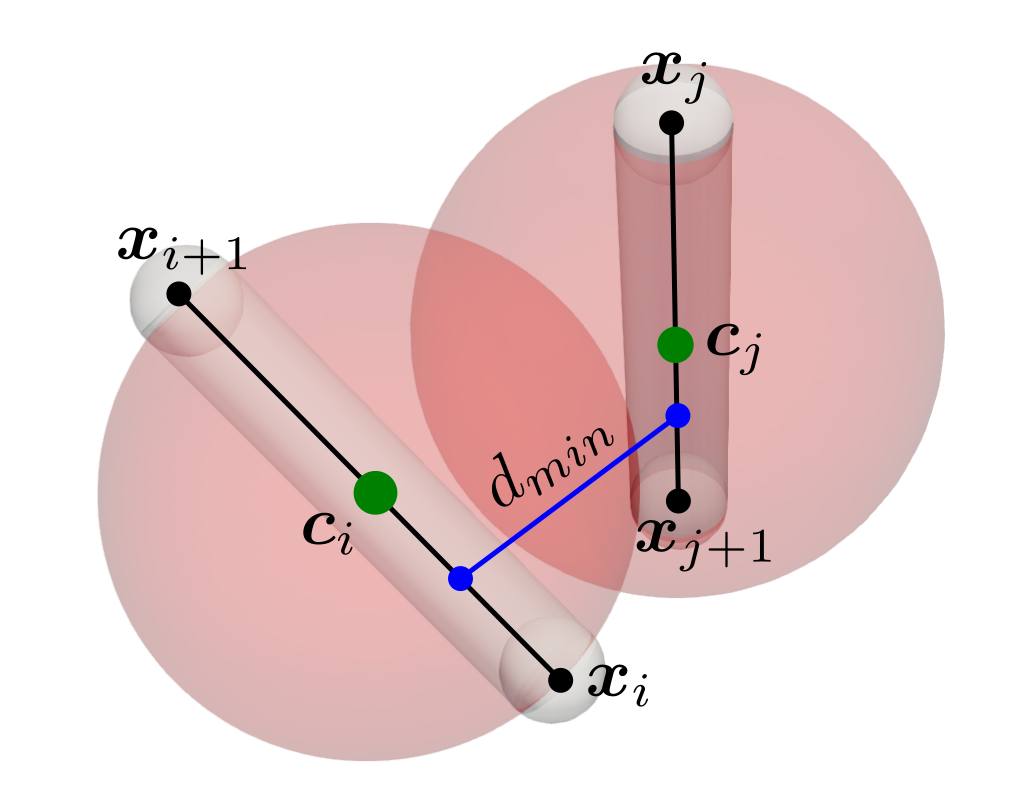}
        \caption{}
        \label{fig:bounding_sphere}
    \end{subfigure}
    \begin{subfigure}{0.4\textwidth}
        \includegraphics[width=\linewidth]{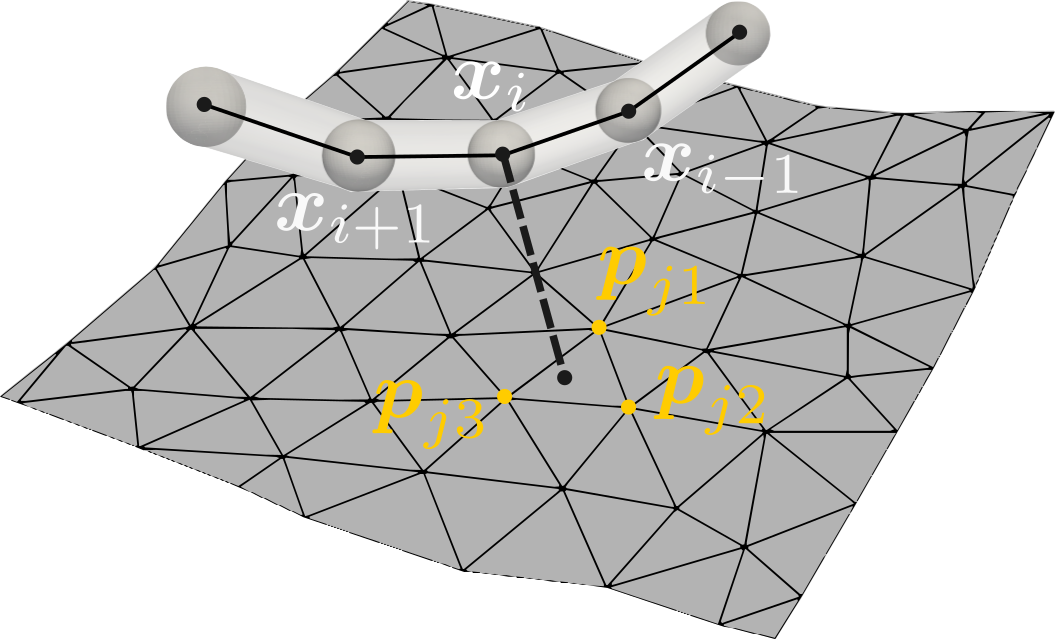}
        \caption{}
        \label{fig:bounding_sphere_triangle}
    \end{subfigure}
    \caption{(\subref{fig:bounding_sphere}) Setup for collision detection between edge segments. (\subref{fig:bounding_sphere_triangle}) Setup for collision detection between an edge and an triangular surface element. By $d_{min}$ we denote the minimal scalar distance between center lines of the contact partners. }
    \label{fig:BoundSphere}
\end{figure}

Our collision detection is structured in two phases. 
\begin{enumerate}
    \item Evaluate a simple necessary condition that allows to find potential collision partners by querying the octree.
    \item For all potential collision partners evaluate the minimum distance between them to discover all collisions and contact points.
\end{enumerate}

We start by formulating a necessary condition for coil self-collisions, see Fig. \ref{fig:BoundSphere} (\subref{fig:bounding_sphere}).
For each edge $\boldsymbol{e}^i$ on the coil, we create a sphere with center $\boldsymbol{c}^i = (\boldsymbol{x}_i+\boldsymbol{x}_{i+1})/2$ and some radius. The radius is then chosen in such a way that whenever another edge center $\boldsymbol{c}_j$ is located in the sphere of $\boldsymbol{c}_i$, we have a necessary condition for a collision between the two edges. We set the spheres radius to the diagonal distance of the triangle with side lengths $\overline{\ell}/2$ and $D_2/2$. Therefore, we can express the condition as
\begin{align}
    \|\boldsymbol{c}_j-\boldsymbol{c}_i\|\leq \sqrt{(\overline{\ell}/2)^2 + D_2^2}.
\end{align}

Next we verify if a collision between edge $\boldsymbol{e}^i$ and the potential partner $\boldsymbol{e}^j$ is actually occurring by calculating the minimum distance vectors $\boldsymbol{d}^{ij}_{min}$ between the segments $s_i = \{ \alpha\boldsymbol{e}^i +\boldsymbol{x}_{i}:\alpha\in[0,1]\}$ and $s_j = \{ \alpha\boldsymbol{e}^j +\boldsymbol{x}_{j}:\alpha\in[0,1]\}$ by the method proposed in \citep{lumelsky1985fast}. Finally, a collision takes actually place if 
\begin{align}
d_{min}^{ij} = \|\boldsymbol{d}^{ij}_{min}  \|\leq D_2.
\end{align}

Next the conditions for wall-collisions are illustrated in Fig. \ref{fig:BoundSphere} (\subref{fig:bounding_sphere_triangle}). We assume that the wall surface is meshed by triangles $T_j\in\mathcal{T}$, which are defined by their vertices $\boldsymbol{p}_{j1}, \boldsymbol{p}_{j2}, \boldsymbol{p}_{j3}$. To formulate a necessary condition for a wall collision, we again consider the center point of an edge $\boldsymbol{c}_i$ and the center-point of a triangle $\boldsymbol{c}_{T_j}=(\boldsymbol{p}_{j1}+\boldsymbol{p}_{j2}+ \boldsymbol{p}_{j3})/3$. Let $r_{ST}$ be the radius and center $\boldsymbol{c}_{T_j}$ of the smallest sphere that contains all $T_i\in\mathcal{T}$ then a sufficient condition for a collision is 
\begin{align}
    \| \boldsymbol{c}_i - \boldsymbol{c}_{T_j} \|\leq D_2/2 + r_{ST}.
\end{align}
To check between each partner whether a collision occurs or not, we first project the point $\boldsymbol{c}_i$ onto the plane defined by the triangle $T_j$ with surface normal $\boldsymbol{n} = (\boldsymbol{p}_{i1}- \boldsymbol{p}_{i2}) \times (\boldsymbol{p}_{i1}- \boldsymbol{p}_{i3})$ and denote this projection by  $\boldsymbol{c}_{i,T_j}$. For the projected point, we check if it is contained in the triangle $\boldsymbol{c}_{i,T_j}\in{T_j}$. If it is contained in the triangle, we calculate the minimum distance by $\|\boldsymbol{c}_{i,T_j}-\boldsymbol{c}_{i}\|$. In all other cases, no collision takes place. Note that our model assumes that no contact partner protrudes into the wall.

\subsection{Contact algorithm: friction model}
Our collision model is based on the one used in \citep{gazzola2018forward} which is a Coulomb stick slip friction contact model. We start by describing the forces that act in normal direction. Assuming a collision takes place between the edges $(i,i+1)$ and $(j,j+1)$, we find the minimal distance between their cylindrical hulls as $d^{ij}_{min}$. Since a collision takes place, we can calculate the overlap by $\epsilon_{ij}=D_2 - d^{ij}_{min}$. Then a collision of the coil segment on edge $(i,i+1)$ with the coil segment of edge $(j,j+1)$ introduces the following force in collision direction
\begin{align}
    (\boldsymbol{F}_{CC,\perp}^{i})_j &= -H(\epsilon_{ij})\cdot \big( k_{sc}\epsilon_{ij} \notag\\
    &+\gamma_{sc}(\dot{\boldsymbol{x}}_{i,i+1} - \dot{\boldsymbol{x}}_{j,j+1})\cdot \boldsymbol{d}^{ij}_{min}\big)\boldsymbol{d}^{ij}_{min}.
\end{align}
Here  $\dot{\boldsymbol{x}}_{i,i+1}$ is the linear interpolation of the nodal velocities onto the point of collision on the edge $(i,i+1)$ and $\dot{\boldsymbol{x}}_{j,j+1}$, respectively.
The model is activated upon collision by the Heaviside function $H(\epsilon_{ij})$. The coefficient $k_{sc}$ stands for the coil-self contact spring constant and $\gamma_{sc}$ for the coefficient of dissipation. Recall that the vector $\boldsymbol{d}_{min}^{ij}$ is the minimum distance vector.

Next we formulate the wall collisions by decomposing the nodal forces and velocities into a wall tangential and wall orthogonal part $(-\nabla_{\boldsymbol{X}}E_{tot})_i+\sum_{j=1}^{n}(\boldsymbol{F}_{CC}^i)_j=\boldsymbol{F}_{\perp} \oplus \boldsymbol{F}_{\|}$ and  $\boldsymbol{v}=\boldsymbol{v}_{\perp} \oplus \boldsymbol{v}_{\|}$. Assuming a collision takes place between a triangle $T_k\in\mathcal{T}$ and a vertex $i$ with intersection width $\epsilon = D_2-d_{min}^{ij}$, then the normal force due to the collision is
\begin{align}
    (\boldsymbol{F}^i_{CW,\perp})_k =  -H(\epsilon)\cdot \big( & \|\boldsymbol{F}_{\perp}\| +k_{w}\epsilon_{ij} \notag\\
    & +\gamma_{w}\dot{\boldsymbol{x}}_{i}\cdot \boldsymbol{n}_{T_k}) \boldsymbol{n}_{T_k}
\end{align}
where analogously to the coil-coil collisions, $k_w$ denotes the coil-wall contact spring constant, $\gamma_w$ is the dissipation coefficient, and $\boldsymbol{n}_{T_k}$ stands for the outer normal vector of the triangle $T_k$.

\begin{figure}
    \centering
    \begin{subfigure}[t]{0.35\textwidth}
    \centering
        \includegraphics[width=\linewidth]{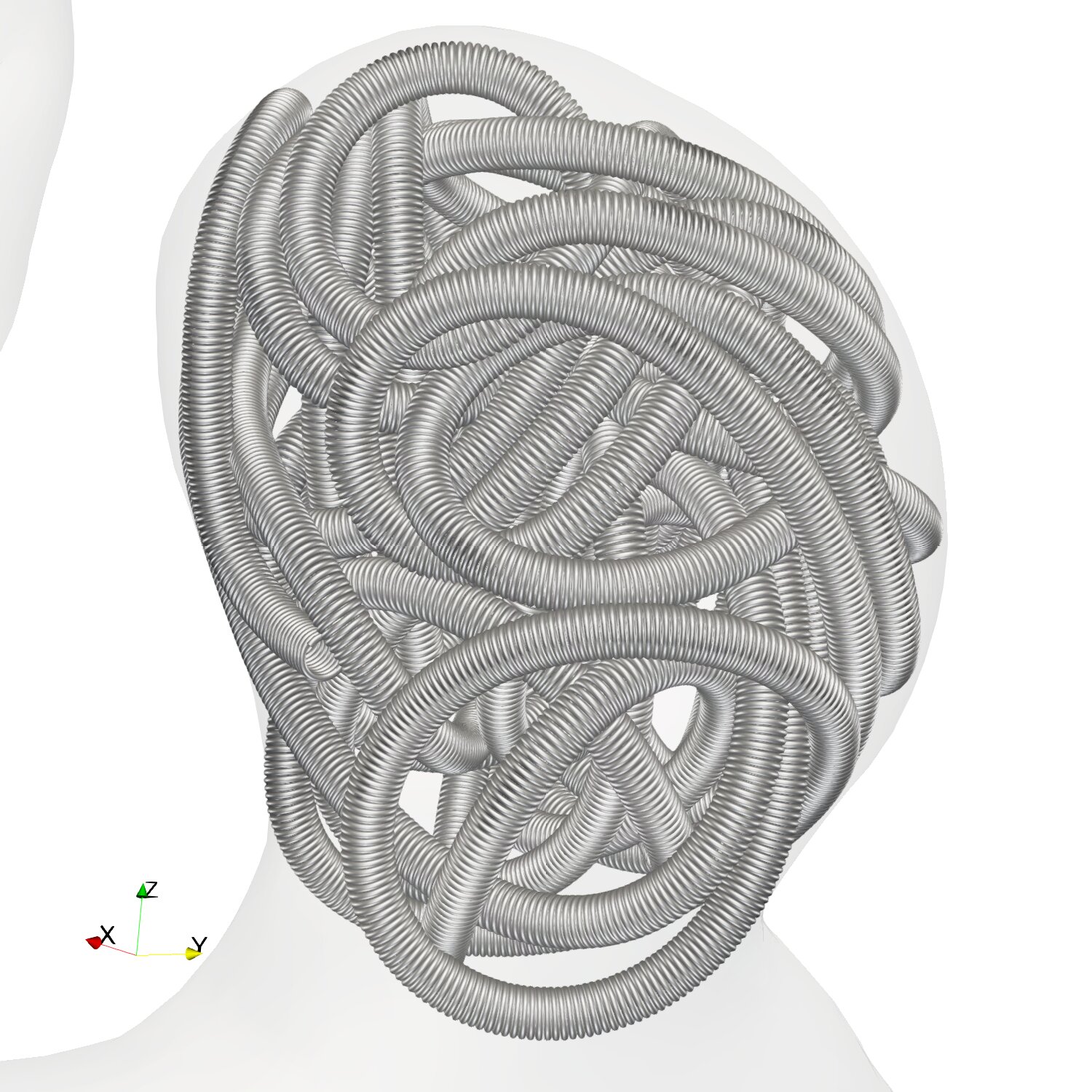}
    \caption{}
    \label{fig:CoilCloseUp1}
    \end{subfigure}
    \begin{subfigure}[t]{0.35\textwidth}
    \centering
        \includegraphics[width=\linewidth]{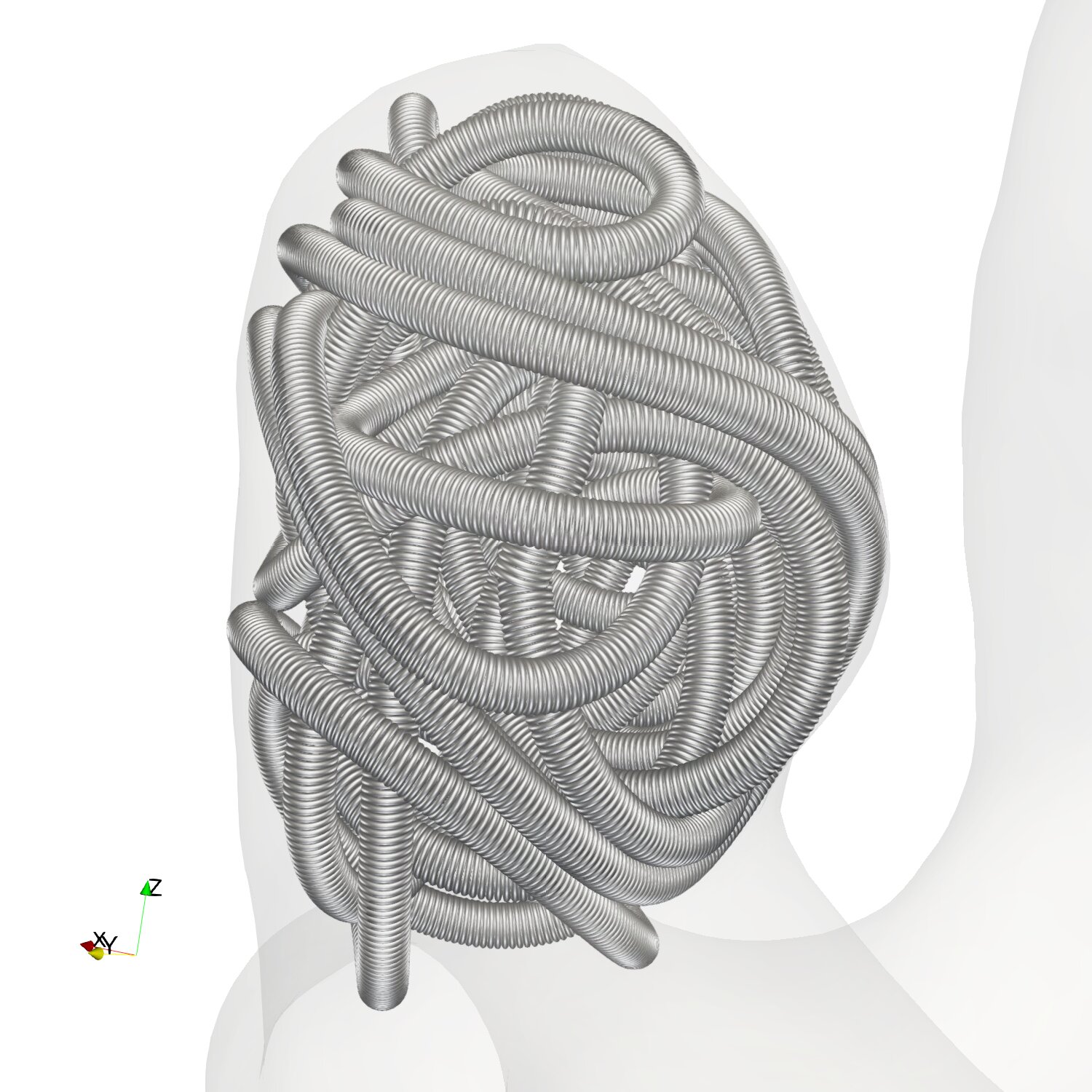}
        \caption{}
        \label{fig:CoilCloseUp2}
    \end{subfigure}
    \begin{subfigure}[t]{0.35\textwidth}
    \centering
        \includegraphics[width=\linewidth]{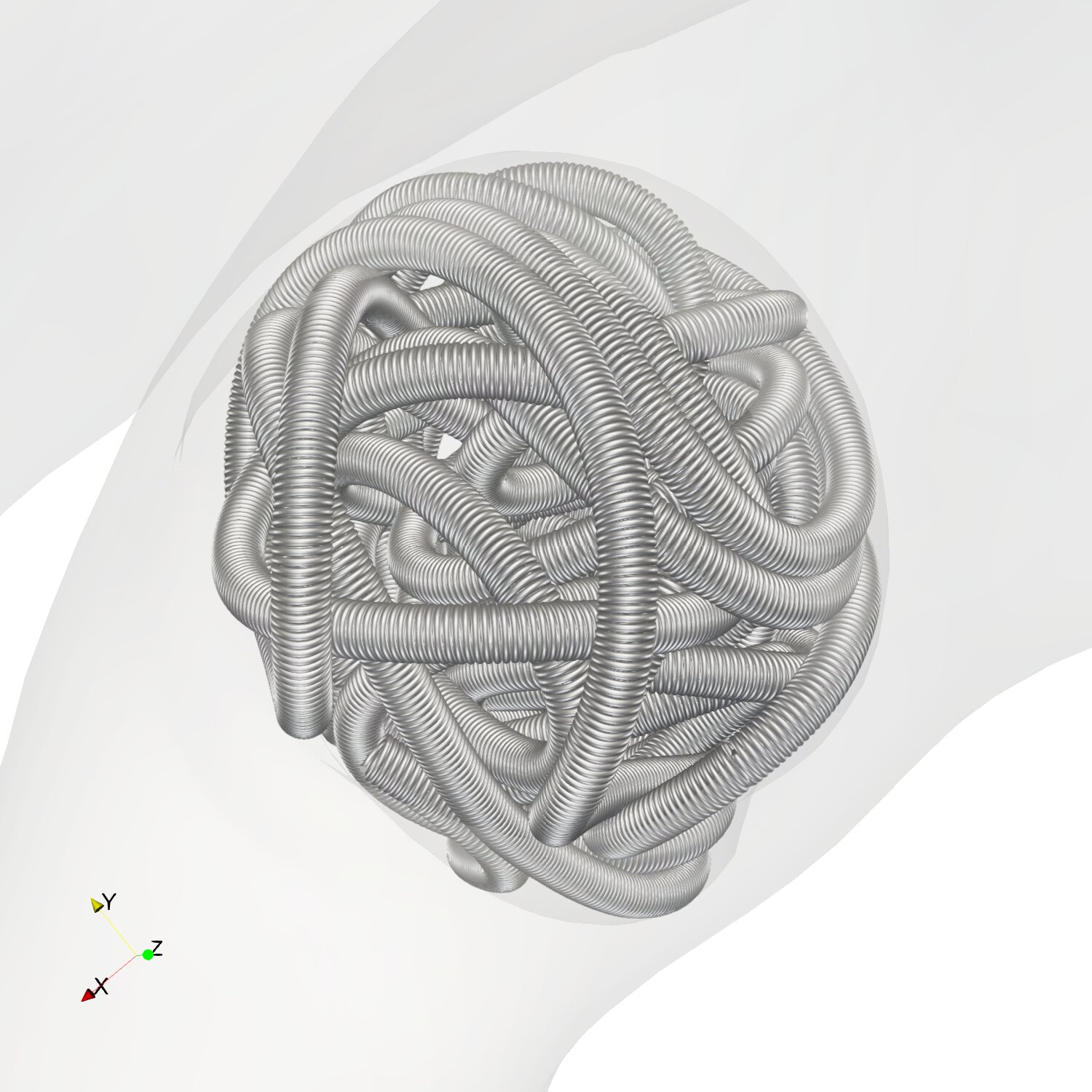}
        \caption{}
        \label{fig:CoilCloseUp3}
    \end{subfigure}
    \caption{Close up view of a coil inserted into the small aneurysm at \SI{20}{\percent} global packing density. (\subref{fig:CoilCloseUp1}), (\subref{fig:CoilCloseUp2}) side views, (\subref{fig:CoilCloseUp3}) view from bottom. }
    \label{fig:CloseUp}
\end{figure}

Next we enrich our contact model by introducing friction in tangential direction. We assume only slip friction in direction of the relative tangential velocities between the edges that are in contact. The relative velocities 
$ \dot{\boldsymbol{x}}_{i,i+1} - \dot{\boldsymbol{x}}_{j,j+1} $ give rise to the relative tangential velocity by 
\begin{align*}
    v_{CC,\|}&=(\dot{\boldsymbol{x}}_{i,i+1} - \dot{\boldsymbol{x}}_{j,j+1})\notag\\
    &-\big(\dot{\boldsymbol{x}}_{i,i+1} - \dot{\boldsymbol{x}}_{j,j+1} \big)\cdot \boldsymbol{d}^{ij}_{min} \frac{\boldsymbol{d}^{ij}_{min}}{\|\boldsymbol{d}^{ij}_{min}\|^2} .
    \end{align*}
    Then the friction force for coil-coil contacts is
\begin{align}
    (\boldsymbol{F}^i_{CC,||})_j =  -\mu_{slip,CC}\|\boldsymbol{F}_{\perp}\| \frac{v_{CC,\|}}{\|v_{CC,\|}\|}
\end{align}
with the slip friction coefficient $\mu_{slip,CC}$. Finally the tangential friction force between wall and coil is
\begin{align}
	&\text{if } \| \boldsymbol{v}_{\|}\| \leq v_\epsilon:\notag\\
	&(\boldsymbol{F}^i_{CW,||})_k = -\min\{\|\boldsymbol{F}_{||}\|, \mu_{stick,CW}\|\boldsymbol{F}_{\perp}\|\} \frac{\boldsymbol{F}_{||}}{\|\boldsymbol{F}_{||}\|}\notag\\
	&\text{else }: \notag\\
	&(\boldsymbol{F}^i_{CW,||})_k = -\mu_{slip,CW}\|\boldsymbol{F}_{\perp}\| \frac{\boldsymbol{v}_{\|}}{\|\boldsymbol{v}_{\|}\|},
\label{EQ:wall_coll}
\end{align}
with $\mu_{slip}, \mu_{stick}$ denoting the coefficients of stick and slip friction, and $v_\epsilon$ being the threshold that decides when the model switches between stick and slip friction. All in all, our forces are thereby given as $\boldsymbol{F}_{ext}=\boldsymbol{F}_{CW}+\boldsymbol{F}_{CC}$.
To conclude this section, we assume that the micro-catheter is modeled as a cylindrical surface without lids following a spline curve that is generated by 3 points in space. By triangulating its surface, we can consider it as rigid object and model its collisions with the coil in the same ways as done in (\ref{EQ:wall_coll}).
This concludes our discussion of the model we use of the endovascular coil embolization. An example for a embolized coil in the small aneurysm is given in Fig. \ref{fig:CloseUp}. Here (\subref{fig:CoilCloseUp1}), (\subref{fig:CoilCloseUp2}) show side views. The view (\subref{fig:CoilCloseUp3}) is from bottom, looking through the aneurysm neck into the aneurysm.

\section{In silico Raymond–Roy type occlusion classification}
In this section, we propose an in silico Raymond-Roy type classification based on the coil deployment. To do so, we follow the classical Raymond-Roy classification idea and introduce four different types of classes. The affiliation to a class is based on local packing densities at the neck and wall regions.
\label{sec:sensitivity-study}
\subsection{Raymond–Roy type occlusion classification}
In clinical practice,  the placement of a coil inside an aneurysm is of major importance. Poor placements can lead to complications such as a coil protruding into the parent artery or complications in the healing process and even aneurysm recurrence \citep{mascitelli2015update, kim2021recurrence}. 

Due to the complicated three dimensional structure of embolized coils a differentiation of a good and poor placement is not an easy task. For this reason, the Raymond Roy Occlusion Classification (RROC) model was developed \citep{roy2001endovascular, mascitelli2015update}. In the RROC two different angiographic views are considered to grade a embolized coil by the following classes
\begin{itemize}
    \item Class I: Complete obliteration, angiographic views from two perspectives show a complete occlusion / obliteration of the aneurysm by the coil
    \item Class II: Similar to Class I, but the aneurysm has a residual neck that is not occluded from the flow
    \item Class IIIa: Occlusion only on the aneurysm walls, but  residual contrast agent shows in the core of the coil
    \item Class IIIb: Regions at the wall of the aneurysm are not correctly occluded.
\end{itemize}
From the perspective of a clinician a coil of Class I is desired to lower the chances of an aneurysm recurrence. Note  that a Class III and Class II have some probability to switch into a Class I over time, since blood clotting enhances the occlusion mechanism significantly. 
In \citep{mascitelli2015update} it is shown statistically that Class IIIa, compared to Class II, has a higher chance to improve to the desired Class I. Moreover, looking at the case where a complete occlusion never occurs, even after blood clotting takes place, Class IIIb, compared to Class IIIa, has a higher chance to remain incompletely occluded. 

In this paper, we consider a simplified numerical version of the original RROC, namely instead of performing a angiography, we solely analyze the local packing density of a given coil within the aneurysm.

\begin{figure}[htbp]
    \begin{subfigure}[t]{0.4\textwidth}
        \includegraphics[width=0.75\linewidth]{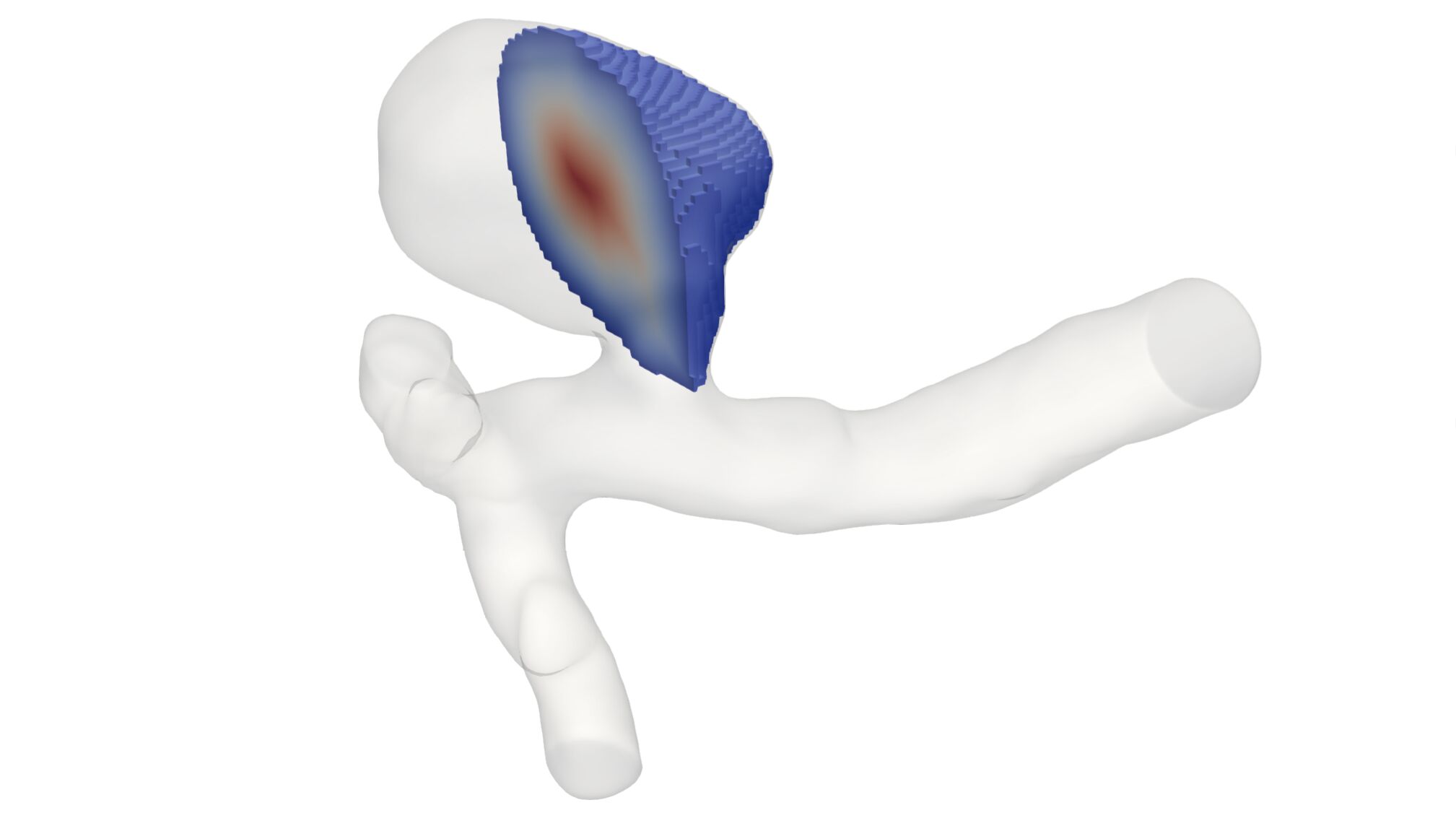}
        \includegraphics[width=0.2\linewidth]{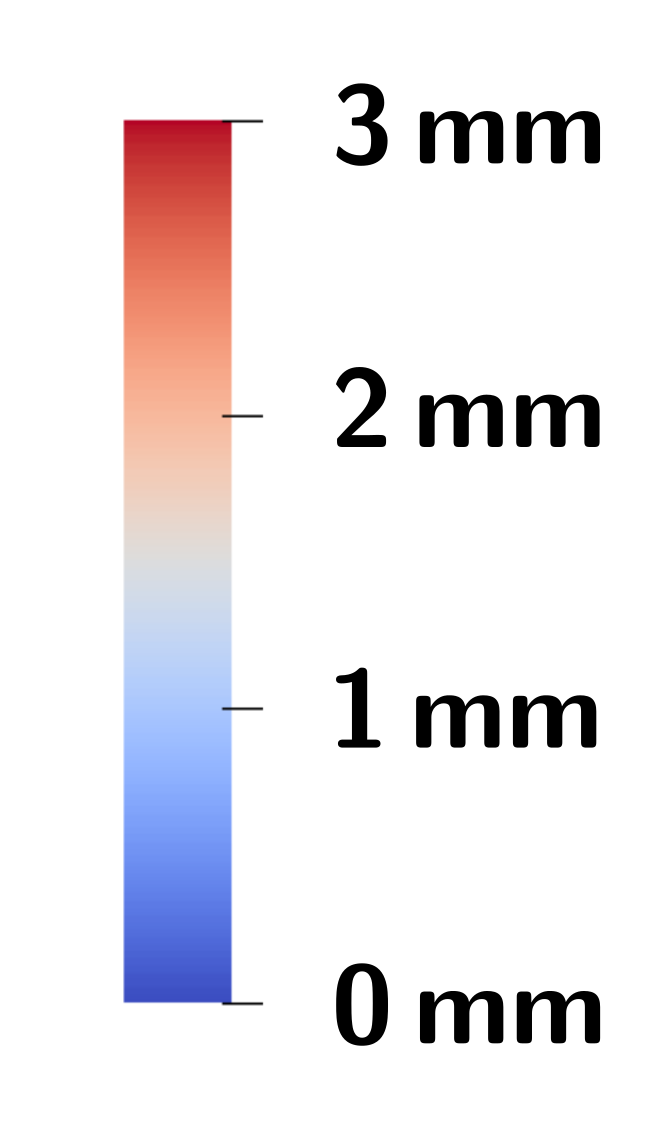}
    \caption{}
    \label{fig:Hom_SDF}
    \end{subfigure}
    \begin{subfigure}[t]{0.4\textwidth}
        \includegraphics[width=0.75\linewidth]{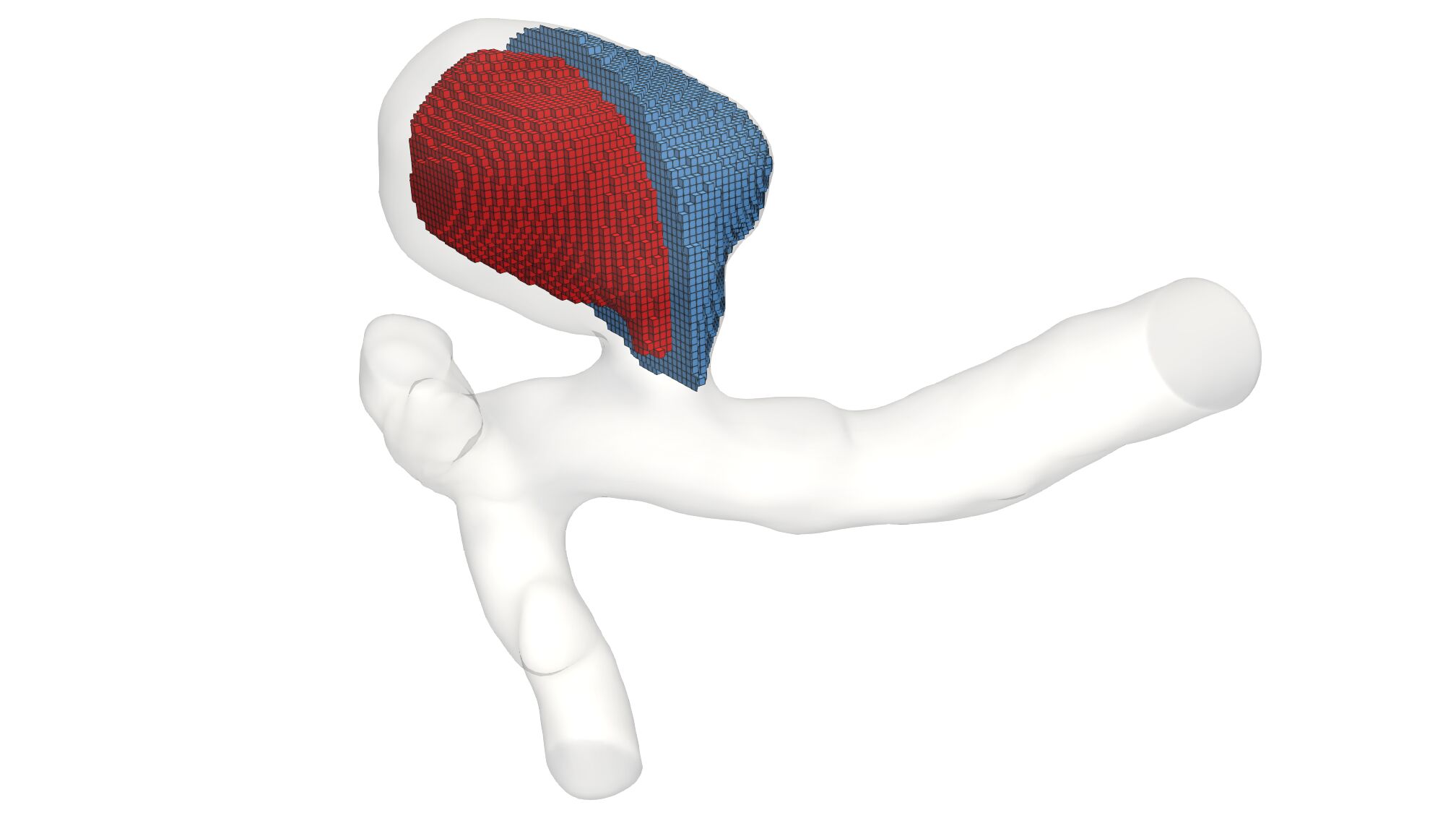}
        \caption{}
        \label{fig:Hom_BoundaryCore}
    \end{subfigure}
    \begin{subfigure}[t]{0.4\textwidth}
        \includegraphics[width=0.75\linewidth]{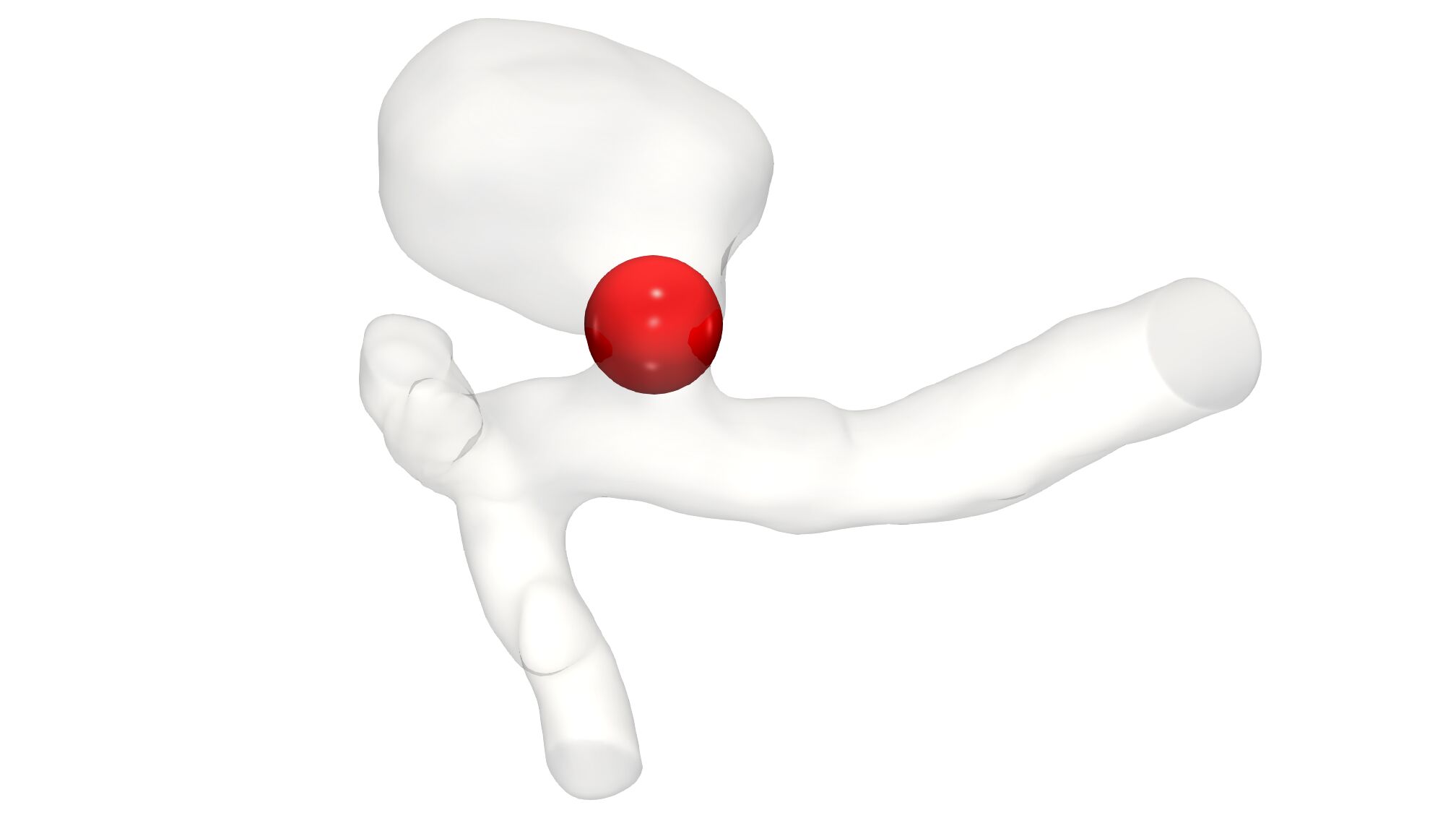}
        \caption{}
        \label{fig:Hom_Sphere}
    \end{subfigure}
    \caption{Setup for the evaluation of the RROC in case of the narrow neck aneurysm. (\subref{fig:Hom_SDF})  A signed distance field \citep{park2019deepsdf} (SDF) w.r.t. the distance to the aneurysm surface is generated. (\subref{fig:Hom_BoundaryCore}) A level-set of the SDF is used to partition the aneurysm into a boundary region (blue) and a core region (red). (\subref{fig:Hom_Sphere}) A sphere is defined at the neck of the aneurysm to analyze the coil distribution within it.}
    \label{fig:Custom_RROC}
\end{figure}

\subsection{Voxelization of coil deployments}
\label{sec:Voxilation}
\begin{figure}[!htb]
\centering
        \includegraphics[width=0.32\textwidth]{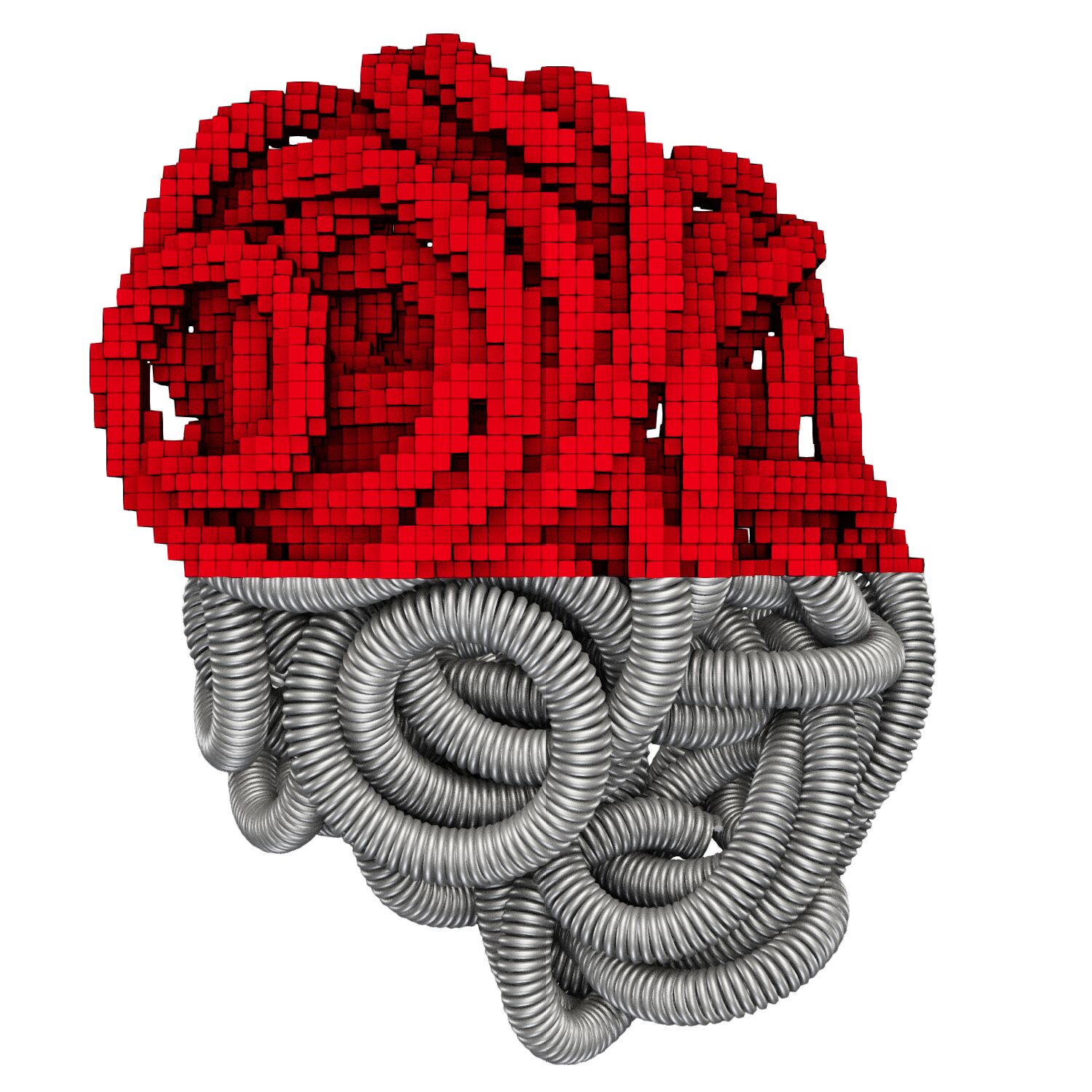}
    \caption{Example of a voxelization of a coil. Only voxels with value 1 (representing coil) are shown.}
    \label{fig:voxel_grid_small}
\end{figure}

For the RROC classification, it is convenient to convert the deployed coil, given as a set of sorted vertices $\boldsymbol{x}=\boldsymbol{x}_0,\boldsymbol{x}_1,...,\boldsymbol{x}_n$ with radius $D_2$, into a more convenient form. We do this by means of voxelization, namely converting the 3D coil into a scalar field that approximates it. Without loss of generality, we set the origin of the Cartesian coordinates system such that all points within the aneurysm have non-negative coordinates and for each coordinate direction there exists one point with zero component.
Let $\Omega_{A}\subset \mathbb{R}^3$ be the space occupied by the aneurysm. Then, we can find a cube $\mathcal{C}_{a}$ with edge length of $a$ such that $\Omega_A \subset \mathcal{C}_a$, meaning it is a bounding box containing the aneurysm. Next we partition the bounding box into small cubes $\mathcal{C}_{N_V}$ of size $a/N_V$ such that
\begin{align}
    \mathcal{C}_a &= \bigcup_{i,j,k=0,1, ..., N_v-1} \mathcal{C}_{N_V}+ 
    \begin{bmatrix} i a/N_V \\ j a/N_V \\ k a/N_V \end{bmatrix} \notag\\
    &=\bigcup_{i,j,k=0,1, ..., N_v-1} \mathcal{C}_{N_V,ijk}.
\end{align}
Each of the small cubes $\mathcal{C}_{N_V,ijk}$ has a central point $c_{N_V,ijk}=a/N_V(i+1/2,j+1/2,k+1/2)^T$. Now assume that we sweep a circle with radius $D_2$ in tangential direction of the coil center line $\boldsymbol{x}$. Then this defines a three dimensional geometry of the coil which we call $\Omega_C$. This now allows to generate the voxelization by a binary mapping $\mathcal{V}:\mathcal{C}_a \rightarrow \{0,1\}$ with 
\begin{align}
    \boldsymbol{y}\in \mathcal{C}_a \mapsto &\begin{cases}
    1 & \text{ if }y\in \mathcal{C}_{N_V,ijk} \text{ and }  c_{N_V,ijk} \in \Omega_C \notag\\
    0 & \text{else}
\end{cases}\\
&\text{ for } i,j,k\in\{0,...,N_V-1\}.
\end{align}
Finally we note that if $N_V \rightarrow \infty$ and $\Omega_A, \Omega_C$ behave sufficiently well then $\int_{\mathcal{C}_a}\mathcal{V}(\mathcal{C}_a)~\textup{d}x\to \int_{\Omega_C}1~\textup{d}x$. Therefore we need to choose $N_V$ in the definition of $\mathcal{V}$ large enough to represent $\Omega_C$ sufficiently well. Before concluding this section, we state how we determine whether a point $\boldsymbol{y}\in \mathcal{C}_a$ is located inside the coil. Introducing the signed distance function (SDF) for a surface $\partial \Omega$ by
\begin{align}
    \mathcal{F}_{\partial \Omega}(\boldsymbol{y}) =  \begin{cases}
    \inf_{\boldsymbol{z}\in \partial \Omega} -\|\boldsymbol{y}-\boldsymbol{z} \| & \text{ if }  \boldsymbol{y}\in \Omega\\
    \inf_{\boldsymbol{z}\in \partial \Omega} \|\boldsymbol{y}-\boldsymbol{z} \| & \text{else}
\end{cases},
\end{align}
we can easily decide  if a point $\boldsymbol{y}$ is located inside the coil $\Omega_C$. This is exactly the case for $\mathcal{F}_{\partial \Omega_C}(\boldsymbol{y})
\leq 0$. Efficient implementations to generate signed distance functions for triangulated surfaces are readily accessible, and in this work we use the algorithm of \citep{park2019deepsdf}. An example of a voxelized coil can be seen in Fig. \ref{fig:voxel_grid_small}. Here and thorough this work we voxelize the coil by  \SI{70}{} voxels in each basis vector direction of the Cartesian coordinate system. In the figure, the voxels that correspond to coil are shown in red on the top section while on the bottom the transition to the actual coil is shown. The full rectangular voxel domain can be used to compare different coil in an aneurysm if we fix the position of the voxels. The idea here is to compare between voxels that correspond to coil for different coils which we describe below in more detail.

\subsection{In silico classification}
\label{insilicioclass}
As stated before, our in silico classification is motivated by the RROC. Our classifier operates by considering the geometrical features of coil and aneurysm. For an extraction of features, we first partition the aneurysm into a core, boundary and neck region. Fig. \ref{fig:Custom_RROC} shows the partitioning of the example of the narrow neck aneurysm. The partitioning is based on an SDF defined with respect to the aneurysm wall $\partial \Omega_A$, see Fig. \ref{fig:Hom_SDF}. The aneurysm is then partitioned into a core and boundary region of equal volume by choosing the level-set of the SDF that generates the interface for separating these regions. In Fig. \ref{fig:Hom_BoundaryCore} we see that the partition can be directly generated on the voxelized domain, allowing us to count the nonzero voxels $\mathcal{V}(\mathcal{C}_a)$ therein and therefore obtaining the approximate coil volume in the respective regions. The neck region of the aneurysm is covered by a sphere as can be seen in Fig. \ref{fig:Hom_Sphere}. Having the center-point of the sphere, we can assess the voxels located inside the sphere and calculate the approximate coil volume contained in the sphere. Having extracted the coil volume at the boundary core and sphere region of the aneurysm enables us to construct our classifier.
To fix a classification scheme, we classify a coil by Tab. \ref{TAB:RROC}.

\begin{table}
\renewcommand{\arraystretch}{1.5}
\centering
\begin{tabular}{|p{0.26\columnwidth}|p{0.26\columnwidth}|p{0.26\columnwidth}|}
\hline
\multicolumn{3}{|c|}{boundary full} \\
\hline
          & sphere full & sphere empty  \\
\hline
core full & I & II  \\
\hline
core empty & IIIa & IIIa  \\
\hline
\multicolumn{3}{|c|}{boundary empty}\\
\hline
          & sphere full & sphere empty \\
\hline
core full & IIIb & IIIb \\
\hline
core empty & fail & fail \\
\hline
\end{tabular}
\caption{The table contains the assignment rules for the classes in our modified RROC. An aneurysm is ranked as core full when its core packing density reaches $\SI{20}{\percent}$ and above and is ranked core empty if its core packing density is below $\SI{20}{\percent}$. It is ranked boundary full if its boundary packing density reaches $\SI{18}{\percent}$ and above and is ranked boundary empty if its packing density is below $\SI{18}{\percent}$. Lastly it is ranked sphere full when the packing density in the sphere is at least $\SI{18}{\percent}$ and sphere empty if the packing density in its sphere is below $\SI{18}{\percent}$. We assume that the cases where we rank coil empty and boundary empty lead to coils that migrated into the parent vessel and therefore are classified as fail.}
\label{TAB:RROC}
\end{table}

The table is motivated by the following assumptions. Having sufficient coil in the sphere blocks blood from flowing into the aneurysm. This then blocks tracer fluid from entering the aneurysm. Combined with a high packing density at the aneurysm boundary and core, no tracer fluid would be visible and Class I is assigned. 
If boundary and core packing densities are large enough, but the sphere is not sufficiently filled, we classify the coil as Class II, because of the possibility that tracer fluid enters the neck region. Whenever the boundary is sufficiently packed  and the core is empty, we classify the coil as  Class IIIa, since in this situation tracer fluid can be trapped in the core.
Having a core that is sufficiently filled by an empty boundary suggests that the vessel wall is only partially occluded from the blood flow, leading to Class IIIb whether the sphere is filled or not. Finally, a case in which neither in the core nor at the boundary there is a large enough coil packing density despite the fact that a sufficiently
large enough coil volume is inserted,  is marked as failed.
In such a case, the coil has at least partially migrated
into the parent artery. 


\section{Simulation of Coil Placements}
\label{sec:SimulationOfCoilPlacements}
This section showcases the capabilities of the DER model in the context of medical coils. At first, we carry out a qualitative validation of the model. Then we test the model by applying different natural coil shapes and deploy them in the three aneurysms given in Fig. \ref{fig:Aneurisms}. The general setup for our simulation parameters is given in Appendix \ref{App:Sim_Params}.

\subsection{Qualitative validation of model correctness}
\label{sec:qualitative_validation}
\begin{figure*}[!htb]
    \centering
    \begin{subfigure}[ht]{0.45\textwidth}
    \centering
        \includegraphics[height=0.85\textwidth]{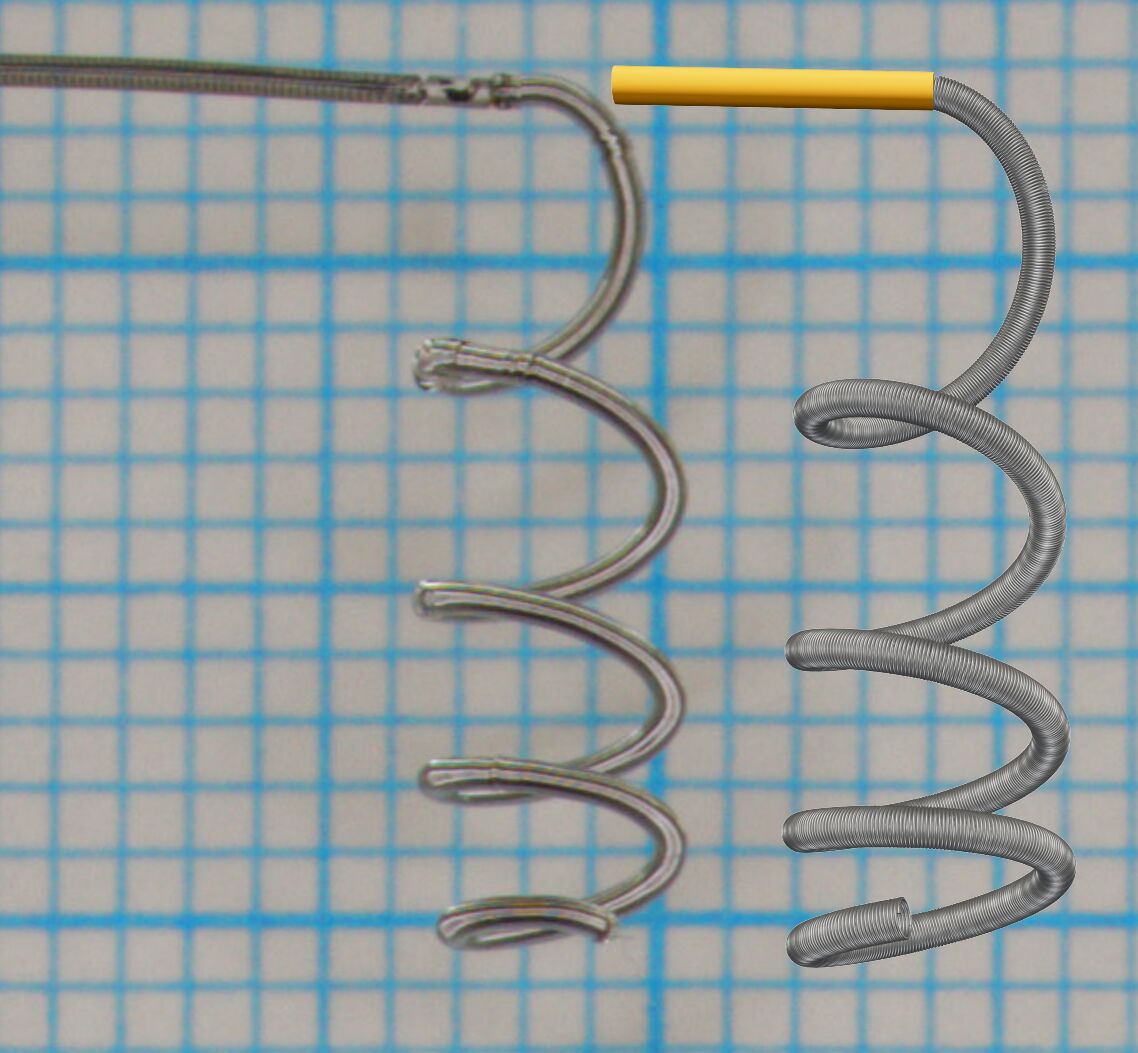}
    \caption{}
    \label{fig:Helix_Validation}
    \end{subfigure}
    \begin{subfigure}[ht]{0.45\textwidth}
    \centering
        \includegraphics[height=0.85\textwidth]{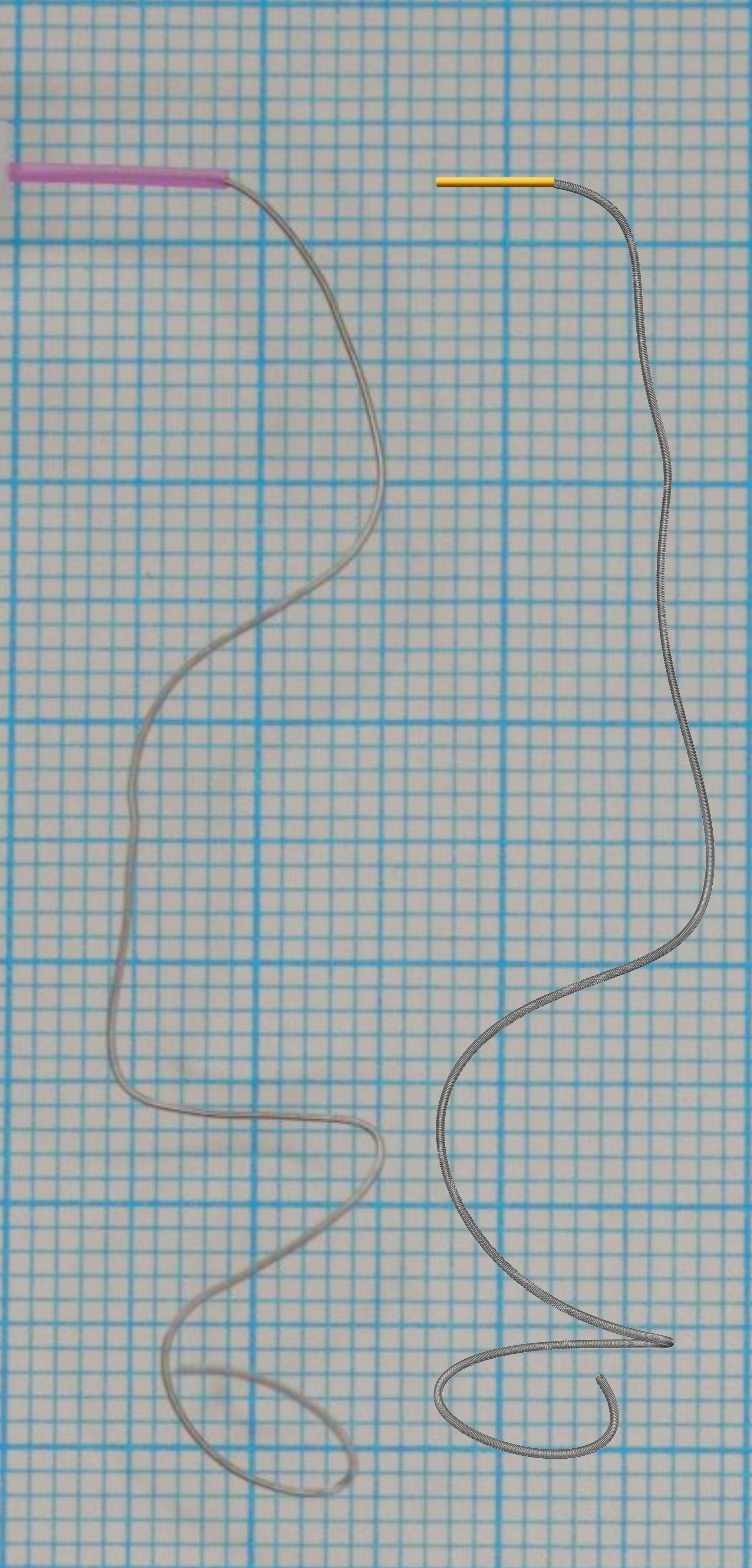}
        \caption{}
        \label{fig:3D_validation}
    \end{subfigure}
    \caption{Qualitative validation of the coil model via coils that are pushed out of the catheter into air while the gravitational force is acting. (\subref{fig:Helix_Validation}) Shows a helix-coil while (\subref{fig:3D_validation}) is a three dimensional framing coil. Simulations are placed on the right besides the experiments on the left within each subfigure.}
    \label{fig:validation}
\end{figure*}

In this subsection, we give an overview of the capabilities of our model by displaying exemplary coil placements. To this end, we start by providing a qualitative validation of the model. In the validation experiment, we fix a micro-catheter in front of a millimeter paper, see Fig. \ref{fig:validation}. A coil is then pushed out of the micro-catheter inducing a bending of the coil, due to its natural shape and the gravitational force acting on it. In a second step, we recreate this experiment virtually by our coil model where an additional gravitational force term is included. 

To make the simulations match the experiments, we parameterize the simulation by the parameters of our test coils. Specifically, for the helix coil simulations, we use $D_1 = \SI{49}{\micro\meter}$, $D_2 = \SI{510}{\micro\meter}$, and $D_3 = \SI{4}{\milli\meter}$. Similarly, for the 3D coil simulations, we utilize $D_1 = \SI{74}{\micro\meter}$, $D_2 = \SI{340}{\micro\meter}$, and $D_3 = \SI{10}{\milli\meter}$. We note that not all parameters are known. In case of the helix coil, the natural shape is only partially known. The diameters $D_2$ and $D_3$ are given in the data-set but the pitch of the helix (spacing between the loops in the $D_3$ diameter) is unknown and assumed to be $1.2\,D_2$. For the three dimensional coil only the diameters $D_3$ and $D_2$ are known, but the natural shape imprinted by the manufacturing process is unknown.  Here we take an educated guess of the natural shape as shown in Fig. \ref{fig:coil_composition} that is motivated by the patent for stable coil designs in \citep{white2008coils}. For the placements, we inversely estimate the unknown parameter $D_1$ such that the length of the coil in the simulation approximately matches the experiment. For the case of the helix coil in Fig. \ref{fig:validation} (\subref{fig:Helix_Validation}), one can see the experiment on the left and the simulation on the right. Since this helix coil is relatively short with a length of $\SI{4}{\centi \meter}$, we have pushed it completely out of the micro-catheter such  that it is only fixed to the delivery wire via the detachment mechanism. Although the position of the helix loops is slightly shifted, the overall shape matches with our simulation. 
Due to the unknown exact natural shape, even in case of the helical coil (since we do not know the pitch of the helix), the length of the coil was rescaled to $\SI{4.6}{\centi \meter}$ to obtain the shown shape.  A validation set up of a three dimensional coil can be seen in Fig.\ref{fig:validation} (\subref{fig:3D_validation}). Here the total length is $\SI{10}{\centi \meter}$. Comparing the experiment on the left to the simulation on the right, we see that even though the shape is unknown the main features, e.g., the loop at the tip and the intermediate bending, are represented in the simulation.

\subsection{Exemplary study of coil-placements}
\begin{figure*}
    \centering
    \begin{subfigure}[b]{0.19\textwidth}
    \centering
        \includegraphics[width=\linewidth]{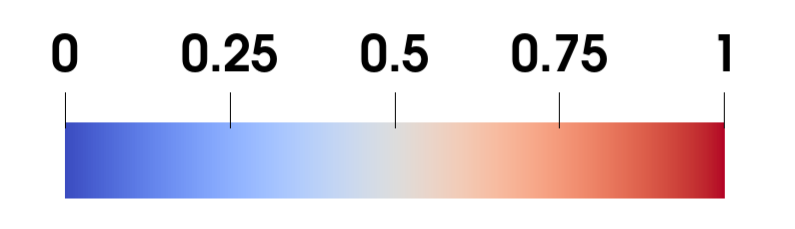}
        \includegraphics[width=\linewidth]{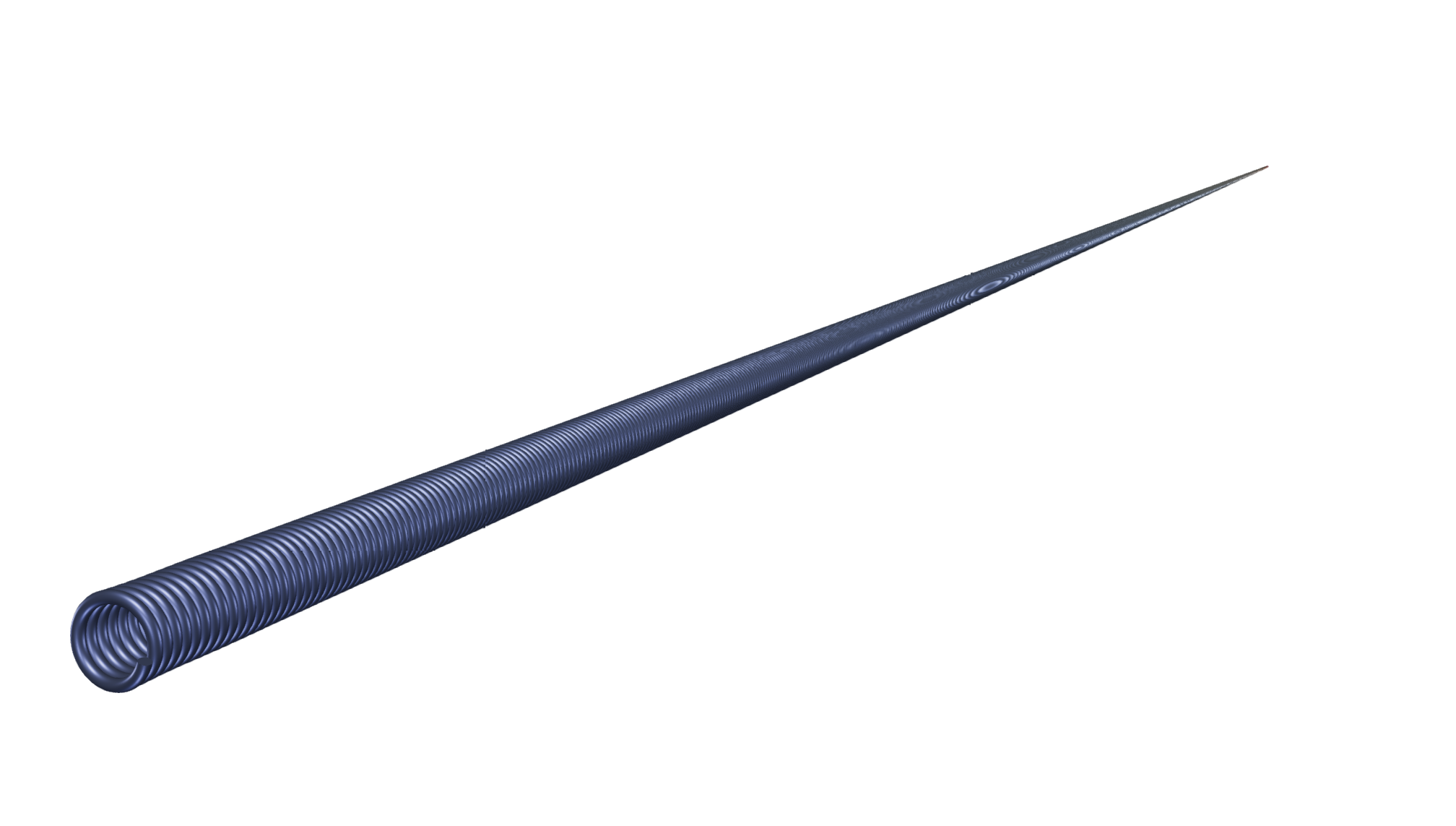}
    \caption{}
    \label{fig:Coil_Straight_ref}
    \end{subfigure}
    \begin{subfigure}[b]{0.19\textwidth}
    \centering
        \includegraphics[width=\linewidth]{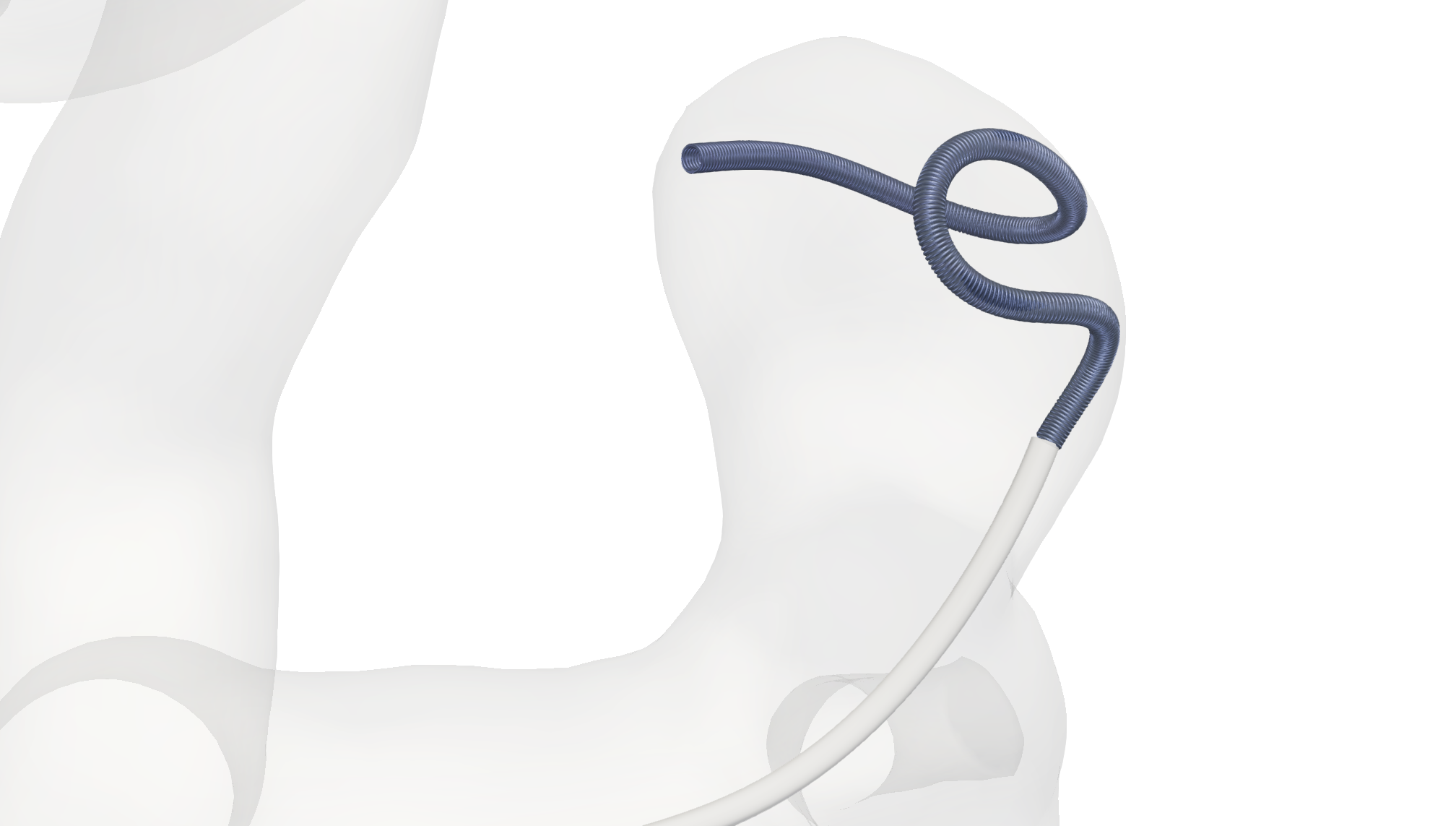}
    \caption{}
    \label{fig:Coil_Straight_1}
    \end{subfigure}
    \begin{subfigure}[b]{0.19\textwidth}
    \centering
        \includegraphics[width=\linewidth]{pictures/Small_Aneurysm_Shape_variation/Coil_Straight_1}
    \caption{}
    \label{fig:Coil_Straight_2}
    \end{subfigure}
    \begin{subfigure}[b]{0.19\textwidth}
    \centering
        \includegraphics[width=\linewidth]{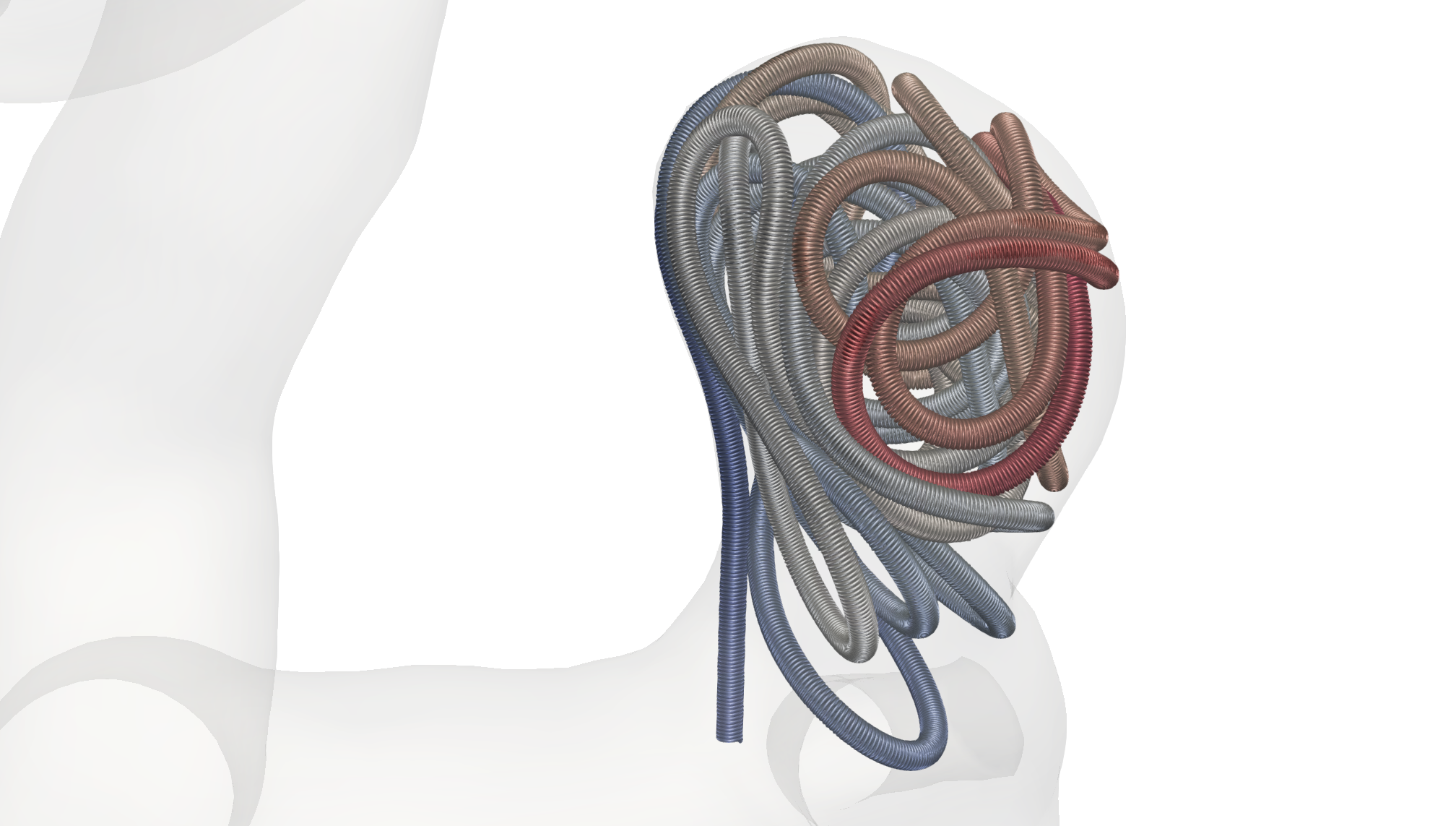}
    \caption{}
    \label{fig:Coil_Straight_3}
    \end{subfigure}
    \begin{subfigure}[b]{0.19\textwidth}
    \centering
        \includegraphics[width=\linewidth]{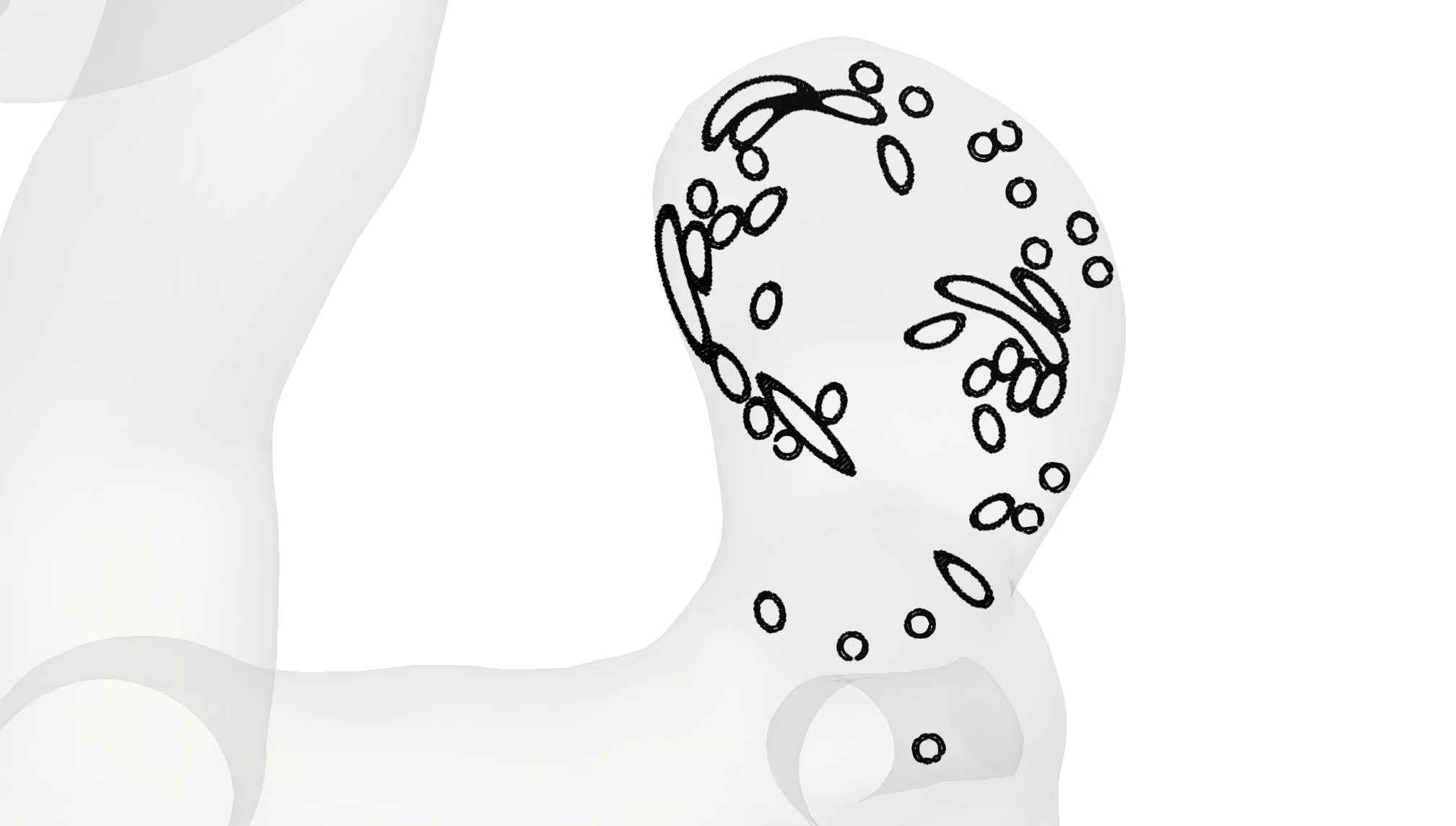}
    \caption{}
    \label{fig:Coil_Straight_cross}
    \end{subfigure}\\
    \begin{subfigure}[t]{0.19\textwidth}
    \centering
        \includegraphics[width=\linewidth]{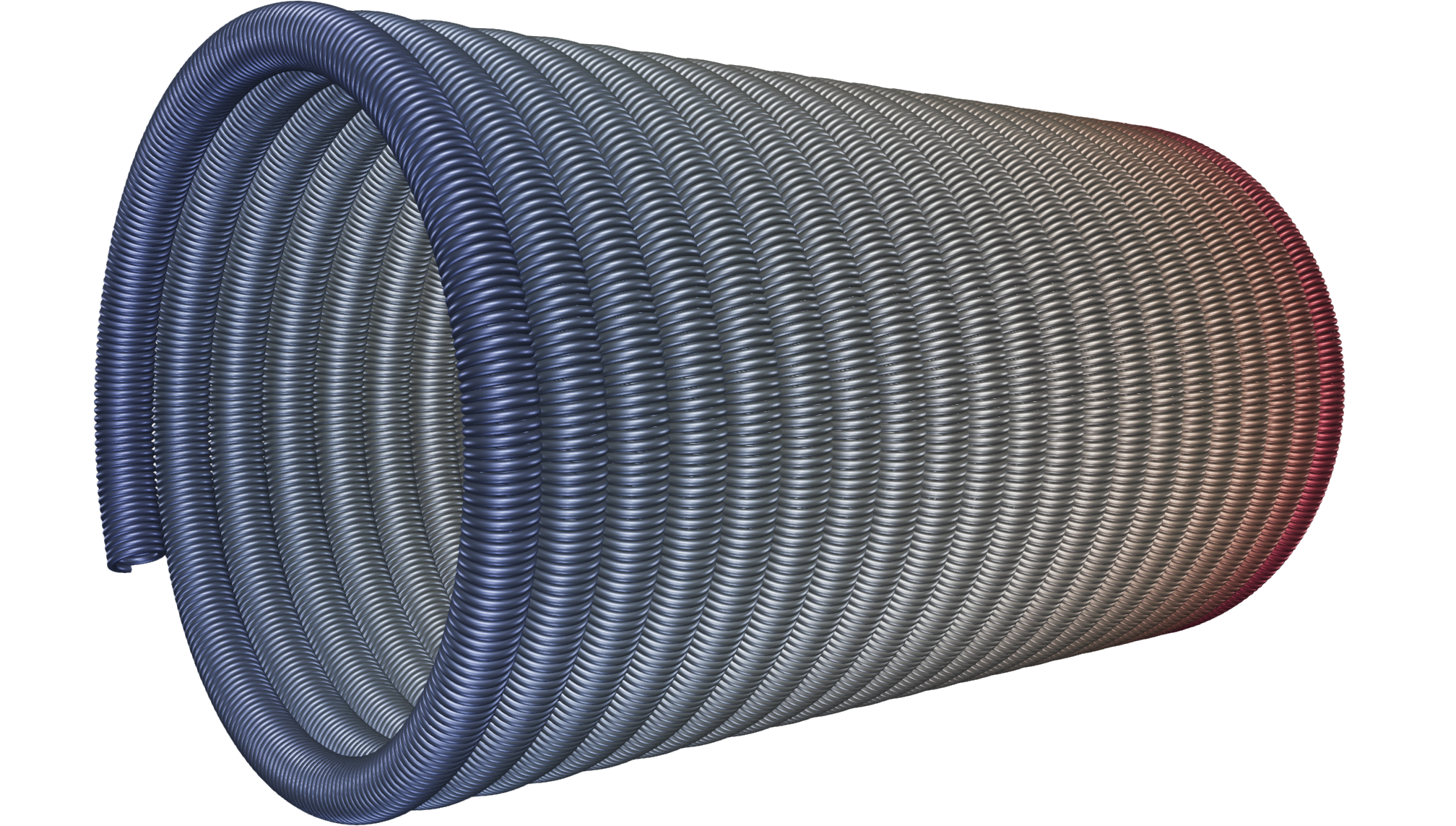}
    \caption{}
    \label{fig:Helix_ref}
    \end{subfigure}
    \begin{subfigure}[t]{0.19\textwidth}
    \centering
        \includegraphics[width=\linewidth]{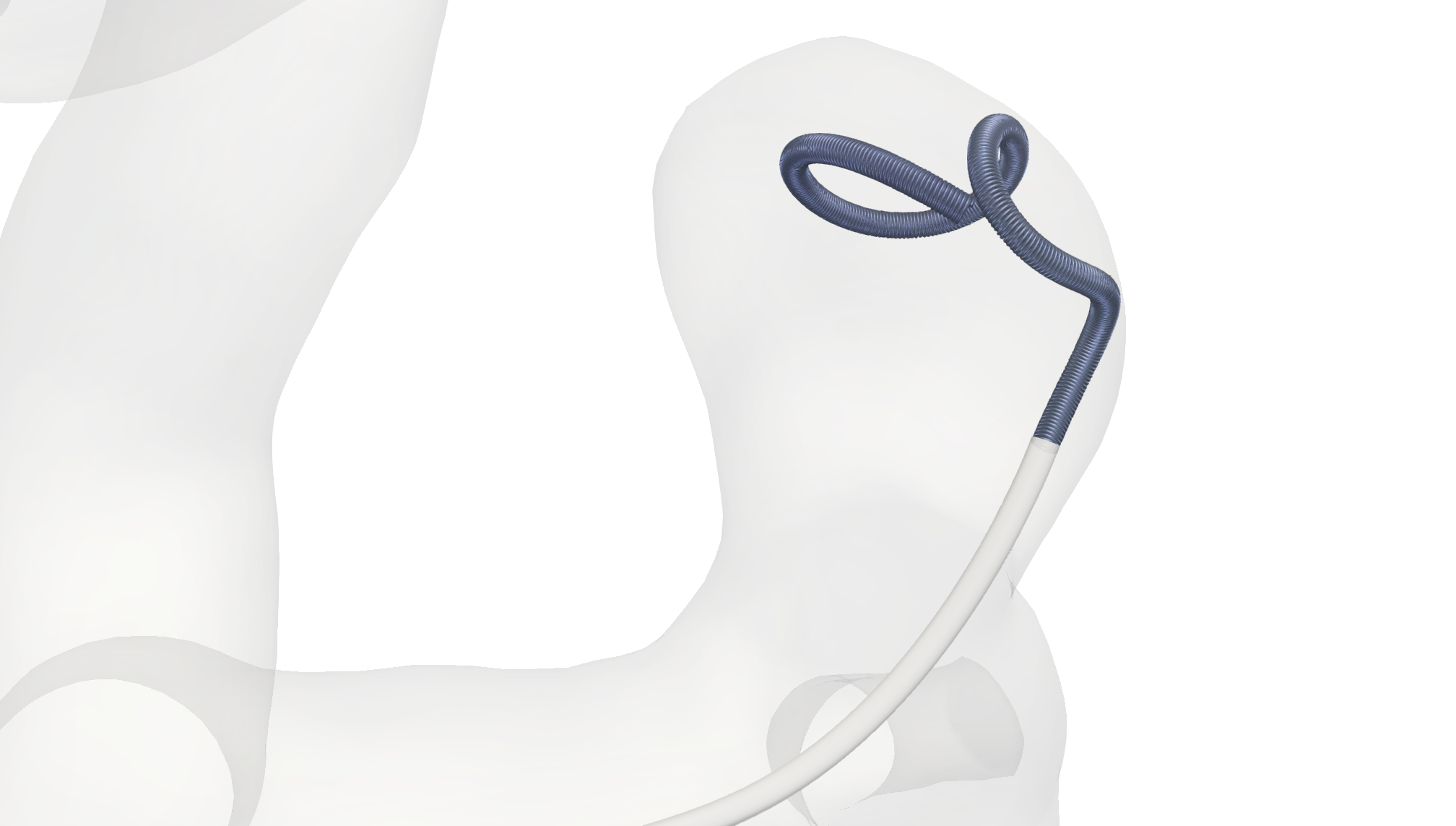}
    \caption{}
    \label{fig:Helix_2}
    \end{subfigure}
    \begin{subfigure}[t]{0.19\textwidth}
    \centering
        \includegraphics[width=\linewidth]{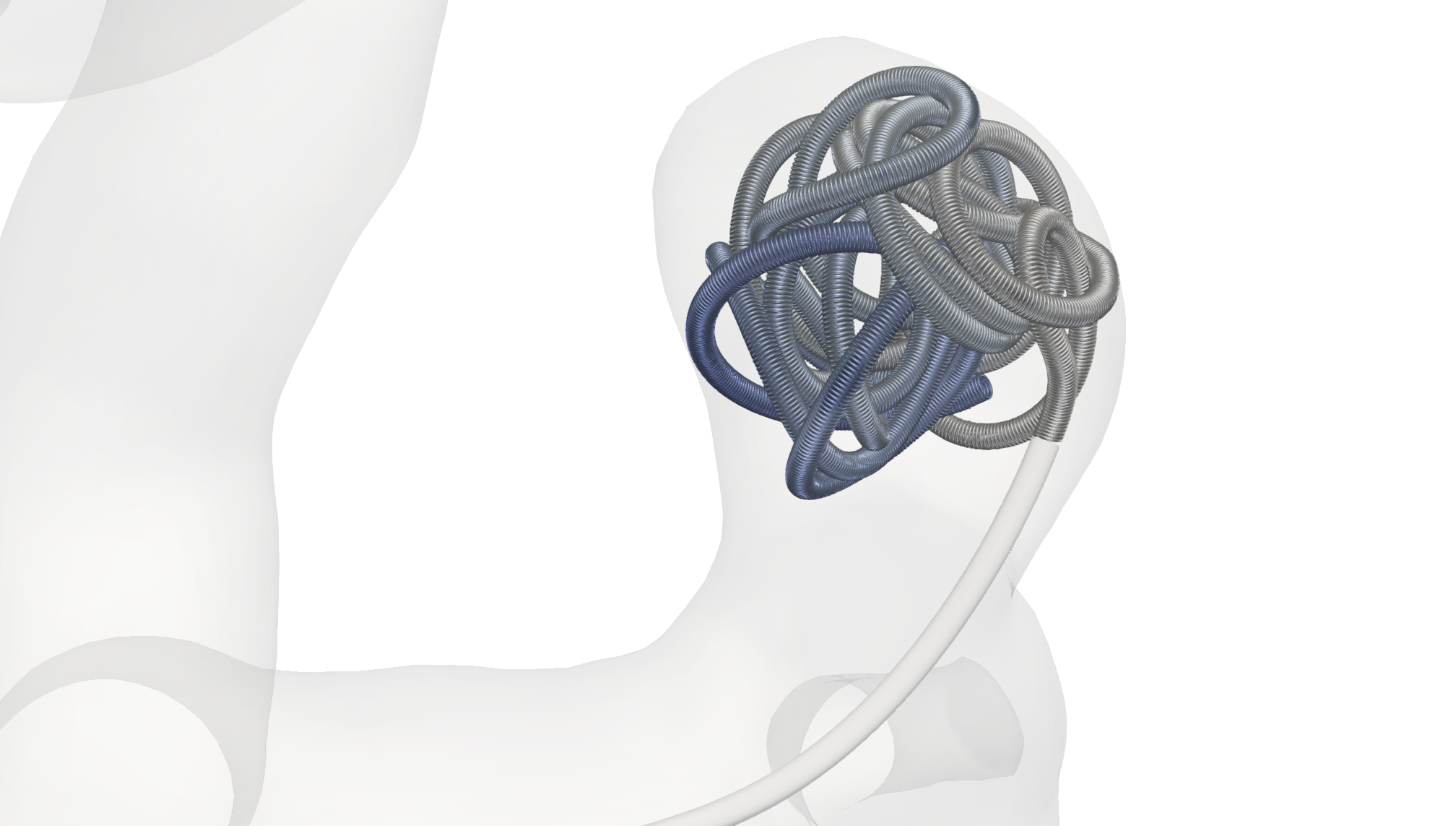}
    \caption{}
    \label{fig:Helix_1}
    \end{subfigure}
    \begin{subfigure}[t]{0.19\textwidth}
    \centering
        \includegraphics[width=\linewidth]{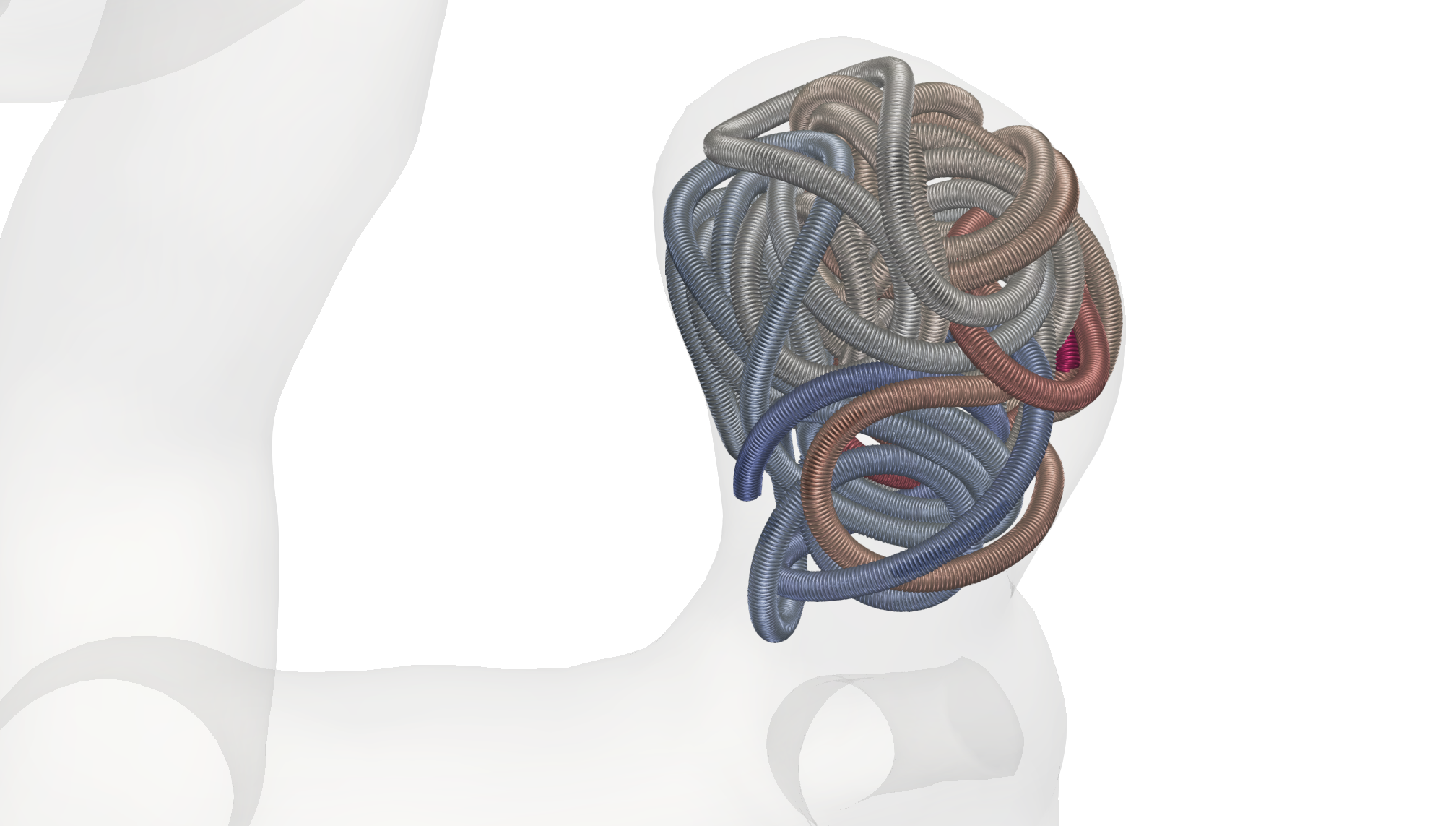}
    \caption{}
    \label{fig:Helix_3}
    \end{subfigure}
    \begin{subfigure}[t]{0.19\textwidth}
    \centering
        \includegraphics[width=\linewidth]{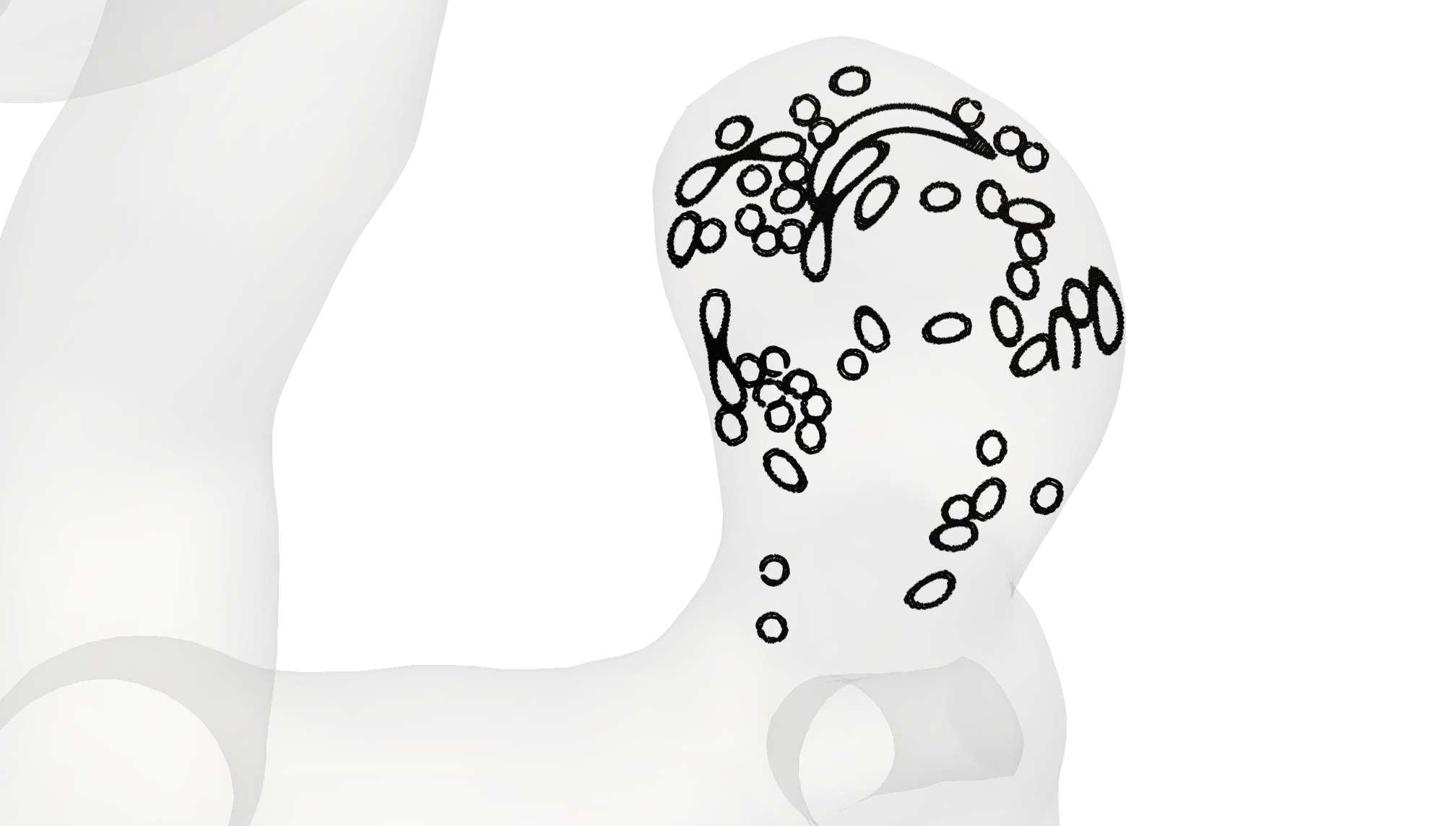}
    \caption{}
    \label{fig:Helix_cross}
    \end{subfigure}\\
    \begin{subfigure}[t]{0.19\textwidth}
    \centering
        \includegraphics[width=\linewidth]{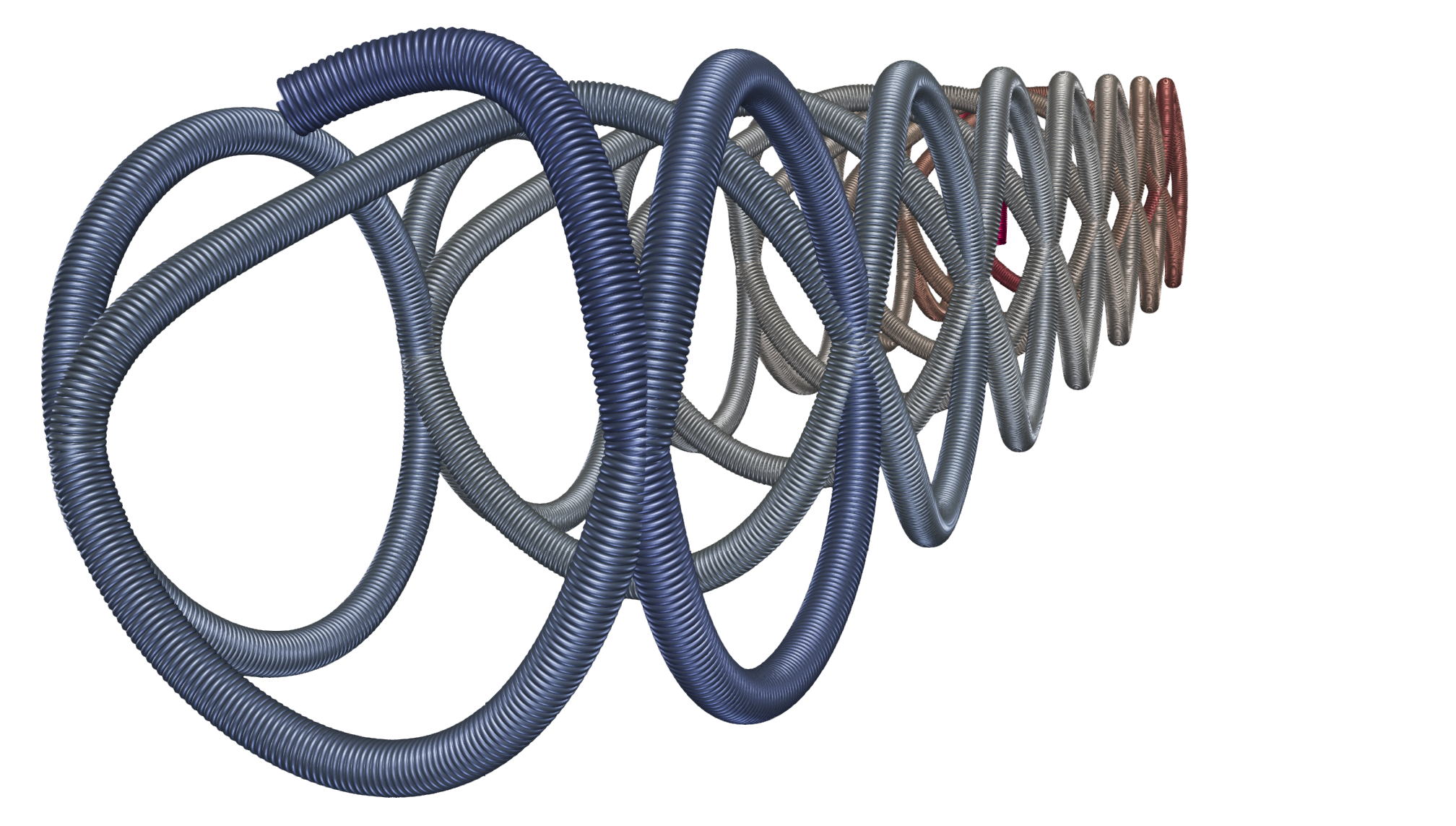}
    \caption{}
    \label{fig:ThreeD_ref}
    \end{subfigure}
    \begin{subfigure}[t]{0.19\textwidth}
    \centering
        \includegraphics[width=\linewidth]{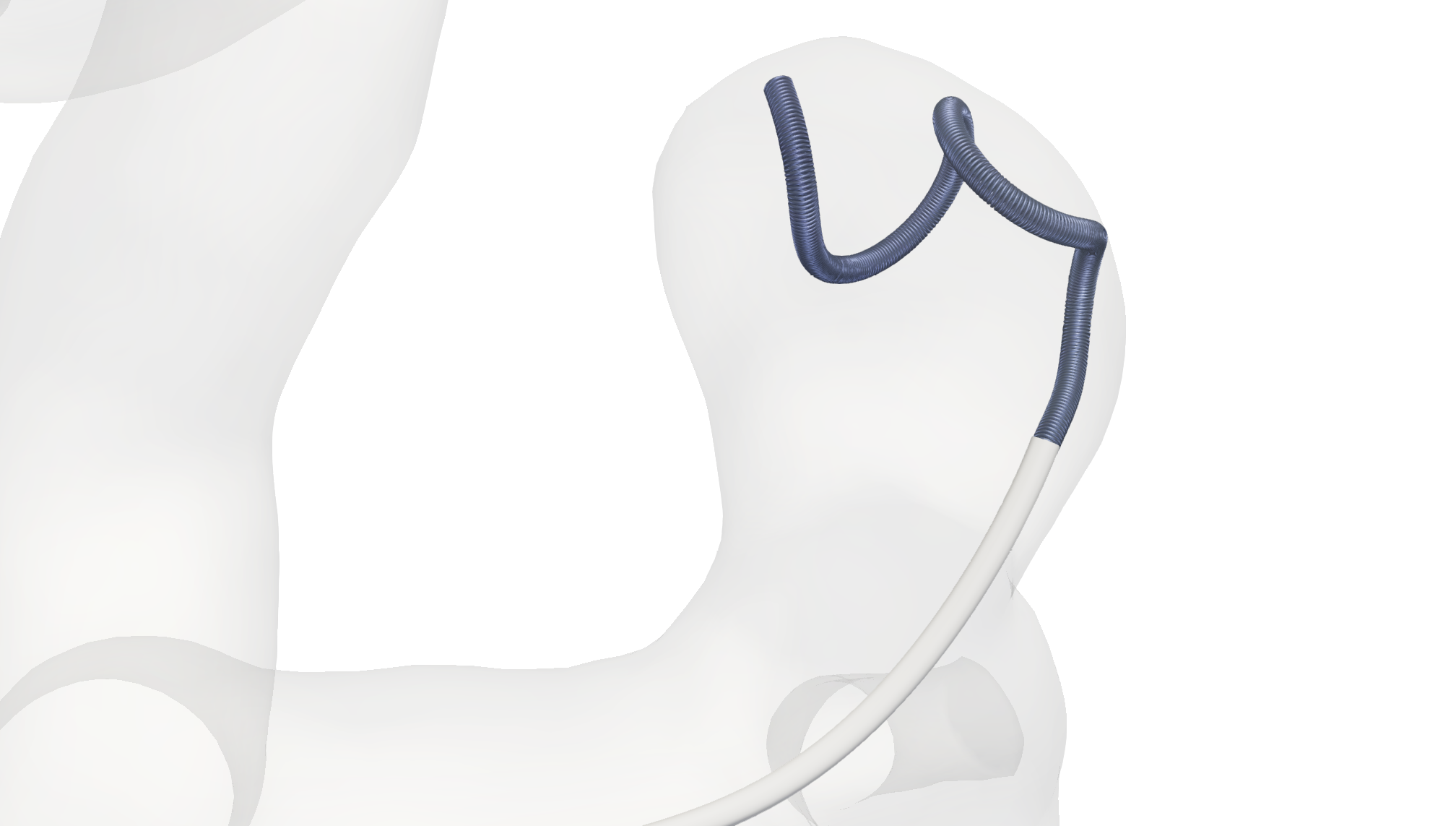}
    \caption{}
    \label{fig:ThreeD_1}
    \end{subfigure}
    \begin{subfigure}[t]{0.19\textwidth}
    \centering
        \includegraphics[width=\linewidth]{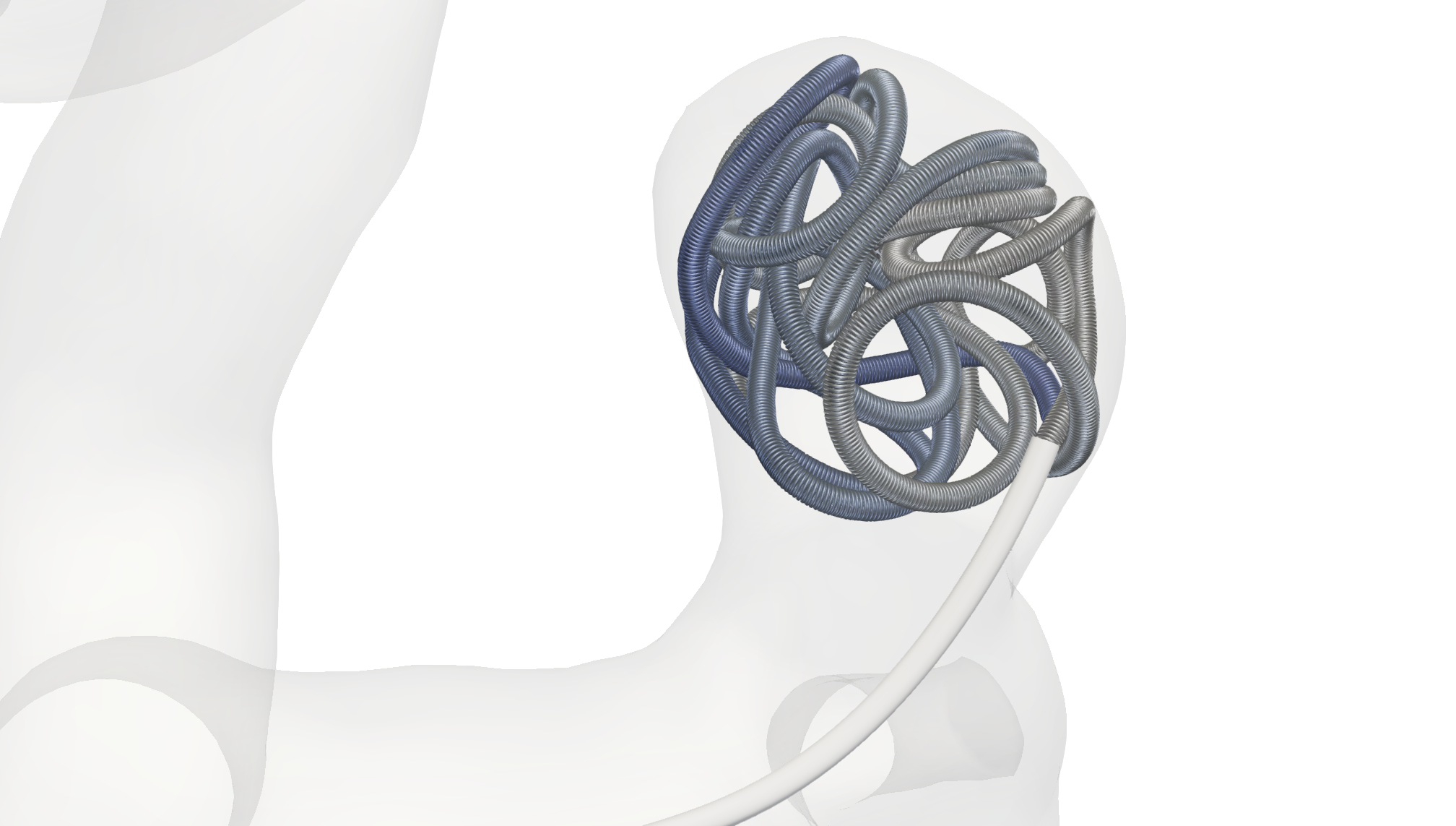}
    \caption{}
    \label{fig:ThreeD_2}
    \end{subfigure}
    \begin{subfigure}[t]{0.19\textwidth}
    \centering
        \includegraphics[width=\linewidth]{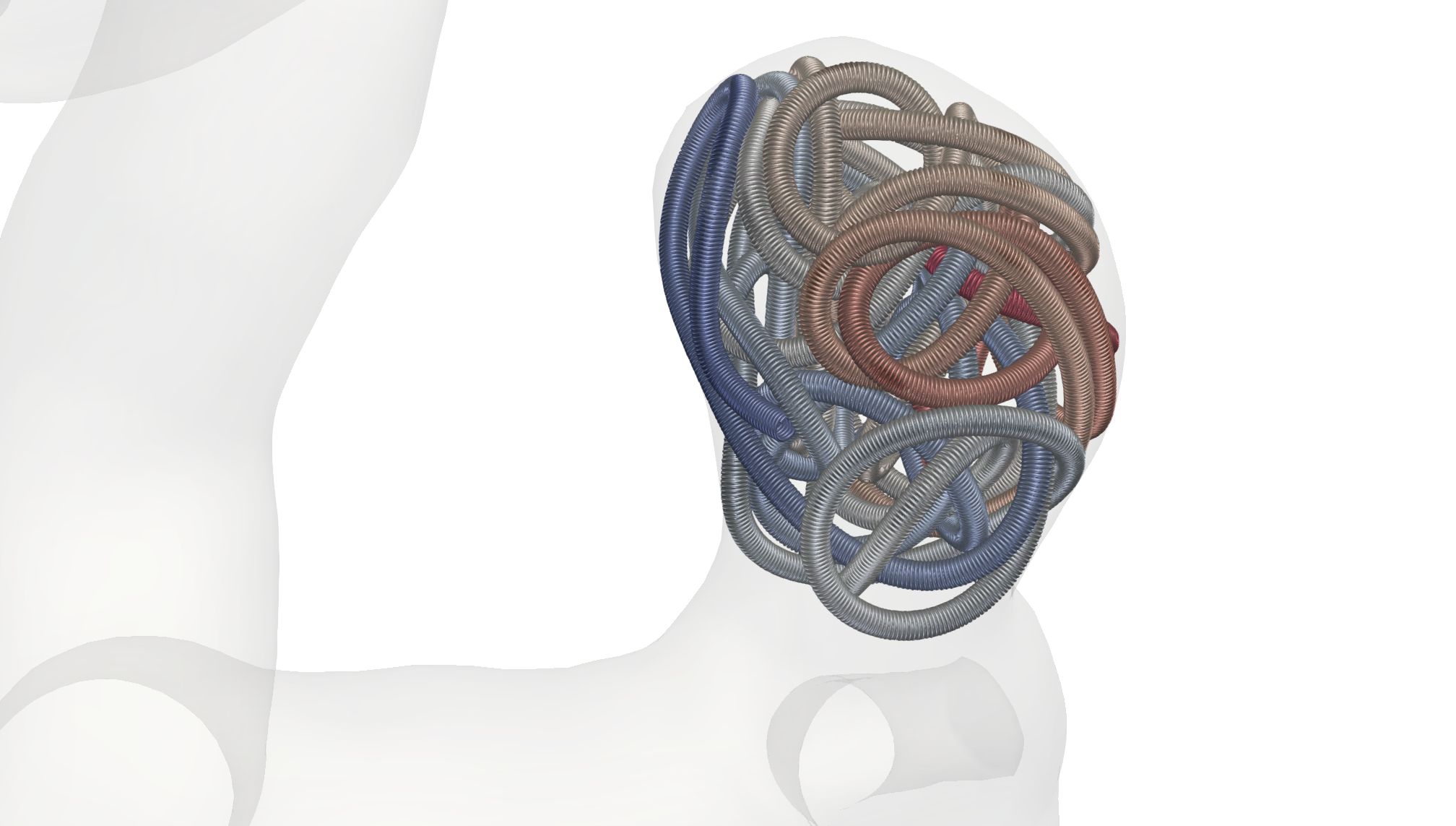}
    \caption{}
    \label{fig:ThreeD_3}
    \end{subfigure}
    \begin{subfigure}[t]{0.19\textwidth}
    \centering
        \includegraphics[width=\linewidth]{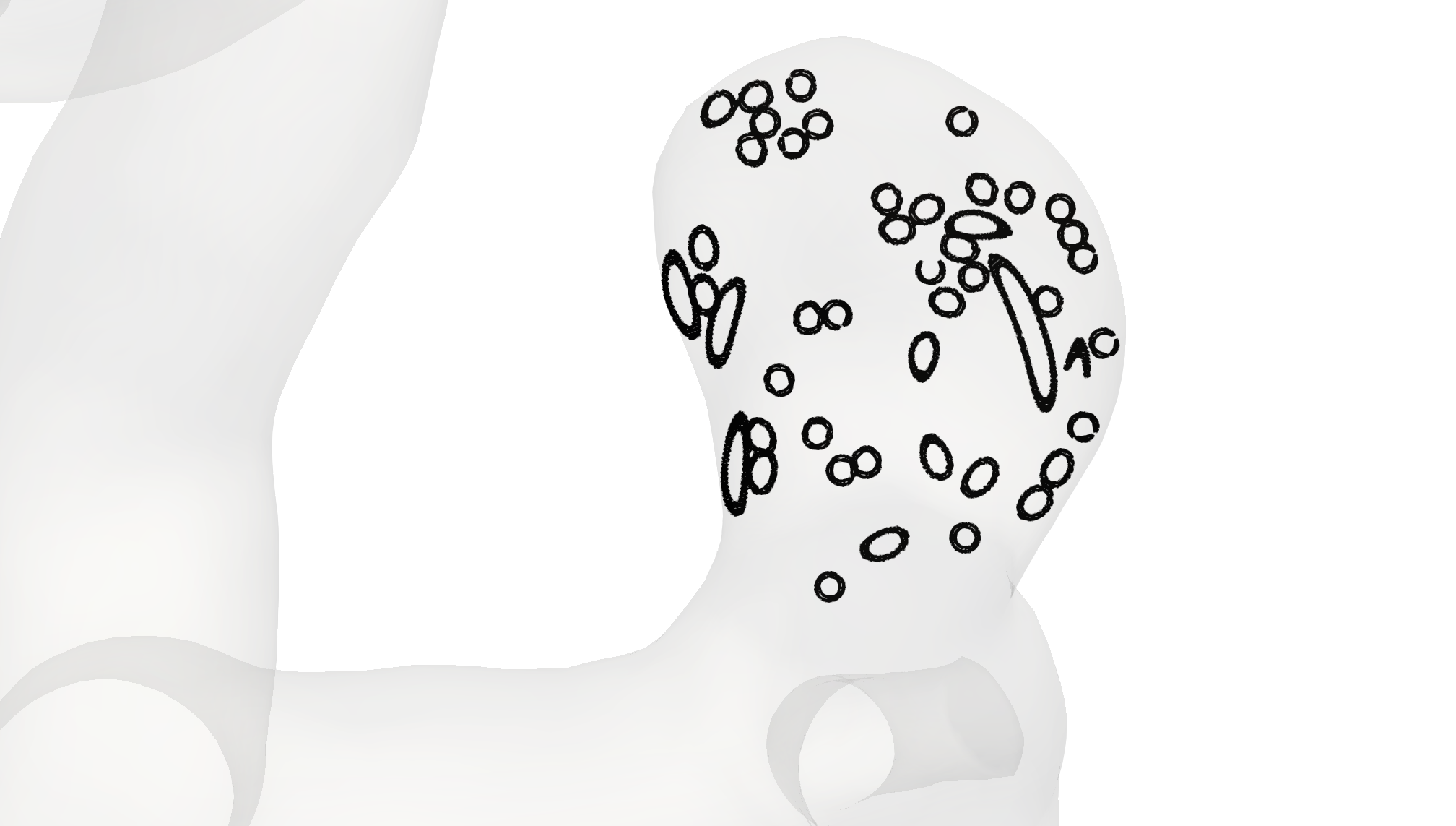}
    \caption{}
    \label{fig:ThreeD_cross}
    \end{subfigure}
    \caption{Insertion of coils with different natural curvature into the small aneurysm. The colors correspond to the coils relative arclength from $0$ to $1$. The coils are inserted until a packing density of $25\%$ is reached. The diameters are chosen as: $D_1=\SI{50}{\micro \meter}$, $D_2=\SI{305}{\micro \meter}$, $D_3=\SI{2}{\milli \meter}$. 
    Row (\subref{fig:Coil_Straight_ref})--(\subref{fig:Coil_Straight_cross}) corresponds to the straight coil, row (\subref{fig:Helix_ref})--(\subref{fig:Helix_cross}) to the helical coil and row (\subref{fig:ThreeD_ref})--(\subref{fig:ThreeD_cross}) to the 3D shaped coil.
    (\subref{fig:Coil_Straight_ref})--(\subref{fig:ThreeD_ref}) (vertically) show the natural shapes of the three coils considered.
    The three center columns show the insertion of each coil when $\SI{5}{\percent}$, $\SI{50}{\percent}$ and $\SI{100}{\percent}$ of the embolization process is completed. 
    In (\subref{fig:Coil_Straight_cross})--(\subref{fig:ThreeD_cross}) (again vertically), one can see a cross section of the fully inserted coil.}
        \label{fig:coil_composition}
\end{figure*}

In this section, we study virtual coil placements by means of our numerical method. We begin by simulating the embolization of coils with different shapes into the small aneurysm (see Fig. \ref{fig:Aneurisms} (top)). We use shapes that are a straight, a helix pre-shaped one and a 3D coil motivated by \citep{white2008coils}; see Fig. \ref{fig:coil_composition}. The point of insertion can be seen on the right of each picture at the tip of the micro-catheter.

The coils are chosen such that a $25\%$ global packing density is reached in the aneurysm (see red zone in Fig. \ref{fig:Aneurisms}). We observe that a straight coil leads to a placement where the coil is protruding into the artery. For the helix and 3D coil, this is not the case. Their $D_3$ diameter is set to $\SI{2}{\milli \meter}$ which is smaller than the aneurysms diameter (compare to Fig. \ref{fig:Aneurisms}) leading to a more uniformly 
distributed placement. In terms of their cross-section distribution, we can see that the distribution of the coil at the neck is the highest for the 3D coil whereas for the straight and helix coil there are still openings at the neck.

\begin{figure*}
    \centering
    \begin{subfigure}[b]{0.24\textwidth}
    \centering
    \includegraphics[width=\linewidth]{pictures/Small_Aneurysm_Shape_variation/Colorbar}
        \includegraphics[width=\linewidth]{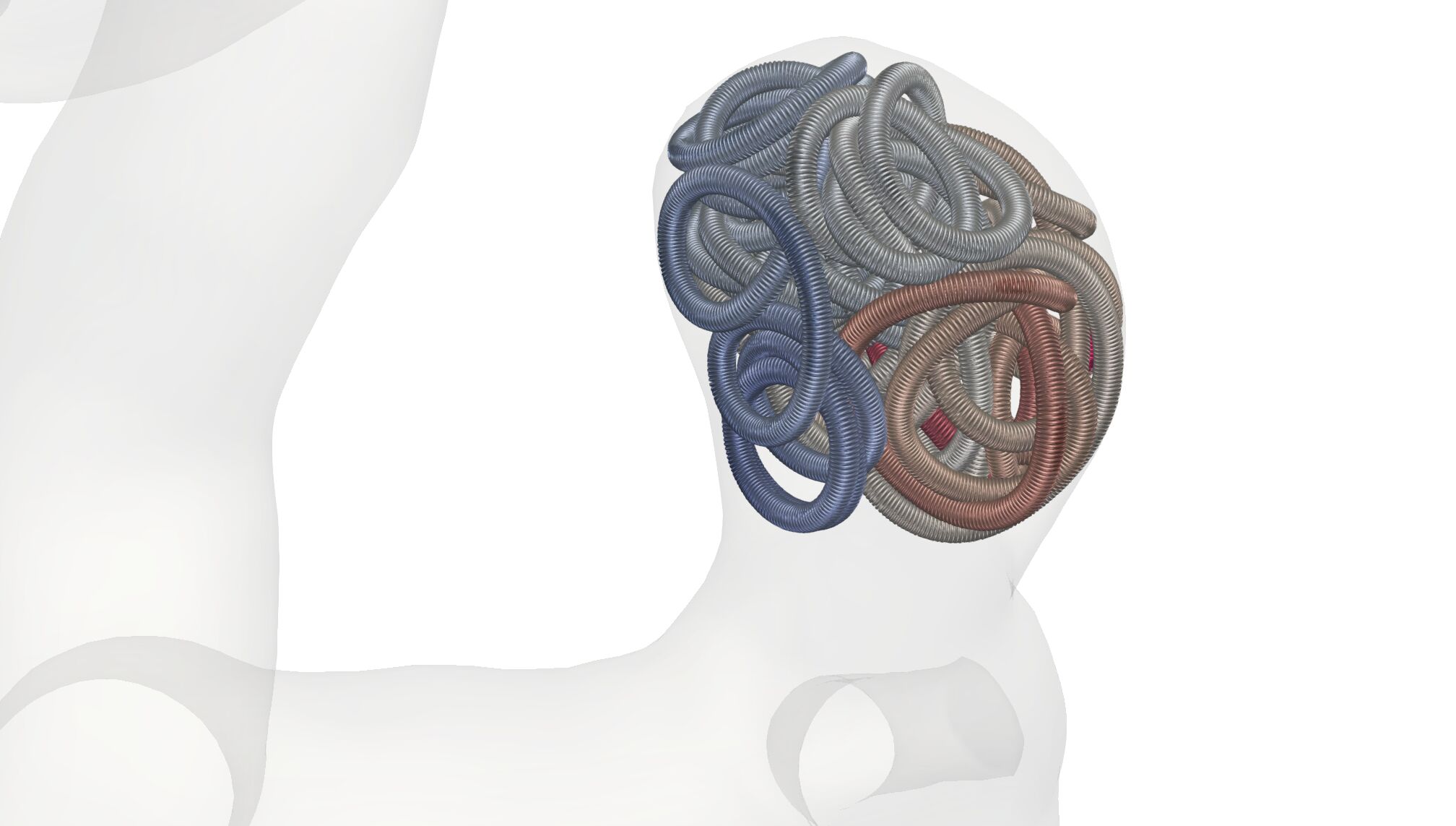}
    \caption{}
    \label{fig:Coil_Small_D2_2mm}
    \end{subfigure}
    \begin{subfigure}[b]{0.24\textwidth}
    \centering
        \includegraphics[width=\linewidth]{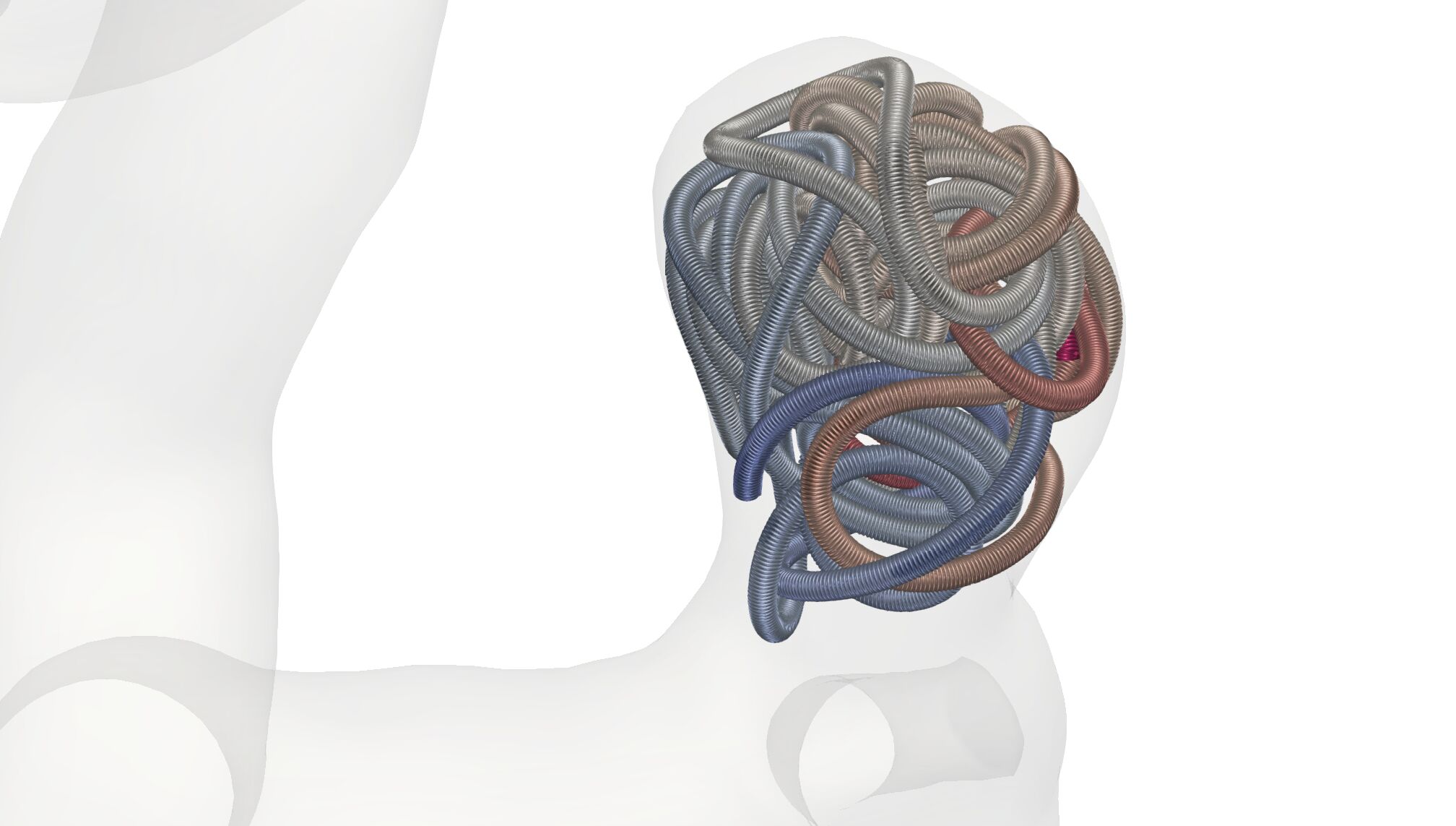}
    \caption{}
    \label{fig:Coil_Small_D2_4mm}
    \end{subfigure}
    \begin{subfigure}[b]{0.24\textwidth}
    \centering
        \includegraphics[width=\linewidth]{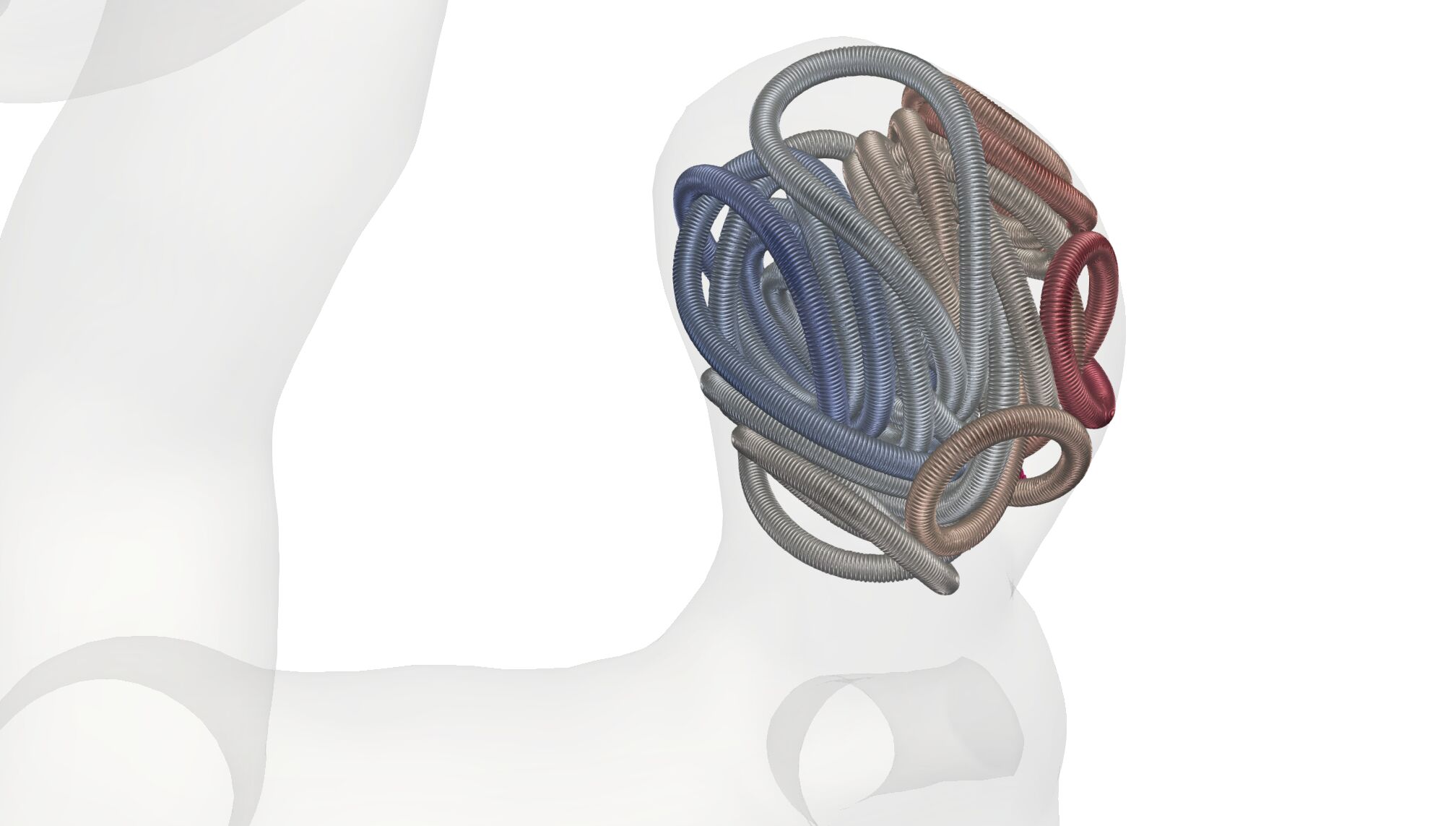}
    \caption{}
    \label{fig:Coil_Small_D2_6mm}
    \end{subfigure}
    \begin{subfigure}[b]{0.24\textwidth}
    \centering
        \includegraphics[width=\linewidth]{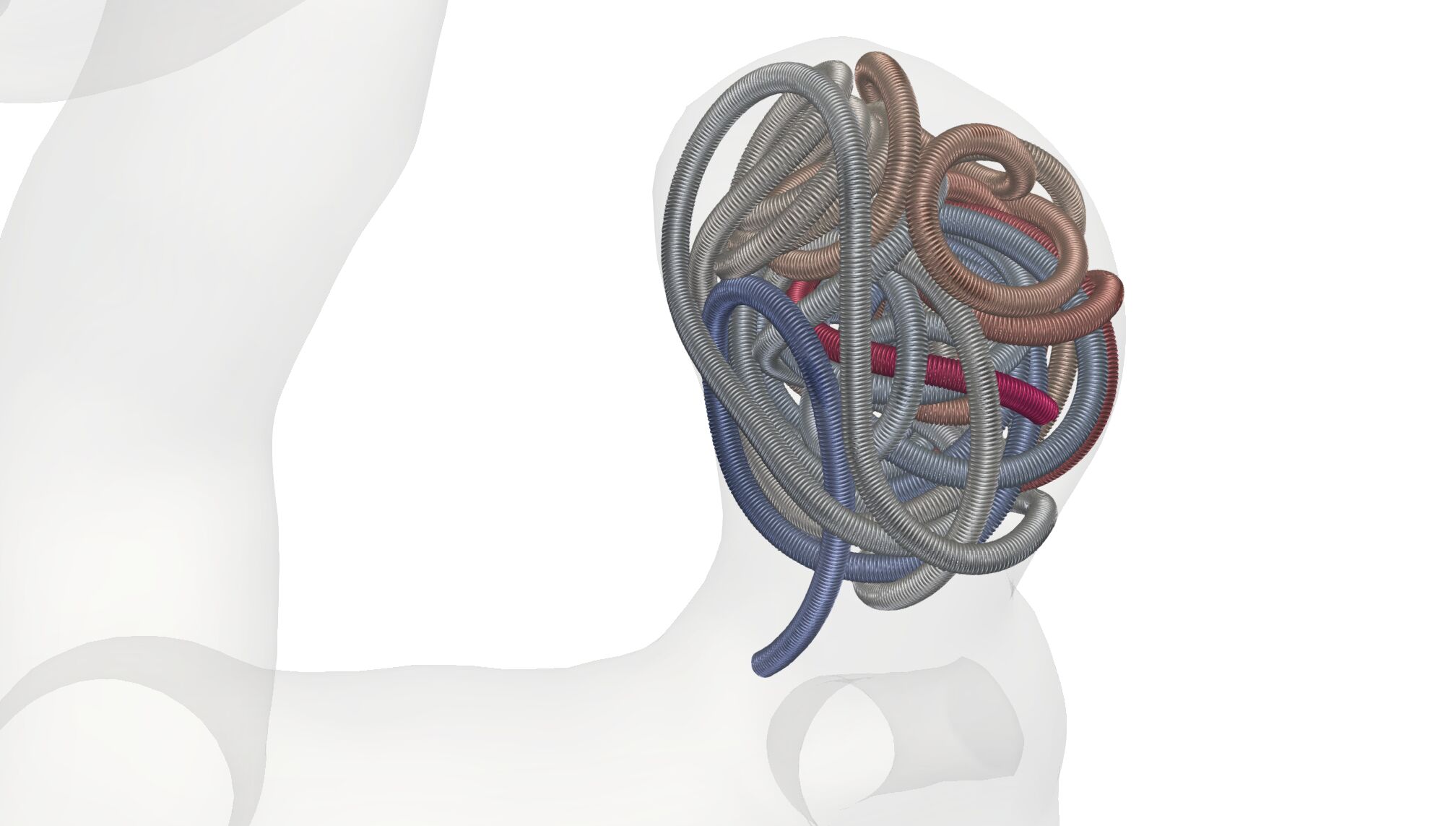}
    \caption{}
    \label{fig:Coil_Small_D2_8mm}
    \end{subfigure}\\
    \begin{subfigure}[b]{0.24\textwidth}
    \centering
        \includegraphics[width=\linewidth]{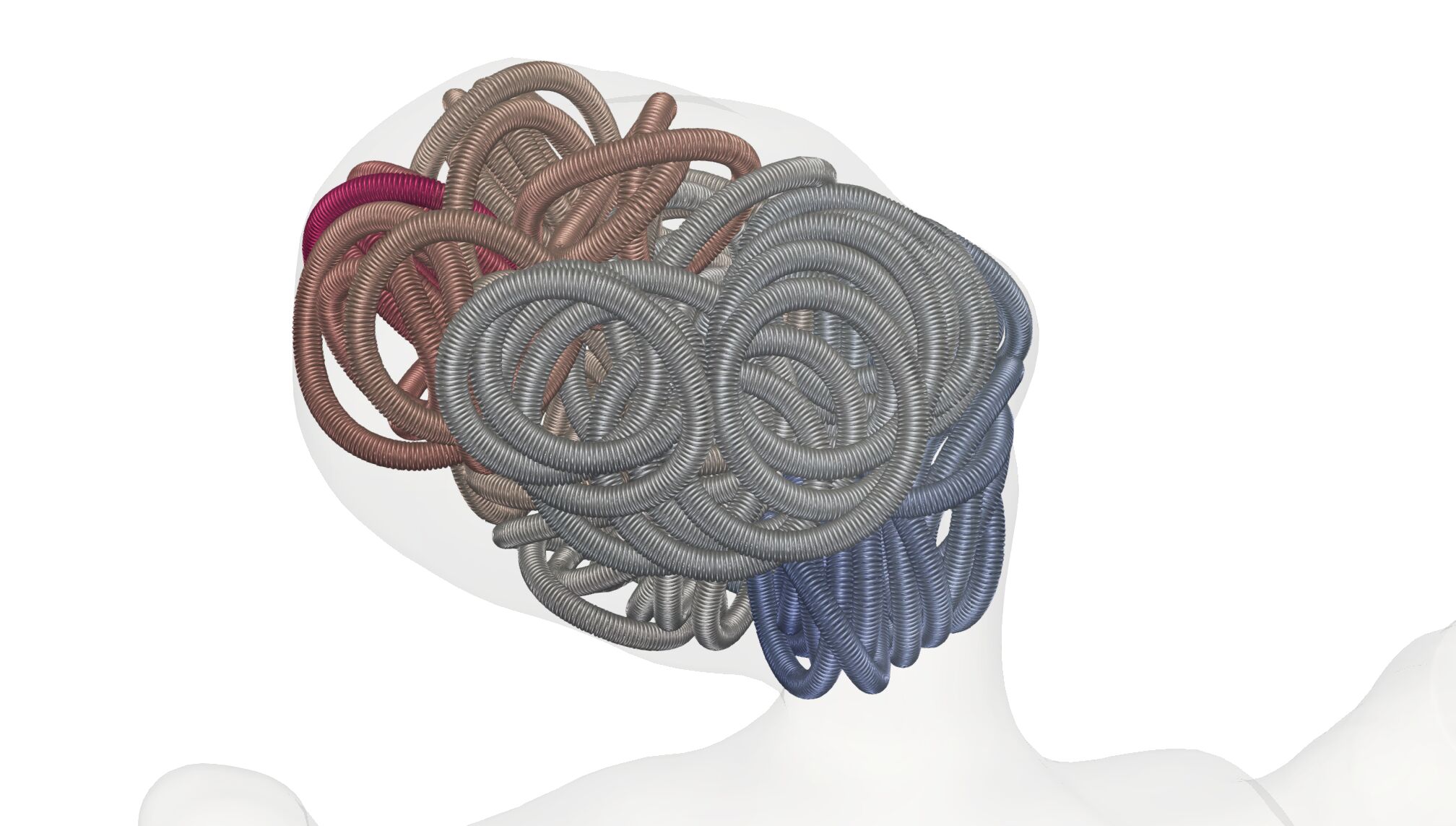}
    \caption{}
    \label{fig:Coil_Narrow_D2_2mm}
    \end{subfigure}
    \begin{subfigure}[b]{0.24\textwidth}
    \centering
        \includegraphics[width=\linewidth]{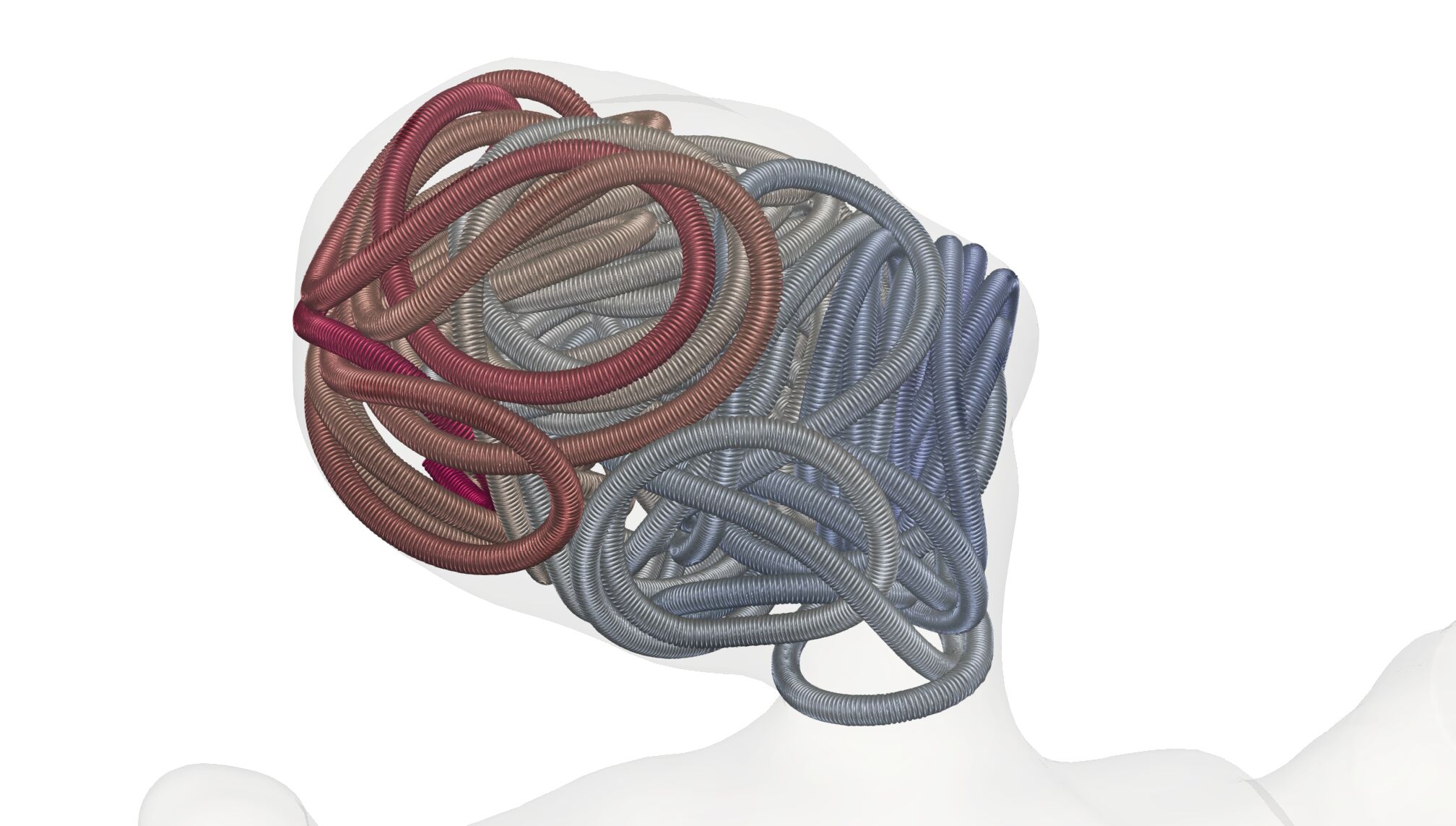}
    \caption{}
    \label{fig:Coil_Narrow_D2_4mm}
    \end{subfigure}
    \begin{subfigure}[b]{0.24\textwidth}
    \centering
        \includegraphics[width=\linewidth]{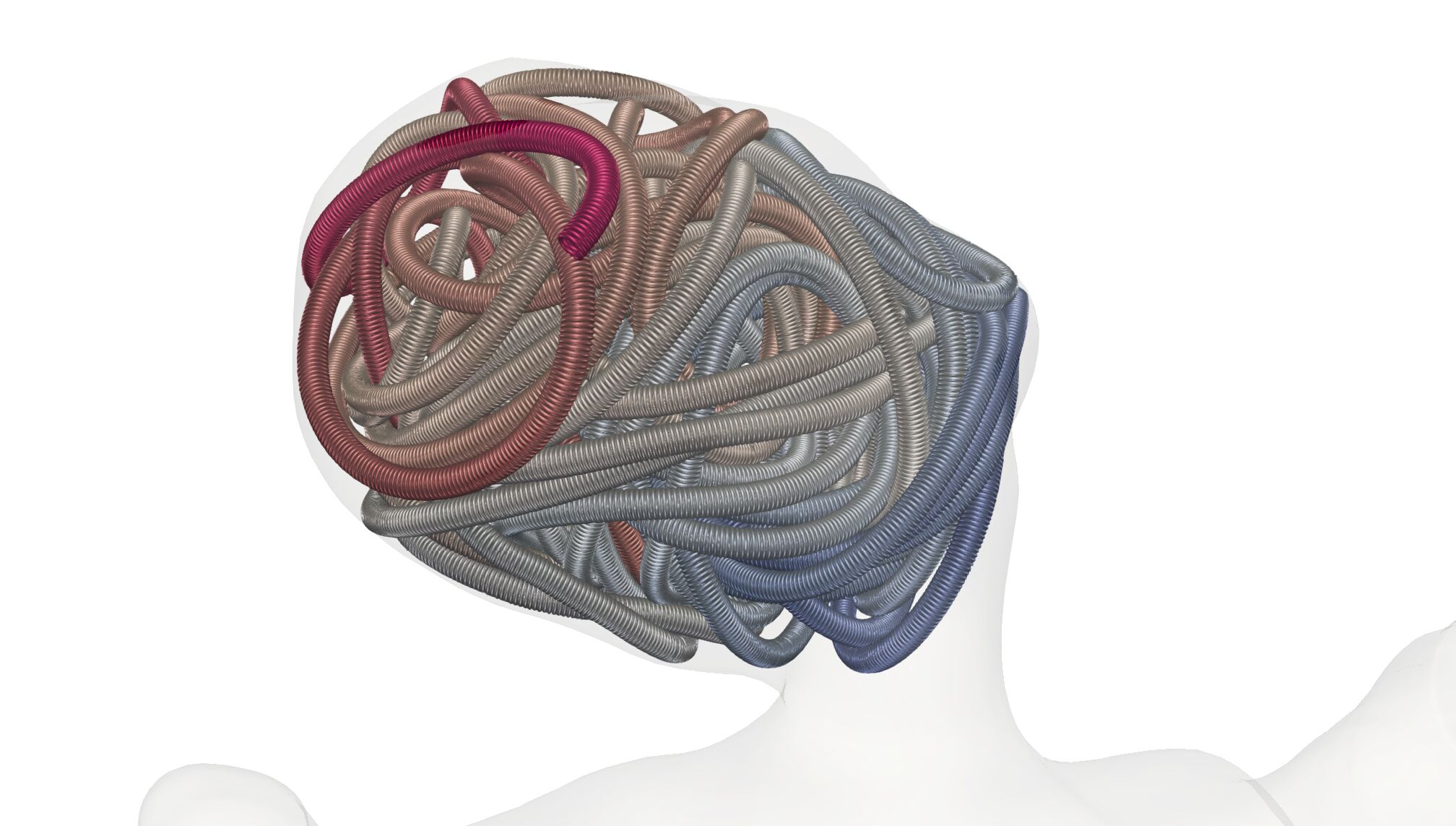}
    \caption{}
    \label{fig:Coil_Narrow_D2_6mm}
    \end{subfigure}
    \begin{subfigure}[b]{0.24\textwidth}
    \centering
        \includegraphics[width=\linewidth]{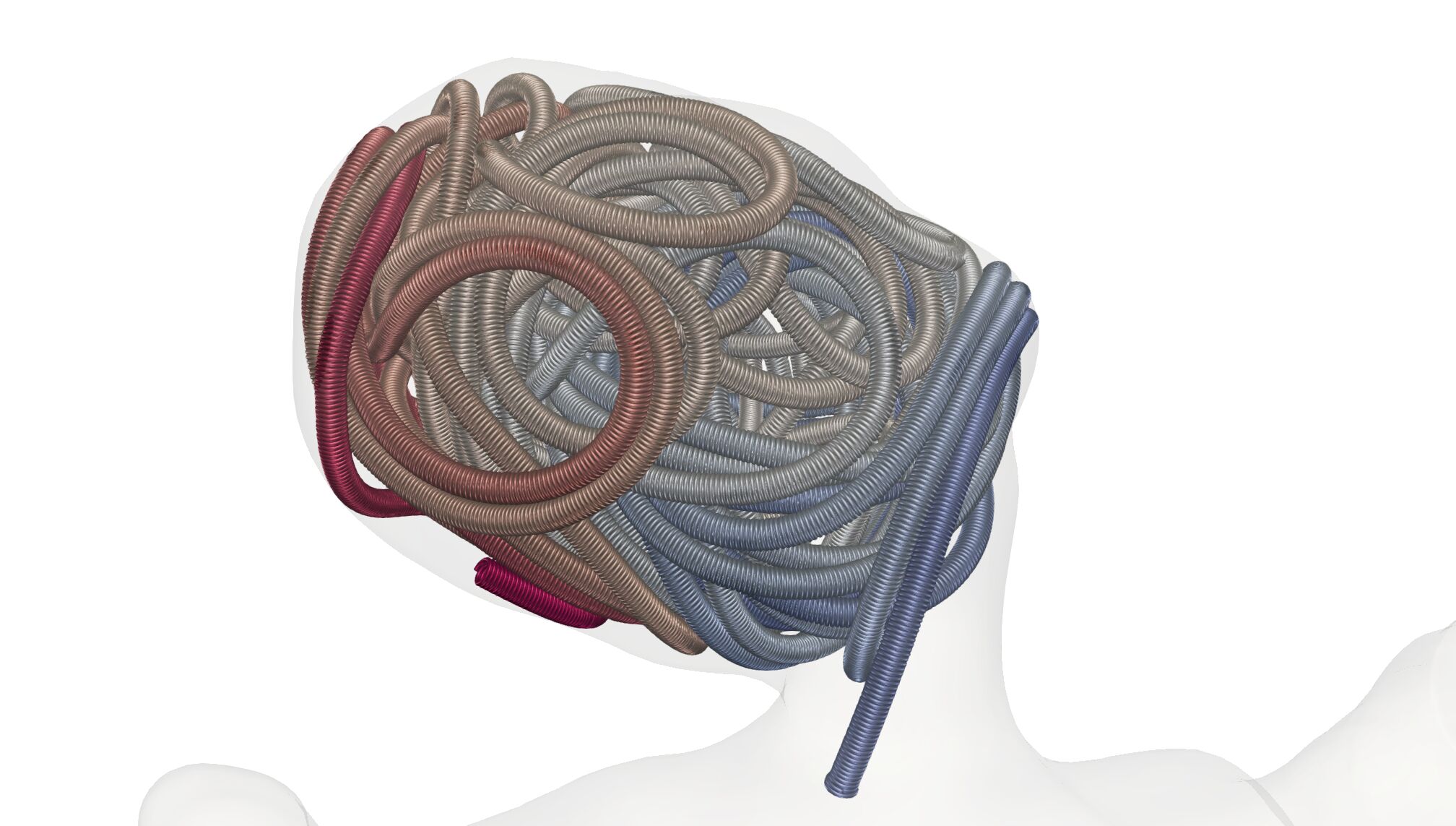}
    \caption{}
    \label{fig:Coil_Narrow_D2_8mm}
    \end{subfigure}\\
    \begin{subfigure}[b]{0.24\textwidth}
    \centering
        \includegraphics[width=\linewidth]{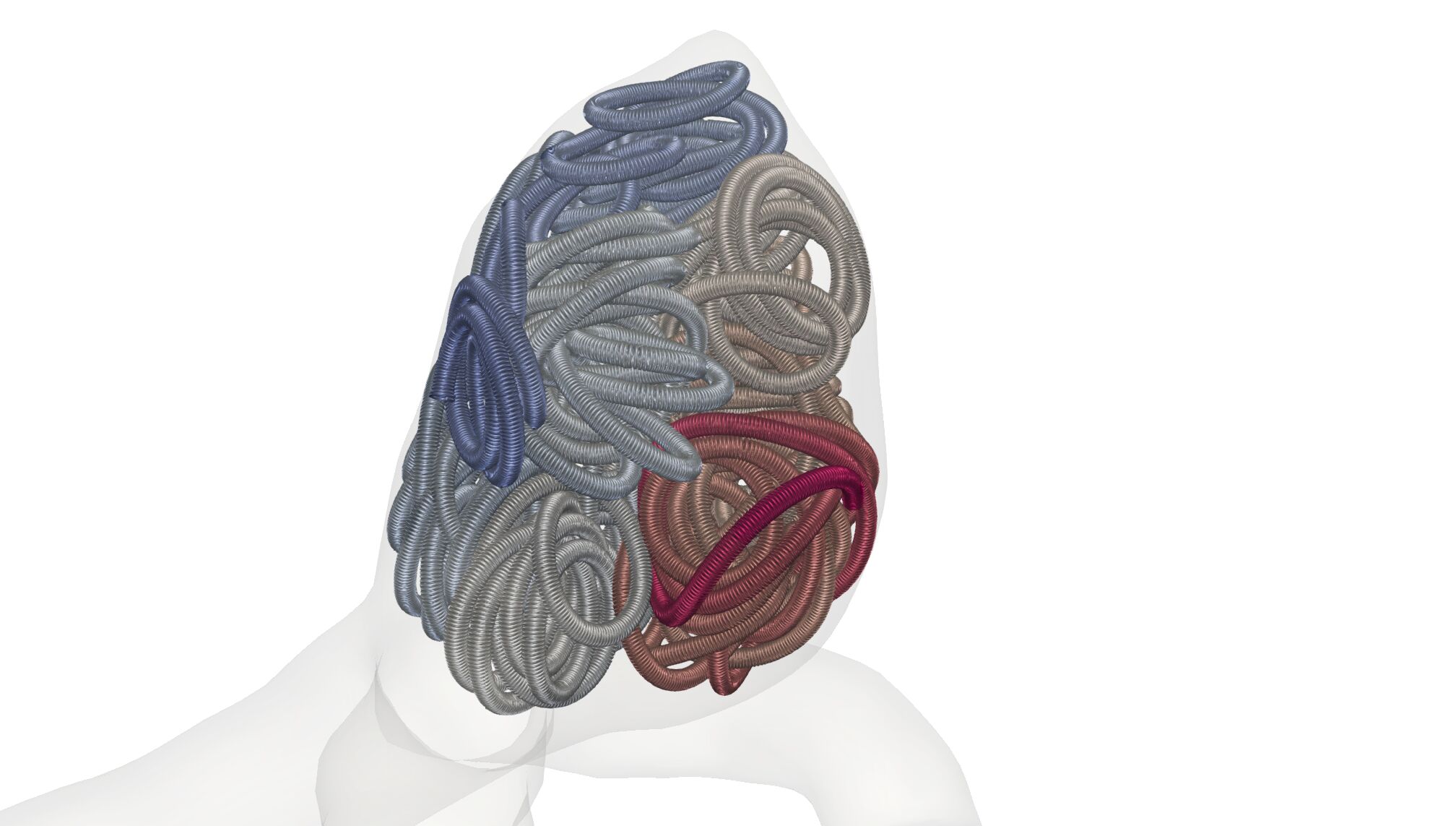}
    \caption{}
    \label{fig:Coil_Bir_D2_2mm}
    \end{subfigure}
    \begin{subfigure}[b]{0.24\textwidth}
    \centering
        \includegraphics[width=\linewidth]{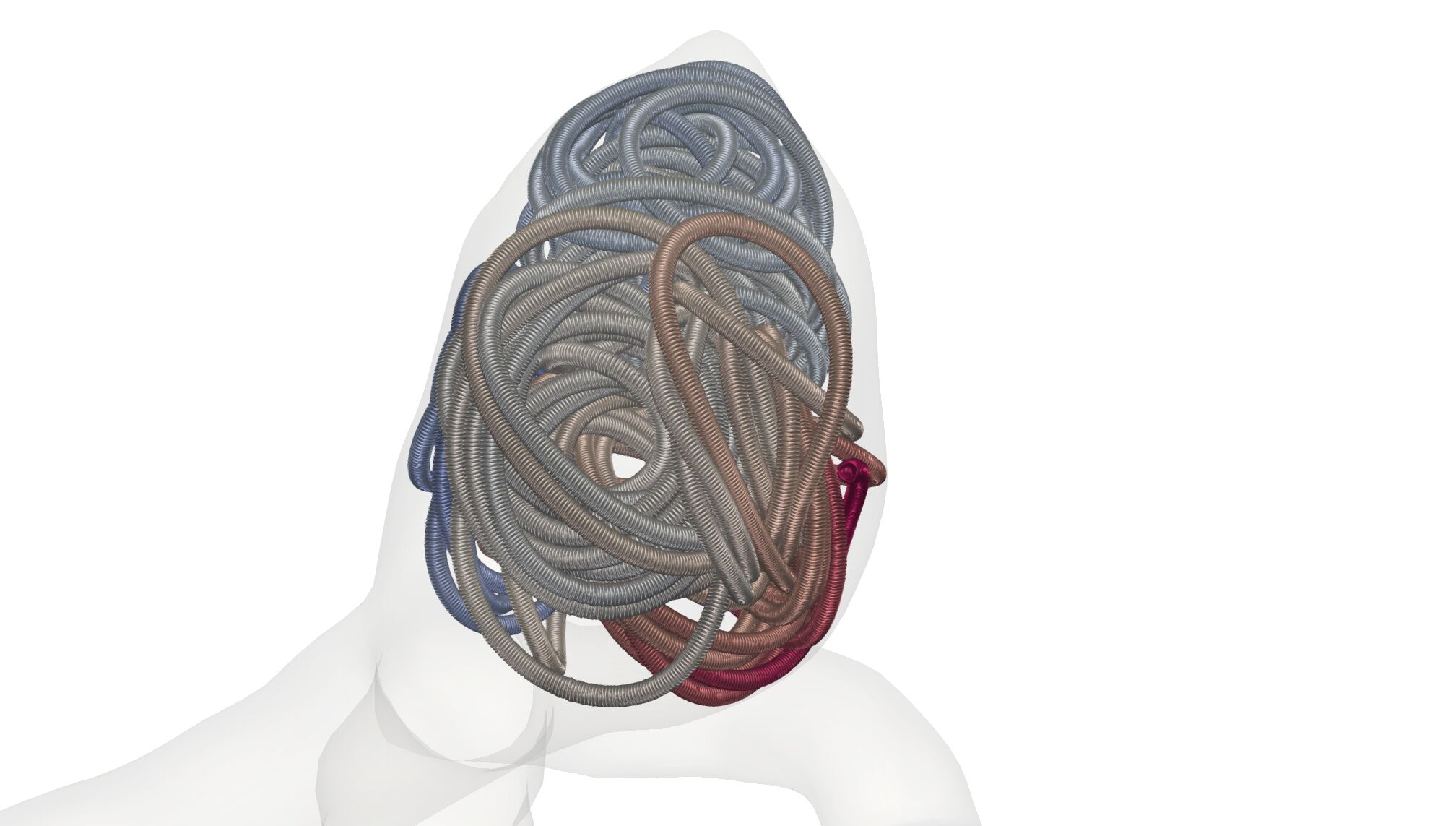}
    \caption{}
    \label{fig:Coil_Bir_D2_4mm}
    \end{subfigure}
    \begin{subfigure}[b]{0.24\textwidth}
    \centering
        \includegraphics[width=\linewidth]{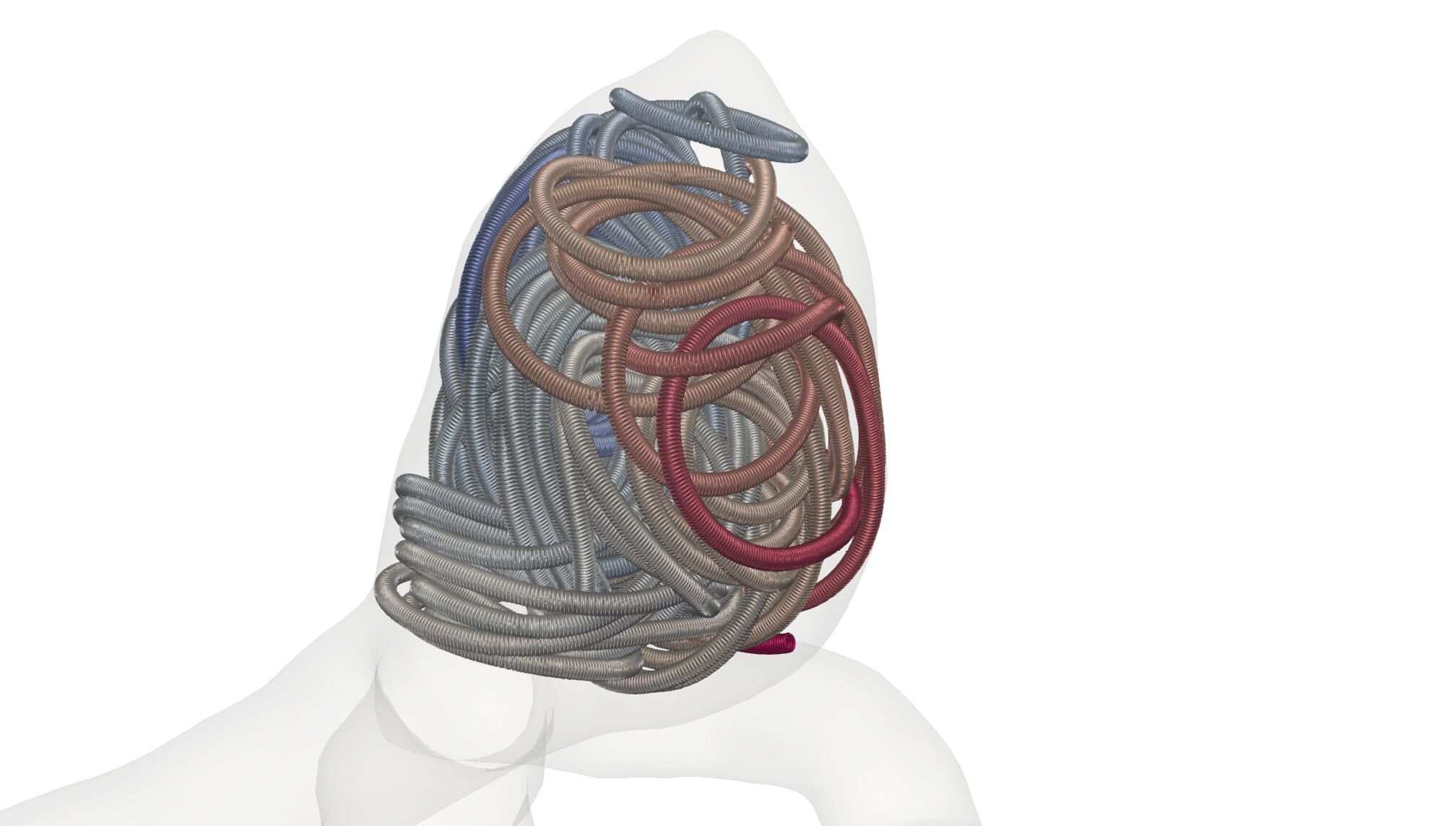}
    \caption{}
    \label{fig:Coil_Bir_D2_6mm}
    \end{subfigure}
    \begin{subfigure}[b]{0.24\textwidth}
    \centering
        \includegraphics[width=\linewidth]{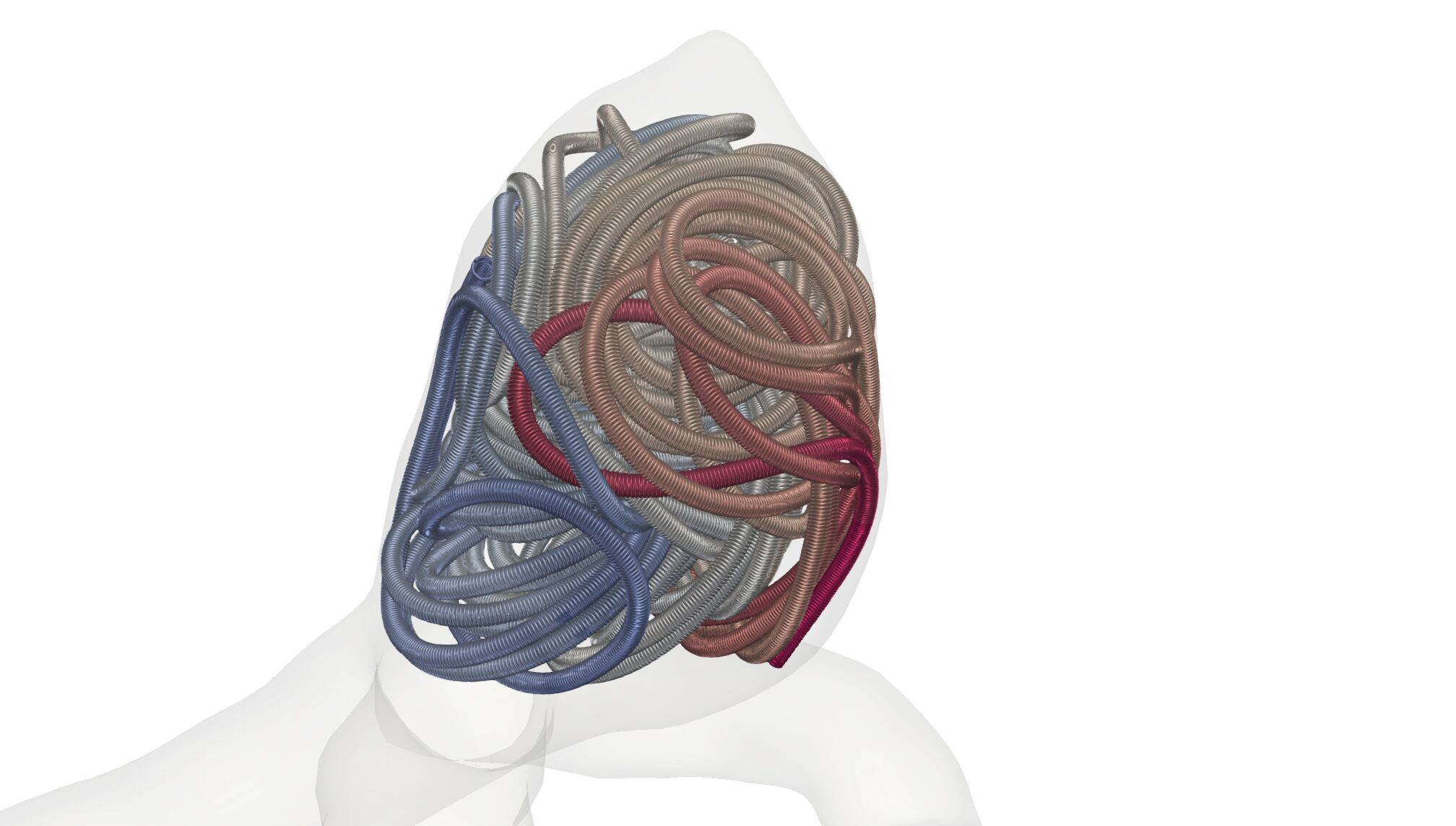}
    \caption{}
    \label{fig:Coil_Bir_D2_8mm}
    \end{subfigure}\\
    \caption{Insertion of helical-coils into the different aneurysm geometries. The colors correspond to the coils relative arclength from $0$ to $1$. Row (\subref{fig:Coil_Small_D2_2mm})--(\subref{fig:Coil_Small_D2_8mm}) corresponds to the small aneurysm geometry, (\subref{fig:Coil_Narrow_D2_2mm})--(\subref{fig:Coil_Narrow_D2_8mm}) the narrow neck aneurysm geometry and (\subref{fig:Coil_Bir_D2_2mm})--(\subref{fig:Coil_Bir_D2_8mm}) to the bifurcation aneurysm geometry. The coils are inserted until a packing density of $25\%$ is reached. The diameters are chosen as: $D_1=\SI{50}{\micro \meter}$, $D_2=\SI{305}{\micro \meter}$ and $D_3$ is chosen from left to right as $\SI{2}{\milli\meter}, \SI{4}{\milli\meter}, \SI{6}{\milli\meter}$ and $\SI{8}{\milli\meter}$.}
        \label{fig:Composition_Helix}
\end{figure*}

Next we show how the different aneurysm geometries (see again Fig. \ref{fig:Aneurisms}) affect the placement of coils.  In Fig. \ref{fig:Composition_Helix} from top to bottom, the three aneurysm geometries are shown while from left to right, we set the $D_3$ diameter of the helix coil to $\SI{2}{\milli\meter}, \SI{4}{\milli\meter}, \SI{6}{\milli\meter}$ and $\SI{8}{\milli\meter}$.

In case of the small aneurysm which can be seen in the top sequence, a smaller $D_3$ radius reduces the probability of the coil protruding into the parent vessel. The three cases on the right show that a higher $D_3$ value increases this chance.
In the middle case, the neck of the aneurysm is much smaller compared to its dome diameter making it a case that can be coiled more easily. Looking at the variations of $D_3$, we can see that up to a $D_3$ diameter of $\SI{8}{\milli\meter}$ there is no protrusion into the main artery. For $D_3 =\SI{2}{mm}$, high local packing densities are possible leading to a poorly (w.r.t. uniformity) distributed coil close to the walls of the aneurysm. Finally, for the bifurcating aneurysm that is located at the bottom of Fig. \ref{fig:coil_composition}, one can see that in all cases, the coil is in proximity of the main vessel. The neck of this aneurysm is relatively large and thus, it is more difficult to insert a coil without protrusion. Setting $D_3 =\SI{6}{\milli \meter}$ yields the best placement, meaning that for this case the protrusion into the parent vessel is not excessive.


\subsection{Analyzing Coil Distributions}
To assess grades of occlusion by the RROC, we proceed as in Fig. \ref{fig:Custom_RROC}. First a SDF is generated. The SDF enables a natural partitioning of the aneurysm by its level sets. In our case, we choose the level set in such a way that the aneurysm core and boundary region are two partitions of equal volume  (see Fig. \ref{fig:Custom_RROC} (\subref{fig:Hom_BoundaryCore}) blue and red region). Having the coil distribution in the boundary region enables us to judge if a coil belongs to the Class IIIb or vice versa the filling of the core region which - if insufficient - leads to a Class IIIa grade. Finally, to assess Class II, we analyze the volume fraction in a sphere at the neck of the aneurysm (see Fig. \ref{fig:Custom_RROC} (\subref{fig:Hom_Sphere})).

To actually calculate the volume fractions, we use the voxelization for each coil as discussed in Section \ref{sec:Voxilation}. To analyze the sensitivity of our numerical coil model with respect to the simulation parameters, we sample the simulation parameters, i.e., the Young's modulus, from a distribution $Z \sim \mathcal{D}$ and generate the respective coil placements with their voxelization $\psi(Z)$. Those can then be used to calculate the empirical mean and standard deviation of the coil distribution via
\begin{align}
    \overline{\psi} & = \frac{1}{N}\sum\limits_{i=1}^N \psi(Z_{i}),  \nonumber \\
    \psi_{\sigma}^2 &= \frac{1}{N}\sum\limits_{i=1}^N (\overline{\psi}-\psi(Z_{i}))^2 .
    \label{EQ:porosity_stats_local}
\end{align}
Note that we assume that the voxelized coil distributions $\psi$ are defined on the same domain and therefore can be summed up in the notation above in a point-wise manner. From this, we calculate the volume of a mean coil distribution in a subregion of our aneurysm domain $V$ and its standard deviation with respect to the voxelized coil distribution $\psi(Z_{i})$ by
\begin{align}
    \overline{\psi}_V & =\int_{V} \overline{\psi}~dx,
    \nonumber \\ 
    \psi_{\sigma, V}^2 &=\frac{1}{N}\sum\limits_{i=1}^N\bigg(\int_{V} \overline{\psi}-\psi(Z_{i})~dx\bigg)^2\label{EQ:porosity_stats_global}
\end{align}
which enables us to generate a volume fraction $\overline{\psi}_V/\widetilde{V}$ with respect to a reference volume $\widetilde{V}$. For the core, boundary and sphere region in the aneurysm (see Fig. \ref{fig:Custom_RROC}), we compute the following volume fractions 
\begin{align}
    \overline{\psi}_{BA}=\frac{\overline{\psi}_{V_\text{B}}}{V_\text{A}},
    \overline{\psi}_{CA}=\frac{\overline{\psi}_{V_\text{C}}}{V_\text{A}}, 
    \overline{\psi}_{AA}=\frac{\overline{\psi}_{V_\text{A}}}{V_\text{A}},
    \overline{\psi}_{SS}=\frac{\overline{\psi}_{V_\text{S}}}{V_\text{S}}
    \label{EQ:volume_fractions}
\end{align}
where $V_{\text{C}}$, $V_{\text{B}}$, $V_{\text{A}}$ and $V_{\text{S}}$ are the volumes of the core, boundary and full aneurysm and the sphere with $V_{\text{A}}=V_{\text{B}}+V_{\text{C}}$. Note that in $V_{\text{S}}$, only the volume of the sphere within the aneurysm is considered.

\begin{figure}[!htb]
    \centering
    \begin{subfigure}[t]{0.22\textwidth}
    \centering
        \includegraphics[width=\linewidth]{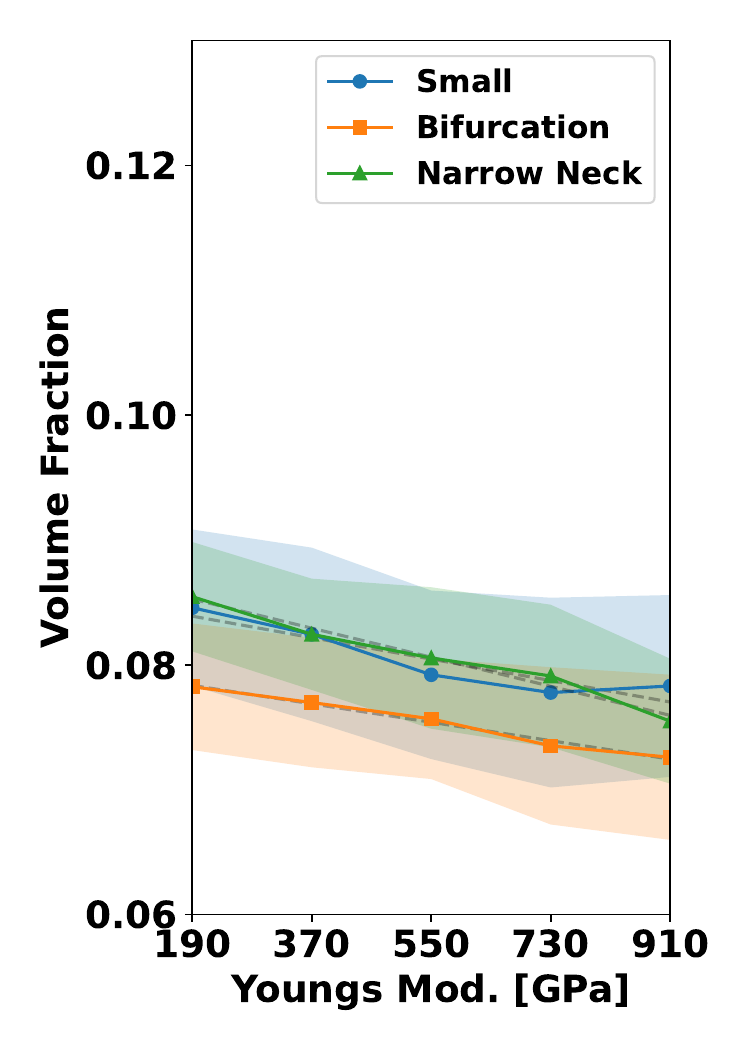}
        \subcaption{}\label{fig:VolumeFraction_Evar_Helix_Boundary}
    \end{subfigure}
    \begin{subfigure}[t]{0.22\textwidth}
    \centering
        \includegraphics[width=\linewidth]{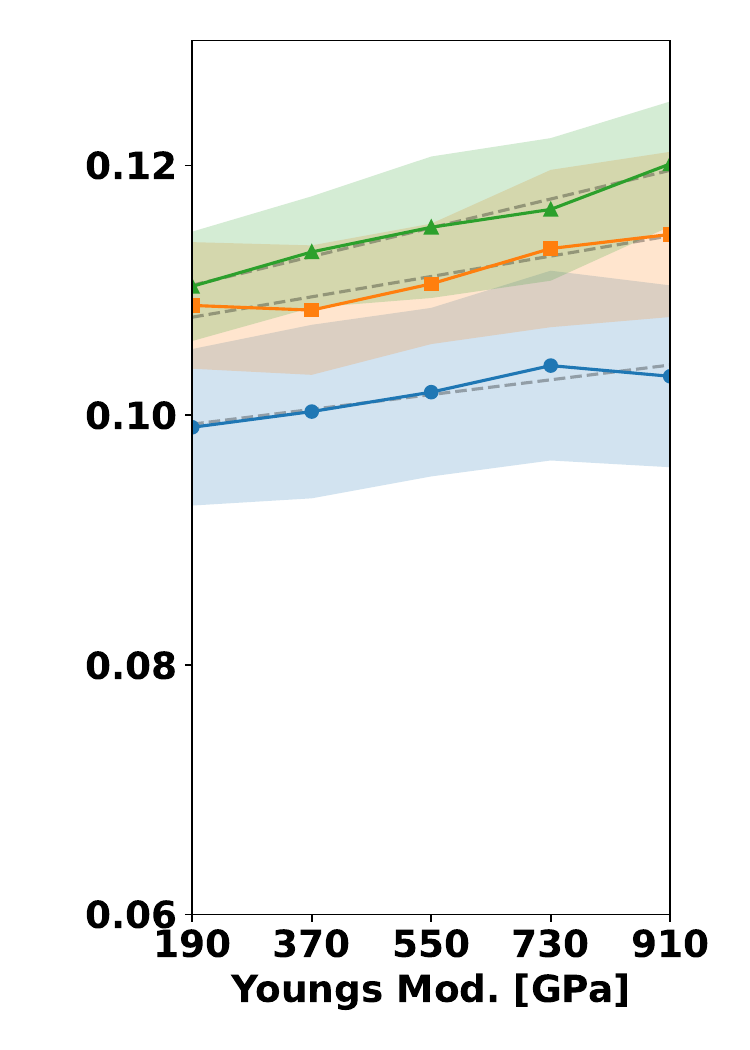}
        \subcaption{}
        \label{fig:VolumeFraction_Evar_Helix_Core}
    \end{subfigure}
    \begin{subfigure}[t]{0.22\textwidth}
    \centering
        \includegraphics[width=\linewidth]{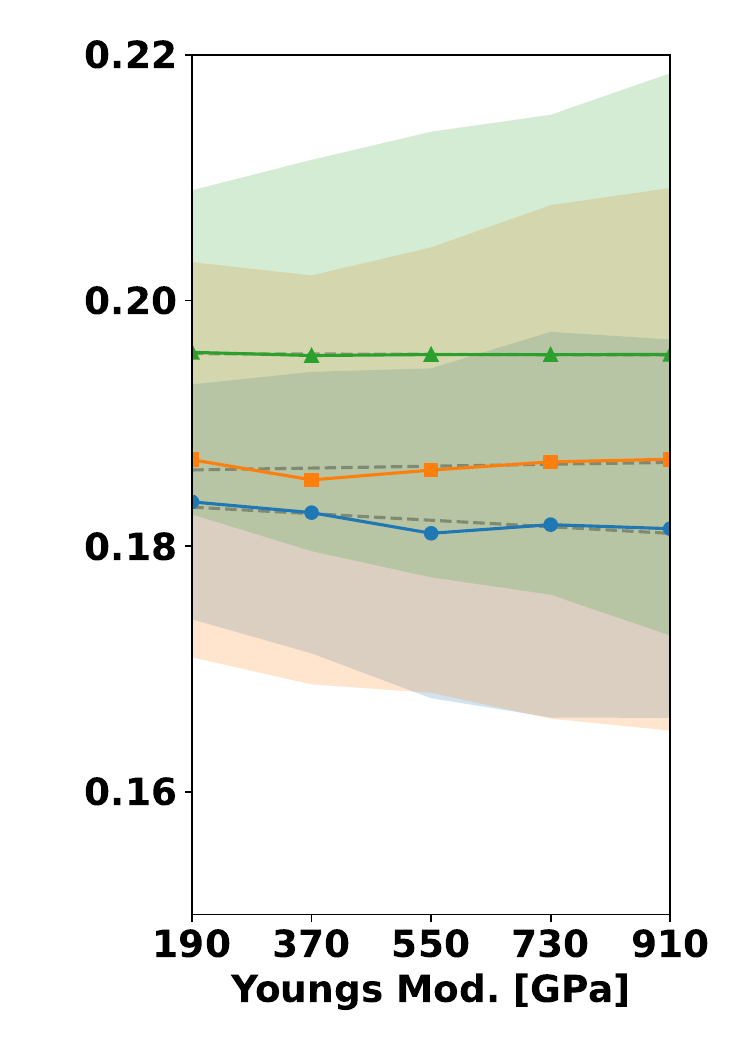}
        \subcaption{}
        \label{fig:VolumeFraction_Evar_Helix_Total}
    \end{subfigure}
    \begin{subfigure}[t]{0.22\textwidth}
    \centering
        \includegraphics[width=\linewidth]{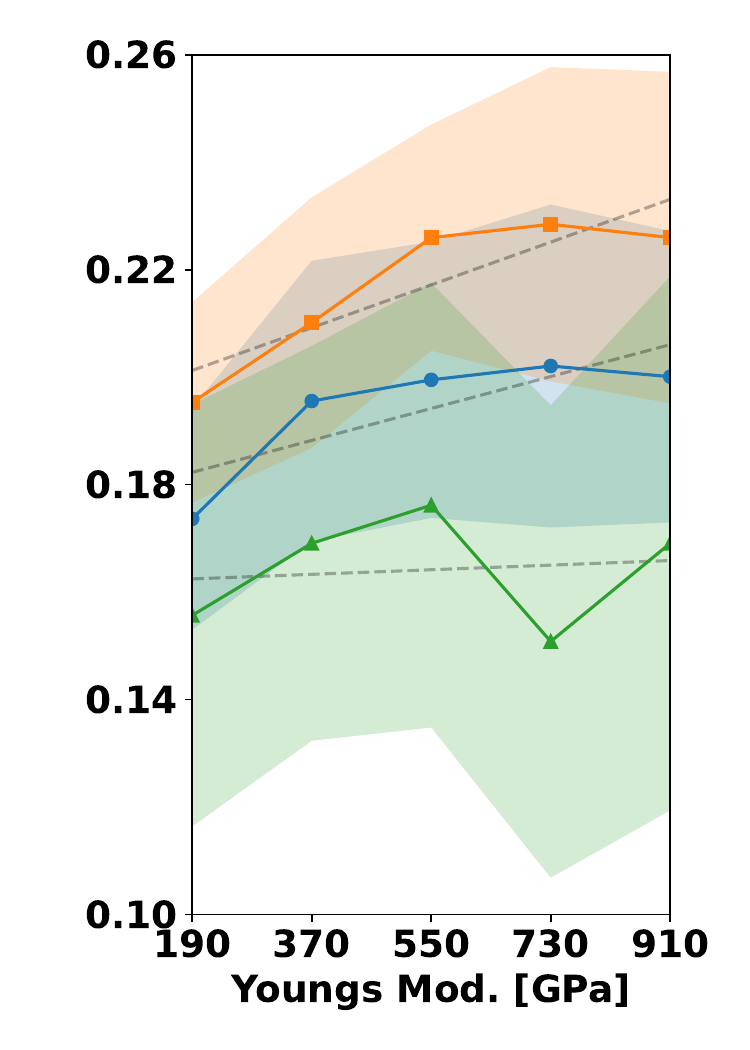}
        \subcaption{}
        \label{fig:VolumeFraction_Evar_Helix_Sphere}
    \end{subfigure}
    \caption{\textbf{Variation of Young's modulus Helix coil}: (\subref{fig:VolumeFraction_Evar_Helix_Boundary}) volume fraction of coil in aneurysm boundary, (\subref{fig:VolumeFraction_Evar_Helix_Core}) volume fraction of coil in aneurysm core, (\subref{fig:VolumeFraction_Evar_Helix_Total}) total volume fraction of coil in aneurysm, (\subref{fig:VolumeFraction_Evar_Helix_Sphere}) volume fraction of coil in the sphere at the aneurysm neck}
    \label{fig:VolumeFraction_Evar_Helix}
\end{figure}

\begin{figure}[!htb]
    \centering
    \begin{subfigure}[t]{0.22\textwidth}
    \centering
        \includegraphics[width=\linewidth]{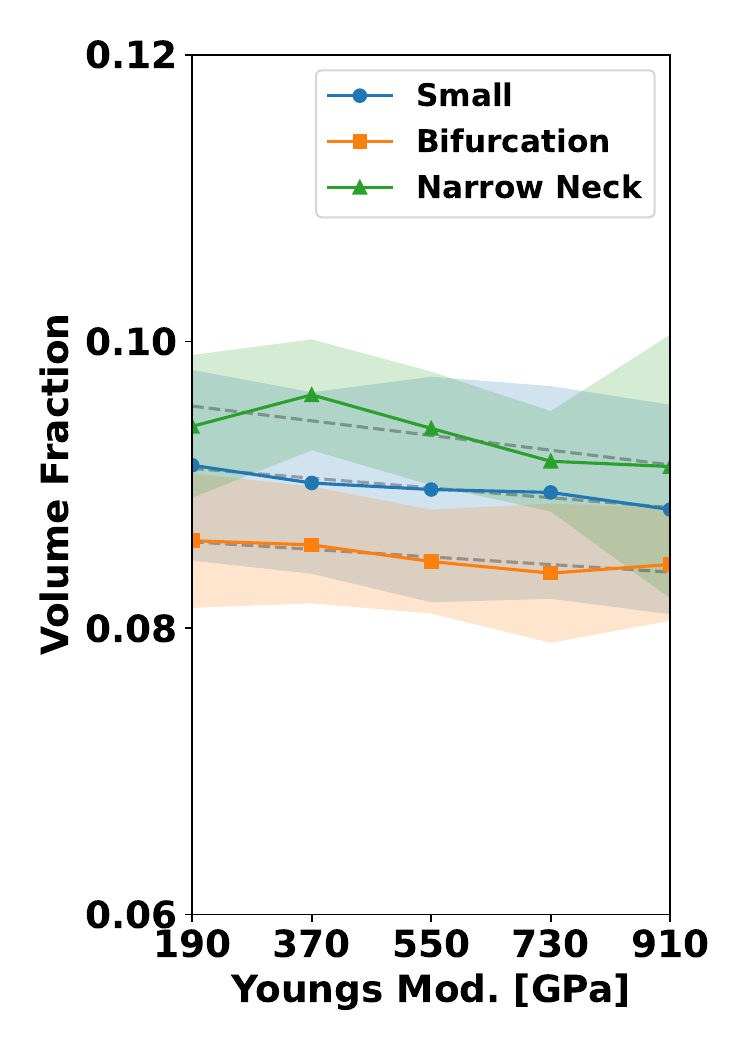}
        \subcaption{}
        \label{fig:VolumeFraction_Evar_ThreeD_Boundary}
    \end{subfigure}
    \begin{subfigure}[t]{0.22\textwidth}
    \centering
        \includegraphics[width=\linewidth]{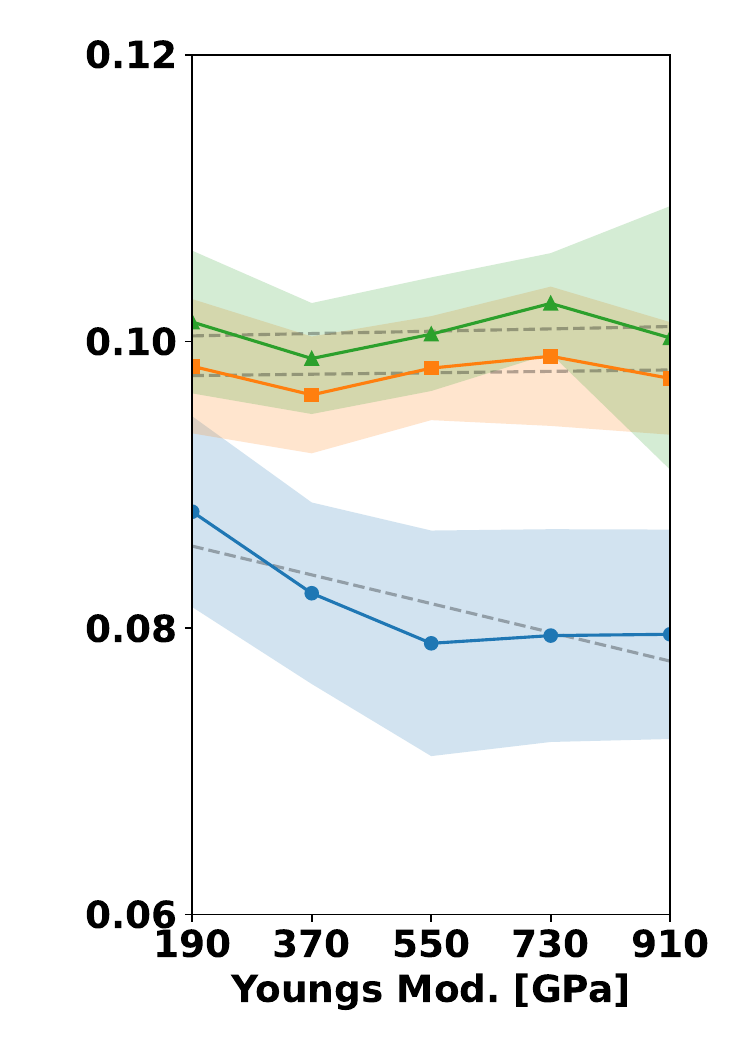}
        \subcaption{}
        \label{fig:VolumeFraction_Evar_ThreeD_Core}
    \end{subfigure}
    \begin{subfigure}[t]{0.22\textwidth}
    \centering
        \includegraphics[width=\linewidth]{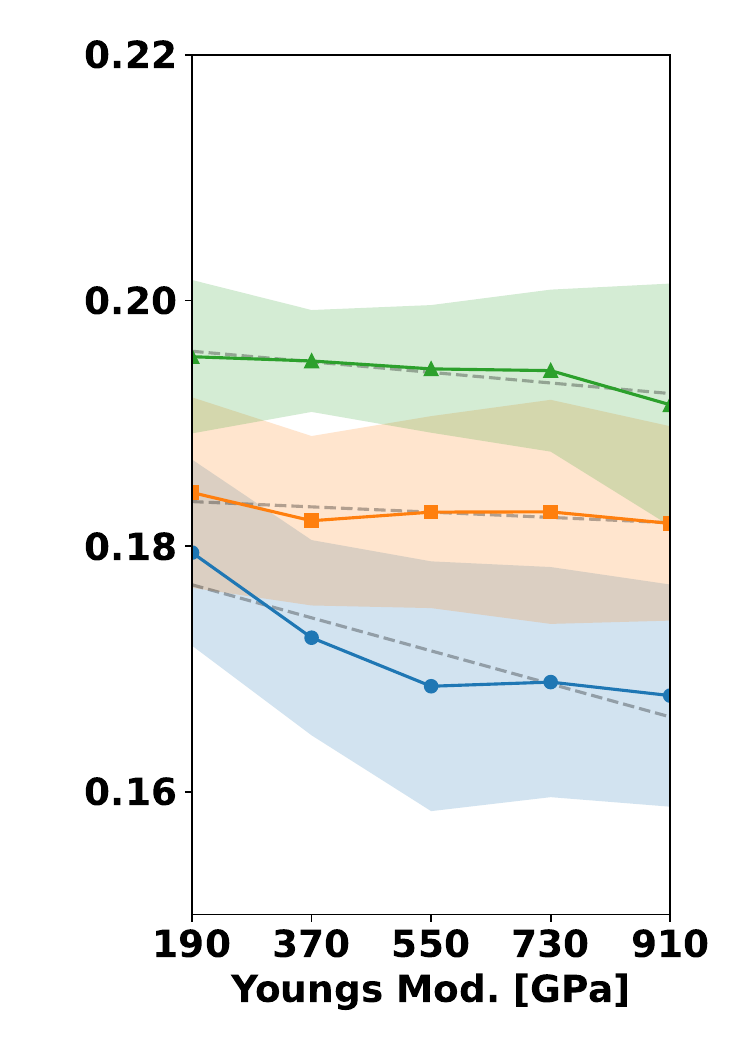}
        \subcaption{}
        \label{fig:VolumeFraction_Evar_ThreeD_Total}
    \end{subfigure}
    \begin{subfigure}[t]{0.22\textwidth}
    \centering
        \includegraphics[width=\linewidth]{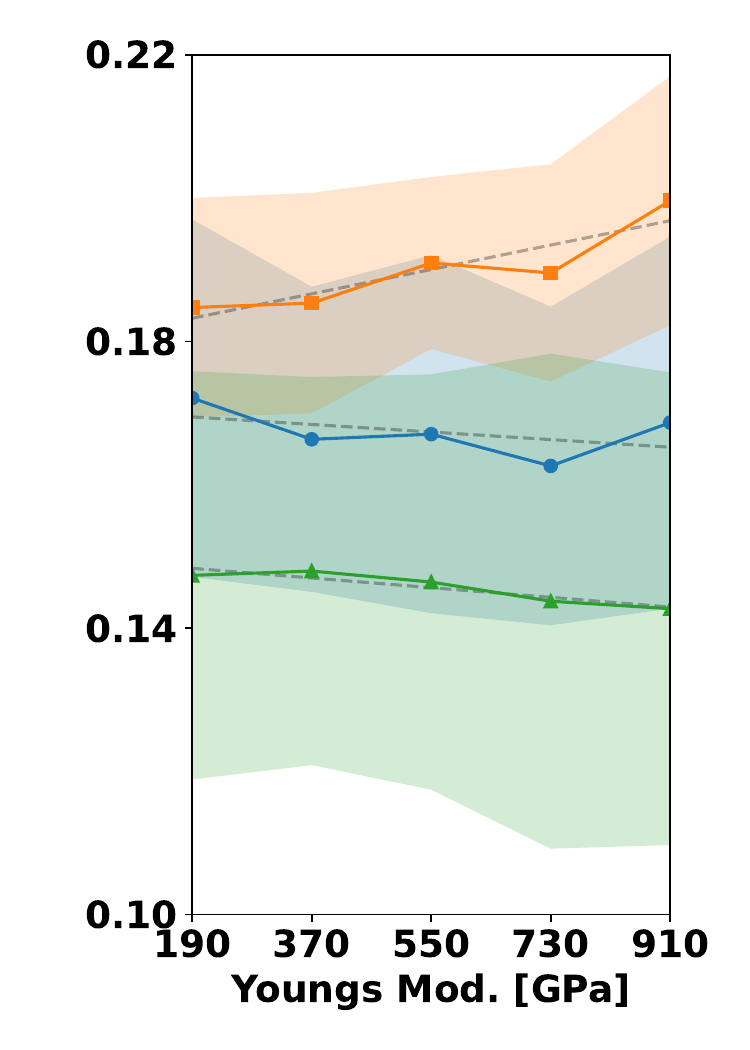}
        \subcaption{}
        \label{fig:VolumeFraction_Evar_ThreeD_Sphere}
    \end{subfigure}
    \caption{\textbf{Variation of Young's modulus 3D coil}: (\subref{fig:VolumeFraction_Evar_ThreeD_Boundary}) volume fraction of coil in aneurysm boundary, (\subref{fig:VolumeFraction_Evar_ThreeD_Core}) volume fraction of coil in aneurysm core, (\subref{fig:VolumeFraction_Evar_ThreeD_Total}) total volume fraction of coil in aneurysm, (\subref{fig:VolumeFraction_Evar_ThreeD_Sphere}) volume fraction of coil in the sphere at the aneurysm neck}
    \label{fig:VolumeFraction_Evar_ThreeD}
\end{figure}

\begin{figure}[!htb]
    \centering
    \begin{subfigure}[t]{0.22\textwidth}
    \centering
        \includegraphics[width=\linewidth]{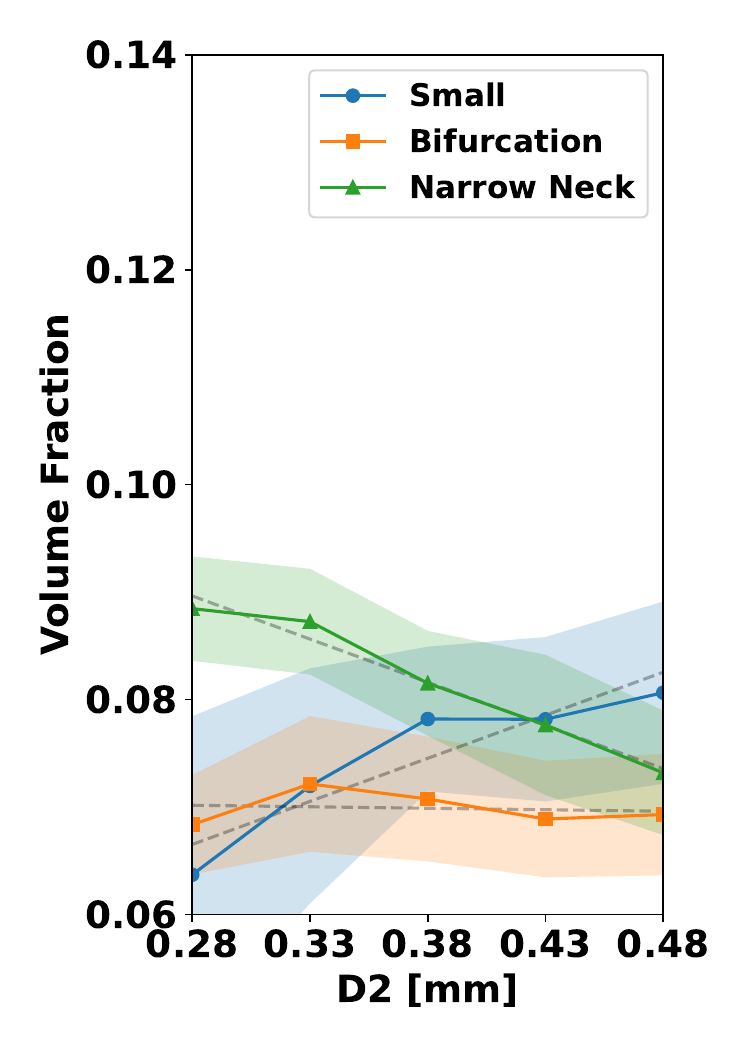}
        \subcaption{}
        \label{fig:VolumeFraction_D2var_Helix_Boundary}
    \end{subfigure}
    \begin{subfigure}[t]{0.22\textwidth}
    \centering
        \includegraphics[width=\linewidth]{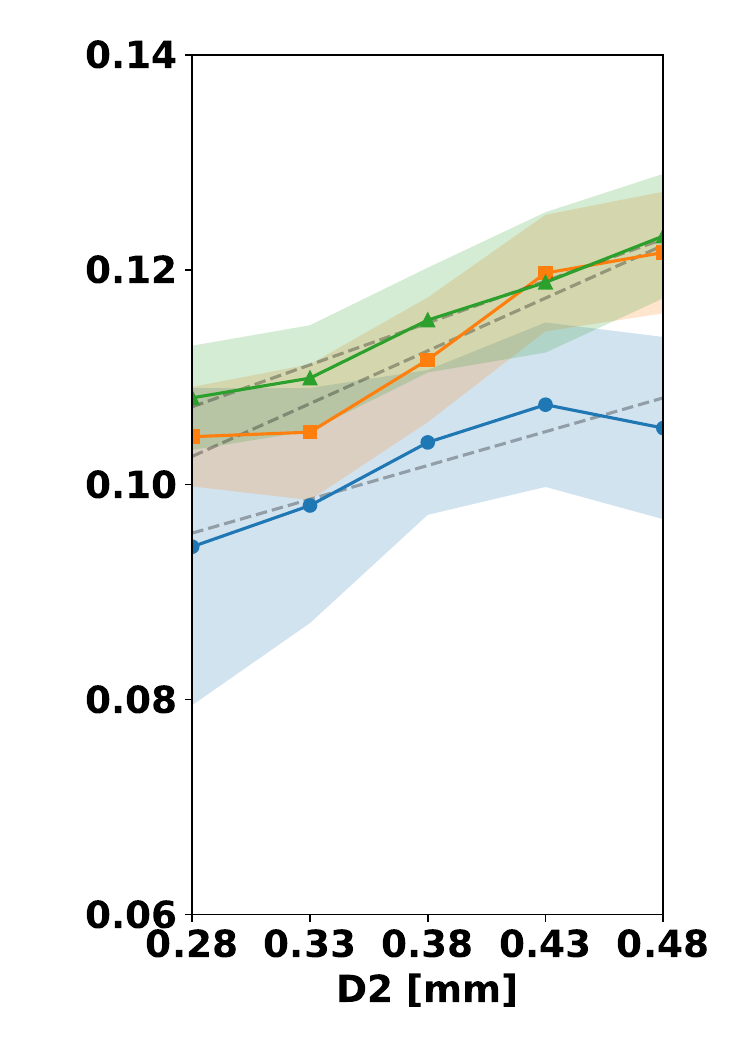}
        \subcaption{}
        \label{fig:VolumeFraction_D2var_Helix_Core}
    \end{subfigure}
    \begin{subfigure}[t]{0.22\textwidth}
    \centering
        \includegraphics[width=\linewidth]{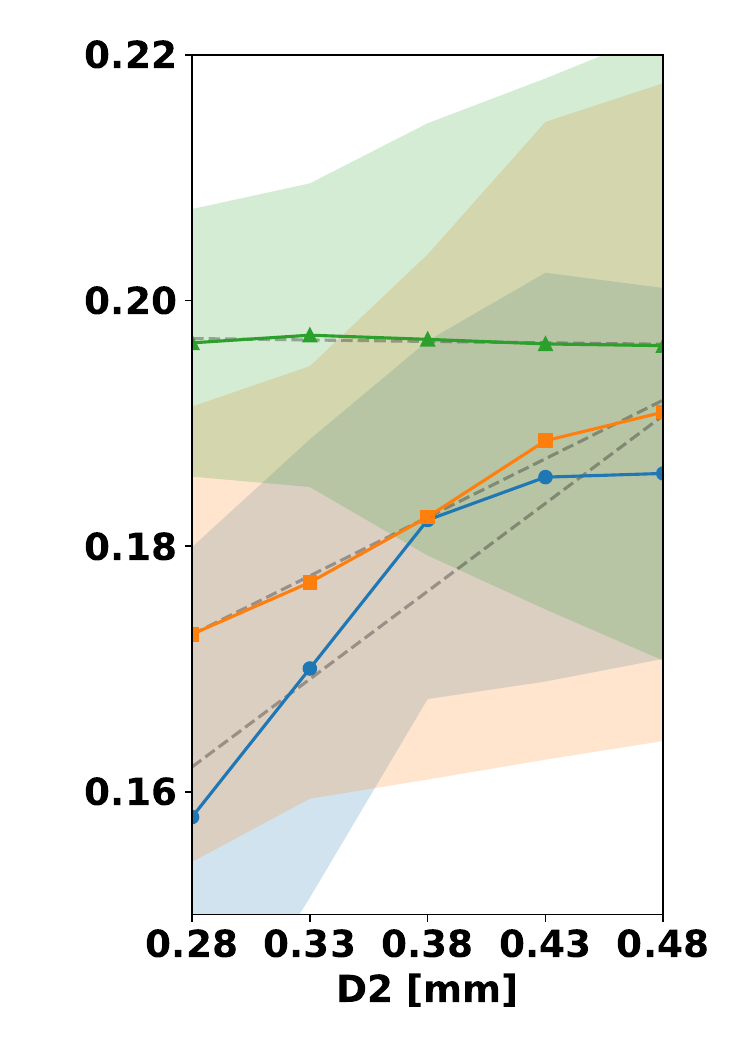}
        \subcaption{}
        \label{fig:VolumeFraction_D2var_Helix_Total}
    \end{subfigure}
    \begin{subfigure}[t]{0.22\textwidth}
    \centering
        \includegraphics[width=\linewidth]{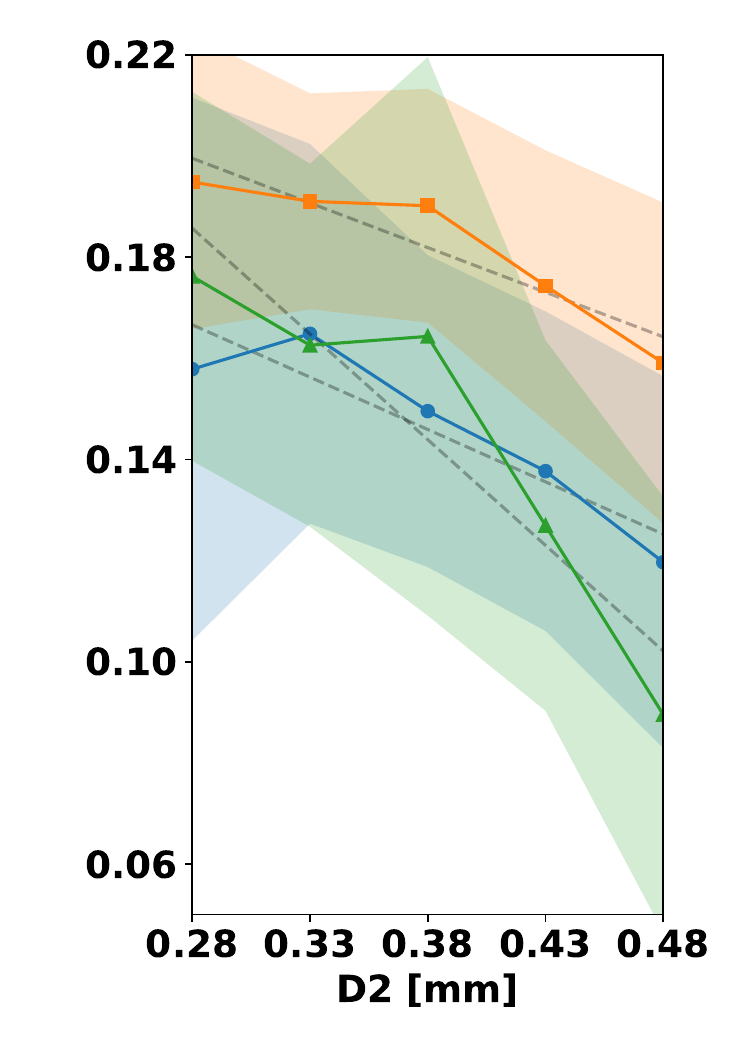}
        \subcaption{}
        \label{fig:VolumeFraction_D2var_Helix_Sphere}
    \end{subfigure}
    \caption{\textbf{Variation of $D_2$ Helix coil}: (\subref{fig:VolumeFraction_D2var_Helix_Boundary}) volume fraction of coil in aneurysm boundary, (\subref{fig:VolumeFraction_D2var_Helix_Core}) volume fraction of coil in aneurysm core, (\subref{fig:VolumeFraction_D2var_Helix_Total}) total volume fraction of coil in aneurysm, (\subref{fig:VolumeFraction_D2var_Helix_Sphere}) volume fraction of coil in the sphere at the aneurysm neck}
    \label{fig:VolumeFraction_D2var_Helix}
\end{figure}

\begin{figure}[!htb]
    \centering
    \begin{subfigure}[t]{0.22\textwidth}
    \centering
        \includegraphics[width=\linewidth]{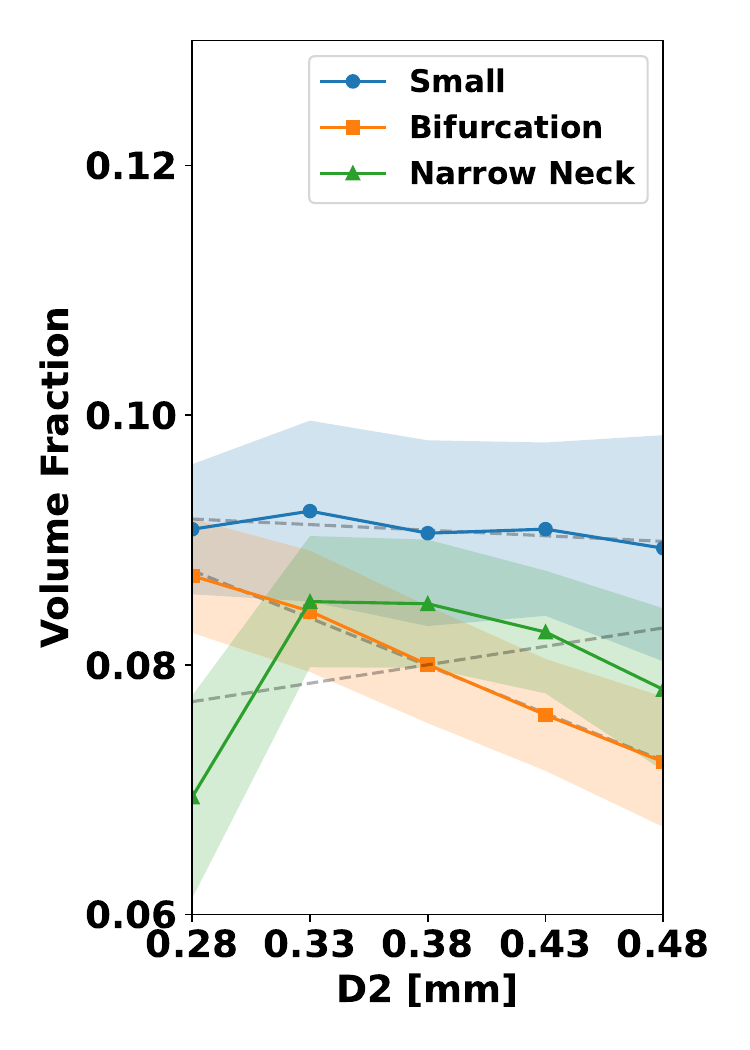}
        \subcaption{}
        \label{fig:VolumeFraction_D2var_ThreeD_Boundary}
    \end{subfigure}
    \begin{subfigure}[t]{0.22\textwidth}
    \centering
        \includegraphics[width=\linewidth]{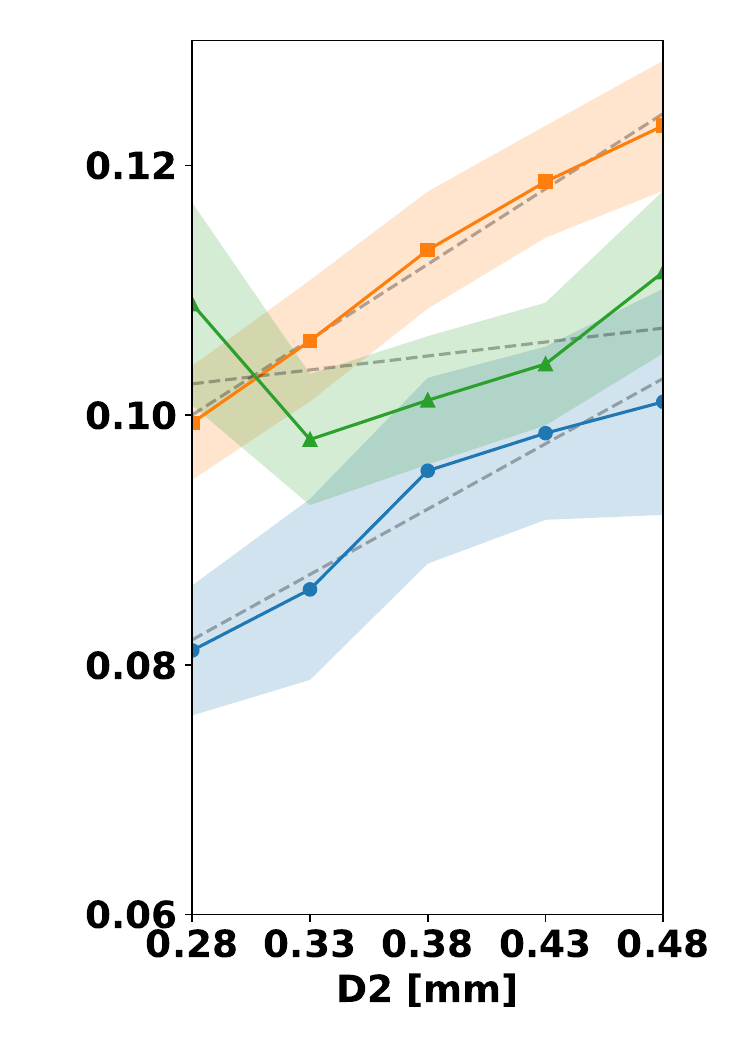}
        \subcaption{}
        \label{fig:VolumeFraction_D2var_ThreeD_Core}
    \end{subfigure}
    \begin{subfigure}[t]{0.22\textwidth}
    \centering
        \includegraphics[width=\linewidth]{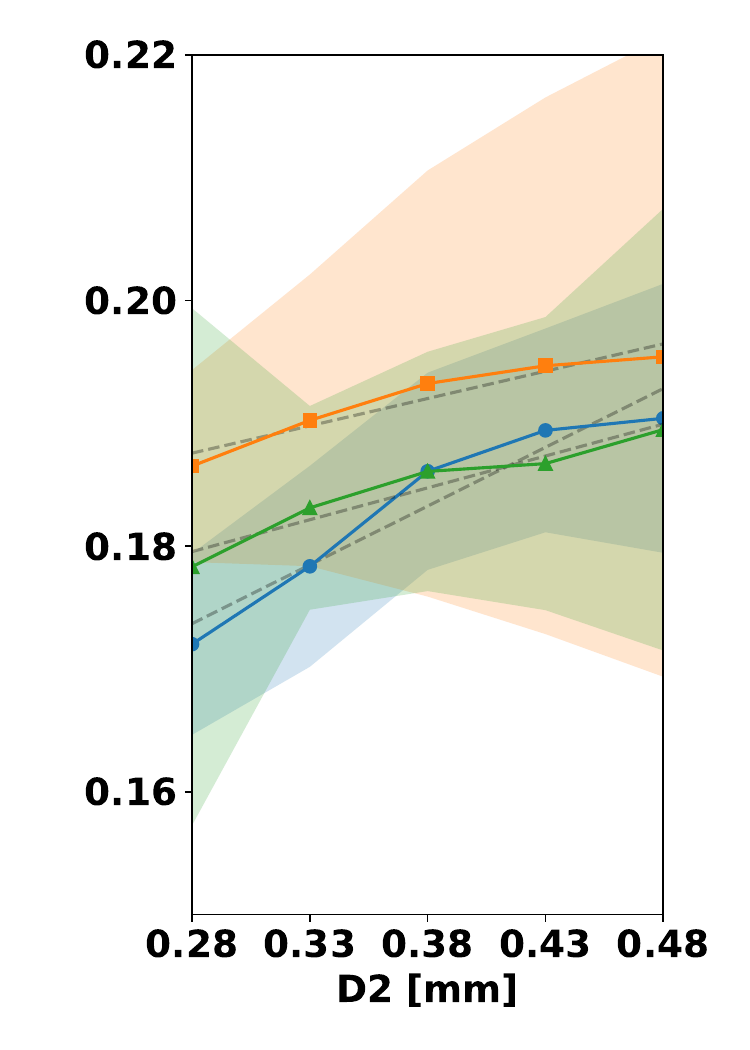}
        \subcaption{}
        \label{fig:VolumeFraction_D2var_ThreeD_Total}
    \end{subfigure}
    \begin{subfigure}[t]{0.22\textwidth}
    \centering
        \includegraphics[width=\linewidth]{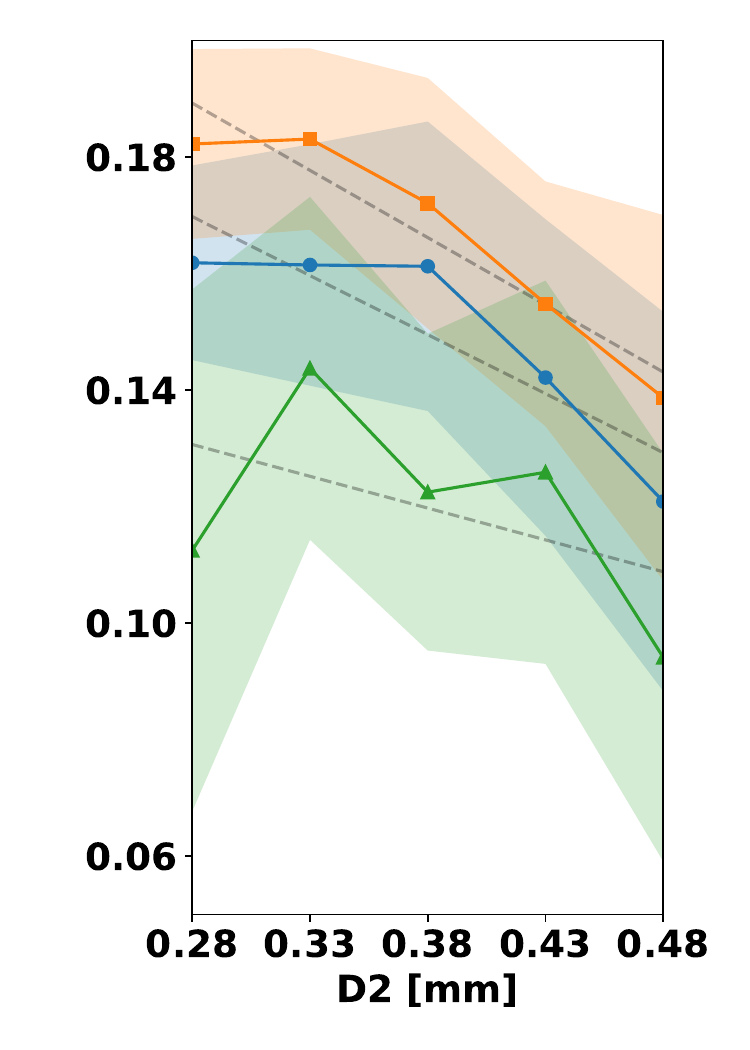}
        \subcaption{}
        \label{fig:VolumeFraction_D2var_ThreeD_Sphere}
    \end{subfigure}
    \caption{\textbf{Variation of $D_2$ 3D coil}: (\subref{fig:VolumeFraction_D2var_ThreeD_Boundary}) volume fraction of coil in aneurysm boundary, (\subref{fig:VolumeFraction_D2var_ThreeD_Core}) volume fraction of coil in aneurysm core, (\subref{fig:VolumeFraction_D2var_ThreeD_Total}) total volume fraction of coil in aneurysm, (\subref{fig:VolumeFraction_D2var_ThreeD_Sphere}) volume fraction of coil in the sphere at the aneurysm neck}
    \label{fig:VolumeFraction_D2var_ThreeD}
\end{figure}

\begin{figure}[!htb]
    \centering
    \begin{subfigure}[t]{0.22\textwidth}
    \centering
        \includegraphics[width=\linewidth]{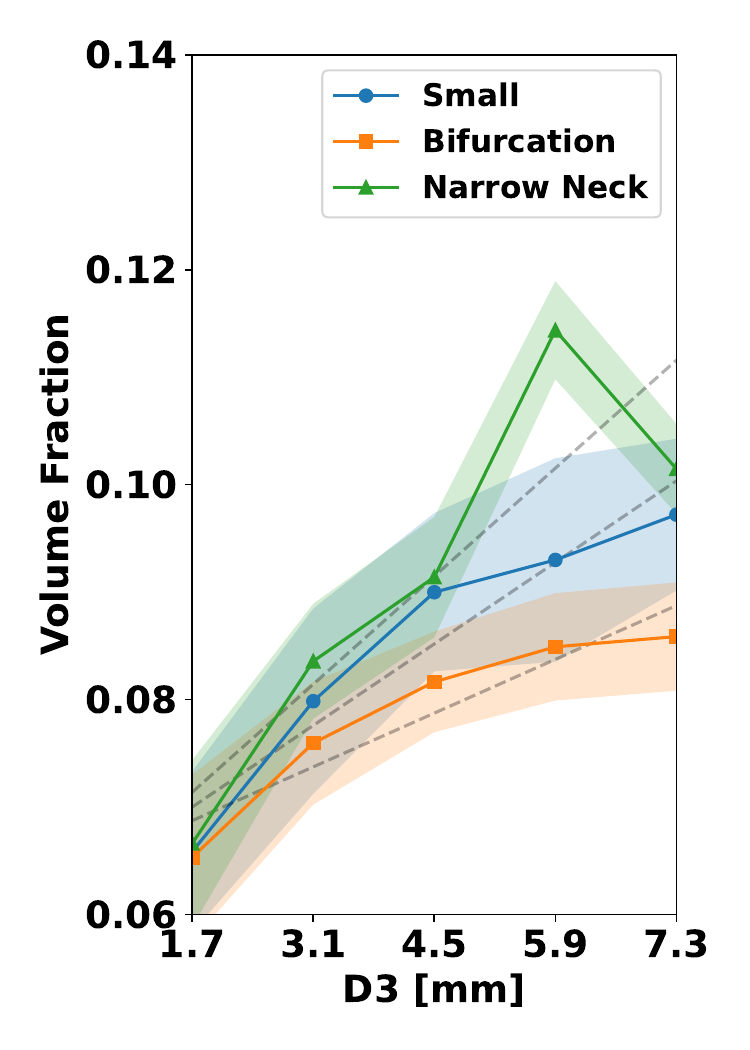}
        \subcaption{}
        \label{fig:VolumeFraction_D3var_Helix_Boundary}
    \end{subfigure}
    \begin{subfigure}[t]{0.22\textwidth}
    \centering
        \includegraphics[width=\linewidth]{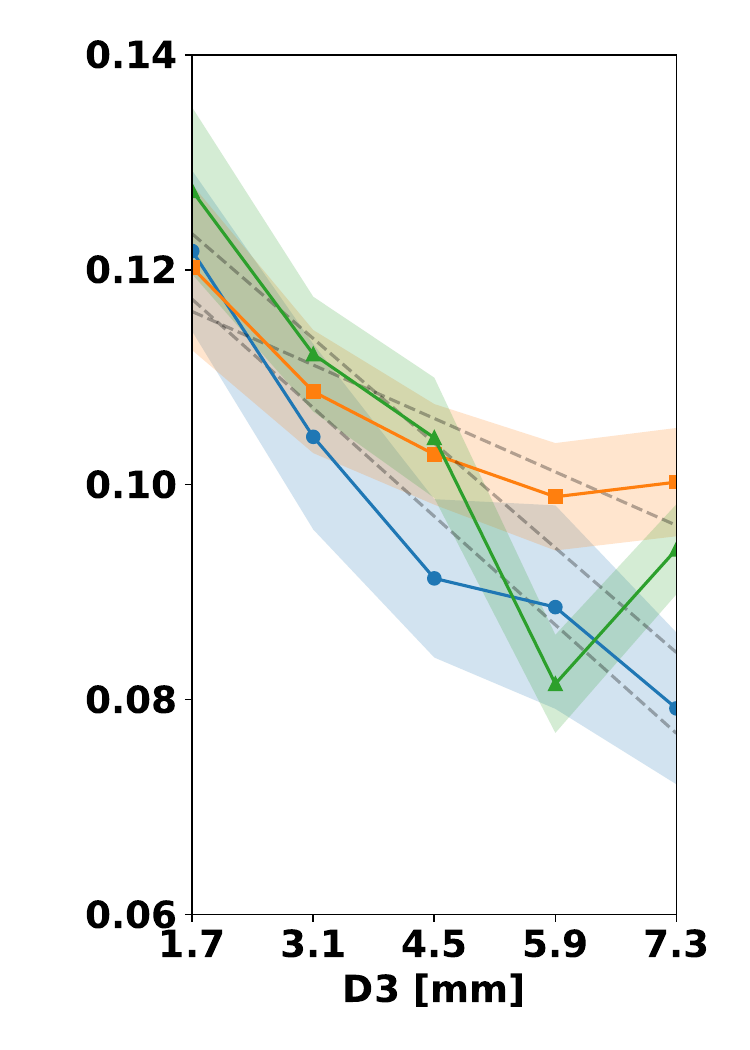}
        \subcaption{}
        \label{fig:VolumeFraction_D3var_Helix_Core}
    \end{subfigure}
    \begin{subfigure}[t]{0.22\textwidth}
    \centering
        \includegraphics[width=\linewidth]{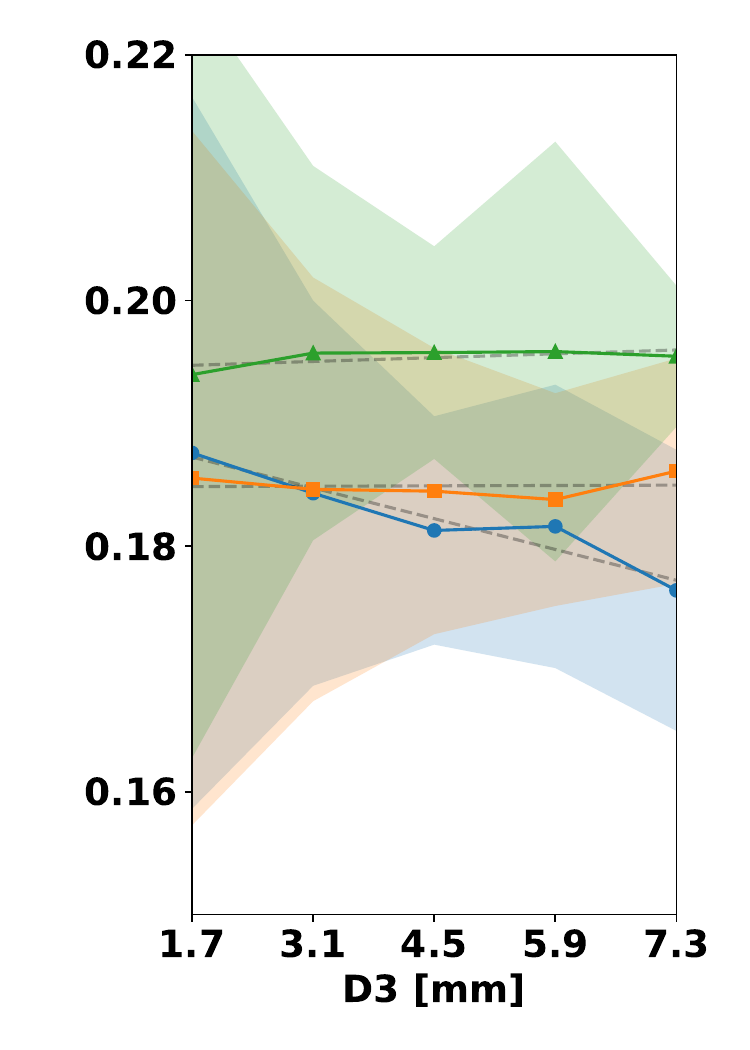}
        \subcaption{}
        \label{fig:VolumeFraction_D3var_Helix_Total}
    \end{subfigure}
    \begin{subfigure}[t]{0.22\textwidth}
    \centering
        \includegraphics[width=\linewidth]{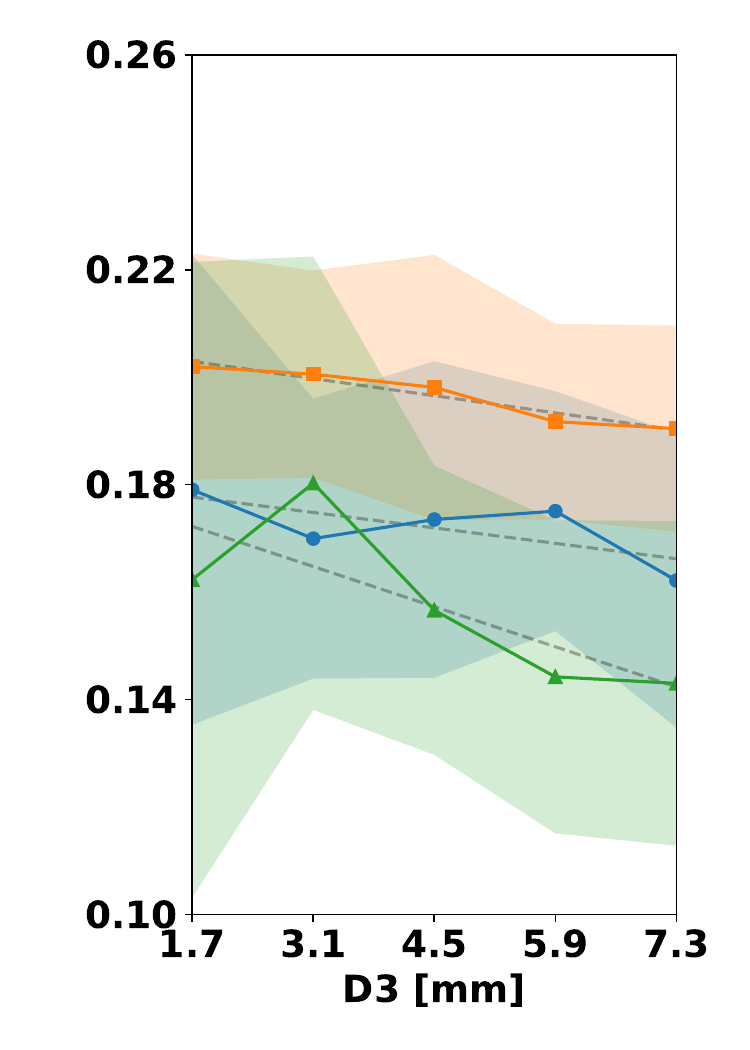}
        \subcaption{}
        \label{fig:VolumeFraction_D3var_Helix_Sphere}
    \end{subfigure}
    \caption{
    \textbf{Variation of $D_3$ Helix coil}: (\subref{fig:VolumeFraction_D3var_Helix_Boundary}) volume fraction of coil in aneurysm boundary, (\subref{fig:VolumeFraction_D3var_Helix_Core}) volume fraction of coil in aneurysm core, (\subref{fig:VolumeFraction_D3var_Helix_Total}) total volume fraction of coil in aneurysm, (\subref{fig:VolumeFraction_D3var_Helix_Sphere}) volume fraction of coil in the sphere at the aneurysm neck}
    \label{fig:VolumeFraction_D3var_Helix}
\end{figure}

In the following, we study the influence of the main variables $E_{w}, D_2 , D_3$ of the coil on the volume fractions (\ref{EQ:volume_fractions}). To this end, we proceed in the following way:

\begin{enumerate}
    \item Let $Z$ be a random variable corresponding to $E_{\text{w}}, D_2, D_3$ and sample it uniformly $Z\sim\mathcal{U}[I(Z)]$ within the interval $I(Z)$.
    \item For each sample $Z_i$ simulate the coil under fixed parameters and generate the voxelization $\psi(Z_i)$.
    \item Create a partition of $I(Z)$ by 5 sub intervals $I_1(Z)\cup\cdots\cup I_5(Z)$.
    \item On each sub interval, we compute $\overline{\psi}$ by (\ref{EQ:porosity_stats_local}) integrate it on a volume region $\widetilde{V}$(\ref{EQ:porosity_stats_global}) and calculate the volume fractions in (\ref{EQ:volume_fractions}).
\end{enumerate}
In order to compute averages and confidence intervals reliably, we ensure that each sub interval contains at least 30 samples of $Z$. Each of the cases is simulated with a final packing density of \SI{20}{\percent}. For the characteristic lengths, we set $D_1$ to \SI{50}{\micro\meter}, $D_2$ to \SI{500}{\micro \meter} and $D_3$ to \SI{4}{\milli \meter} when not otherwise stated.
We depict our findings in Figs. \ref{fig:VolumeFraction_Evar_Helix}--\ref{fig:VolumeFraction_D3var_Helix}. The plots show on the abscissa the center of the 5 intervals $I_1(Z),...,I_5(Z)$ and on the ordinate the averaged volume fraction on the corresponding interval. The figures are sorted such that $\overline{\psi}_{BA}$ corresponds to (a), $\overline{\psi}_{CA}$ is (b), $\overline{\psi}_{AA}$ is (c) and $\overline{\psi}_{SS}$ can be found in (d). Colors correspond to the aneurysm, with blue being the small, orange being the bifurcation and green being the narrow neck aneurysm. Confidence intervals are given in the same colors as faded areas in the background.

We start by showing the influence of a variation in the Young's modulus on the volume fractions with $E_{w}\sim \mathcal{U}[\SI{1e2}{\giga\pascal},\SI{1e3}{\giga\pascal}]$.

\textbf{Variation of Young’s modulus for a Helix coil Fig. \ref{fig:VolumeFraction_Evar_Helix}:} For $\overline{\psi}_{BA}$ in (\subref{fig:VolumeFraction_Evar_Helix_Boundary}) a slight decrease can be observed when $E_w$ is increased. This leads to an increase in  $\overline{\psi}_{CA}$ in (\subref{fig:VolumeFraction_Evar_Helix_Core}) when $E_w$ is increased. Considering $\overline{\psi}_{AA}$ in (\subref{fig:VolumeFraction_Evar_Helix_Total}) the volume fraction is relative constant for all three aneurysms. The highest value is observed in the narrow neck aneurysm and the lowest in the small aneurysm. Looking at $\overline{\psi}_{SS}$ in (\subref{fig:VolumeFraction_Evar_Helix_Sphere}), we can see that the volume in the sphere increases for larger $E_w$ but stagnates when reaching about \SI{550}{\giga\pascal}.

\textbf{Variation of Young's modulus for a 3D coil Fig. \ref{fig:VolumeFraction_Evar_ThreeD}:}
In this case, $\overline{\psi}_{BA}$ is relatively constant when changing $E_w$ in (\subref{fig:VolumeFraction_Evar_ThreeD_Boundary}). For $\overline{\psi}_{CA}$ in (\subref{fig:VolumeFraction_Evar_ThreeD_Core}), we observe for the small aneurysm a downwards trend when increasing $E_w$ to a value of $\SI{550}{\giga\pascal}$. The same trend of the decrease in the volume fraction for the small aneurysm is visible in $\overline{\psi}_{AA}$ while it is relatively constant for the other two geometries. Lastly we note that there is a slight increase in $\overline{\psi}_{SS}$ when increasing $E_w$ which is approximately constant in the other geometries.

We summarize that the variation of $E_w$ has a greater impact on the helix coil where it causes the coil to migrate from the boundary region into the core region. Moreover, increasing $E_w$ leads to an increase in coil volume at the neck when a helix coil is used. The helix coil is more densely packed in the core region while the 3D coil is more densely packed in the boundary region. For the helix coil, the packing in the neck region is higher than in the 3D coil case as one can see by comparing the results in (\subref{fig:VolumeFraction_Evar_ThreeD_Sphere}).

Next we focus on the variation of $D_2$ where the sampling interval is chosen as $D_2\sim \mathcal{U}[\SI{0.255}{\milli\meter},\SI{0.505}{\milli\meter}]$.

\textbf{Variation of the $D_2$ diameter for a Helix coil Fig. \ref{fig:VolumeFraction_D2var_Helix}:} 
Increasing $D_2$ leads to an increase in $\overline{\psi}_{BA}$ as can be seen in (\subref{fig:VolumeFraction_D2var_Helix_Boundary}) for coils pushed into the small aneurysm while a decrease in this volume fraction is caused in case of the narrow neck aneurysm. For the bifurcation aneurysm there is no clear trend visible. Considering $\overline{\psi}_{CA}$, an increase in volume fraction is  observed when increasing $D_2$. The volume fraction $\overline{\psi}_{AA}$ increases by approximately \SI{3}{\percent} in the small aneurysm and by \SI{2}{\percent} in the bifurcation aneurysm. For the narrow neck aneurysm, it is nearly constant. Analyzing  $\overline{\psi}_{SS}$ in (\subref{fig:VolumeFraction_D2var_Helix_Sphere}) one can see that the volume fraction of coil in the sphere is significantly impacted by $D_2$. When the coil radius is increased, the volume fraction in the sphere is lowered by up to \SI{3}{\percent} in each of the aneurysms.

\textbf{Variation of the $D_2$ diameter for a 3D coil Fig. \ref{fig:VolumeFraction_D2var_ThreeD}:} For the 3D coil a increase of $D_2$ lead to a decrease in $\overline{\psi}_{BA}$ for the small aneurysm and narrow neck aneurysm. Note that the lowest values of $D_2$ caused the smallest volume fraction in the core region for the narrow neck aneurysm. The impact of $D_2$ on $\overline{\psi}_{BA}$ in the small aneurysm is not significant. In (\subref{fig:VolumeFraction_D2var_ThreeD_Core}), one can observe that the core volume fraction depends on $D_2$ and rises as $D_2$ is increased. Additionally, a small $D_2$ in (\subref{fig:VolumeFraction_D2var_ThreeD_Sphere}) leads to a  decrease of roughly  $\SI{2}{\percent}$ in the volume fractions of  the coil within the aneurysm when compared to the volume fractions at the highest $D_2$ value. For the volume fraction in the sphere region, a higher radius $D_2$ decreases the volume fraction in the sphere region when considering the bifurcation and small aneurysm case. For the narrow neck aneurysm, a trend is less clearly visible.
Concluding our findings for the variation in $D_2$, we note that $D_2$ has a substantial influence on all volume fractions. For both the helix and the 3D coil, an increase in $D_2$ lead to an overall decrease in the volume fraction $\overline{\psi}_{BA}$ in the narrow neck aneurysm. For the small aneurysm packed with a helix coil,  a small $D_2$ value causes a lower volume fraction in the boundary region. For the bifurcation aneurysm, a 3D coil results in a smaller volume fraction at the boundary when compared to the helix coil.  In the core region, an increase in $D_2$ leads to an increase in $\overline{\psi}_{CA}$. Additionally, we observe that a smaller value of $D_2$ leads to possible coil migrating into the parent vessel direction and results in large variations in sphere and total aneurysm volume. Finally, we see for both coils that a higher volume fraction $\overline{\psi}_{SS}$ is obtained by a smaller $D_2$.

Finally, we sample the diameter $D_3$ according to $D_3\sim\mathcal{U}[\SI{2}{\milli \meter},\SI{8}{\milli\meter}]$.

\textbf{Variation of $D_3$ for a Helix coil Fig. \ref{fig:VolumeFraction_D3var_Helix}:} For each of the geometries, the volume fraction $\overline{\psi}_{BA}$ increases roughly by $\SI{2}{\percent}$ in (\subref{fig:VolumeFraction_D3var_Helix_Boundary}) and  decreases when considering $\overline{\psi}_{CA}$ in (\subref{fig:VolumeFraction_D3var_Helix_Core}). At a $D_3$ diameter of \SI{5.7}{\milli\meter}--\SI{6.1}{\milli\meter} for the narrow neck aneurysm a peak is visible increasing $\overline{\psi}_{BA}$ by another \SI{2}{\percent}. Note that the confidence interval at the peak is smaller than \SI{1}{\percent}, meaning that it occurs for most of the samples. Looking at $\overline{\psi}_{AA}$ in (\subref{fig:VolumeFraction_D3var_Helix_Total}), one can see that the coil volume fraction  in the interior of the small aneurysm decreases when increasing $D_3$. Finally, we note that an increase of $D_3$ in (\subref{fig:VolumeFraction_D3var_Helix_Sphere}) leads to a slight decrease of $\overline{\psi}_{SS}$ in all geometries.
Therefore, changing $D_3$ can impact the volume of the coil at the boundary and core significantly. This behavior can be observed for framing coils where larger $D_3$ lead to coils that are more present in the boundary region of the aneurysm creating a stabilizing cage for the filling coils.

Our observations can be summarized in the following way.
Changing $E_w$ has less impact than changing the other parameters. The largest changes are observed for the small aneurysm which is to be expected since the coil diameter $D_3$ for the variation of $E_w$ was fixed to \SI{4}{\milli\meter} which is similar to the dome size of the small aneurysm. Therefore, the stiffening of the coil via $E_w$ can cause wires to be less flexible leading to coils that are partly  pushed into the parent vessel. This effect only occurred for the 3D coil.
The influence of $D_2$ on the coil volume fractions is higher compared to the other parameters. Here a large $D_2$ value leads to a more rapid filling of the aneurysm at the cost of less coil being present in the boundary and neck of the aneurysm as our study suggests.
For the helix coil, a higher $D_3$ is beneficial since the coil volume fraction  in the neck region increases.
Finally, we did see by considering variations in $D_3$ that the qualitative behavior of framing coils being more present in the boundary for large values of $D_3$ is reproducible in our study.

\begin{figure*}[htbp]
    \centering
    \begin{subfigure}[t]{0.32\textwidth}
    \centering
        \includegraphics[width=\linewidth]{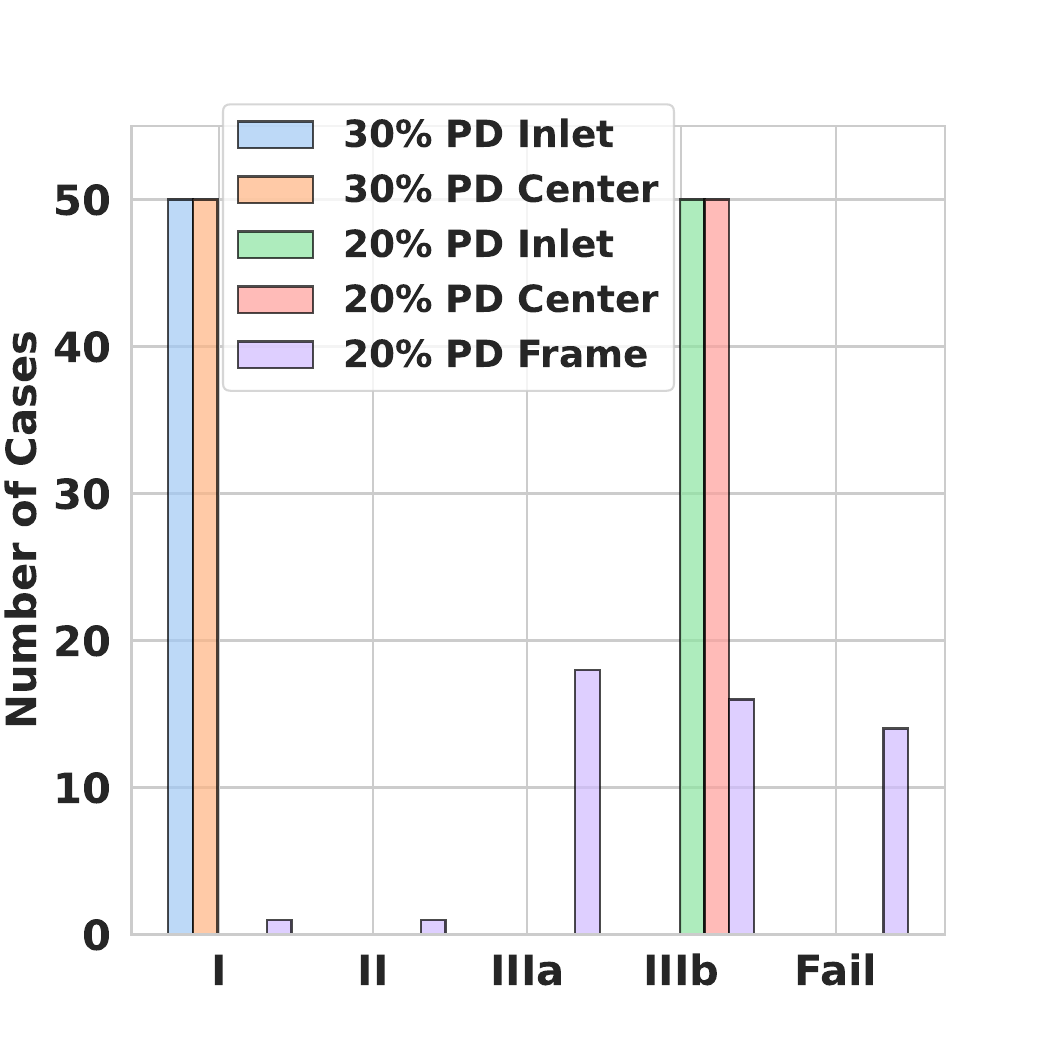}
        \subcaption{}
        \label{fig:classes_Small}
    \end{subfigure}
    \begin{subfigure}[t]{0.32\textwidth}
    \centering
        \includegraphics[width=\linewidth]{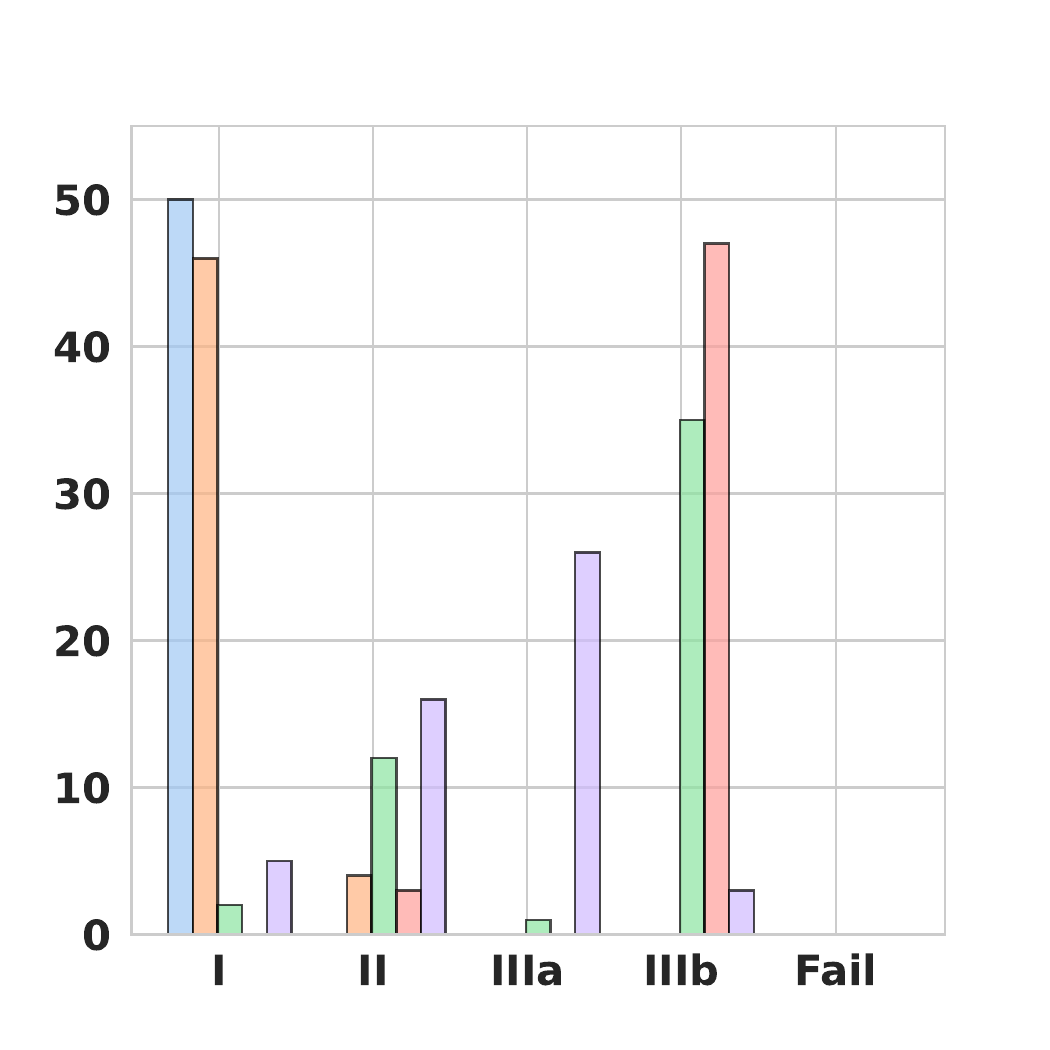}
        \subcaption{}
        \label{fig:classes_Gamm}
    \end{subfigure}
    \begin{subfigure}[t]{0.32\textwidth}
    \centering
        \includegraphics[width=\linewidth]{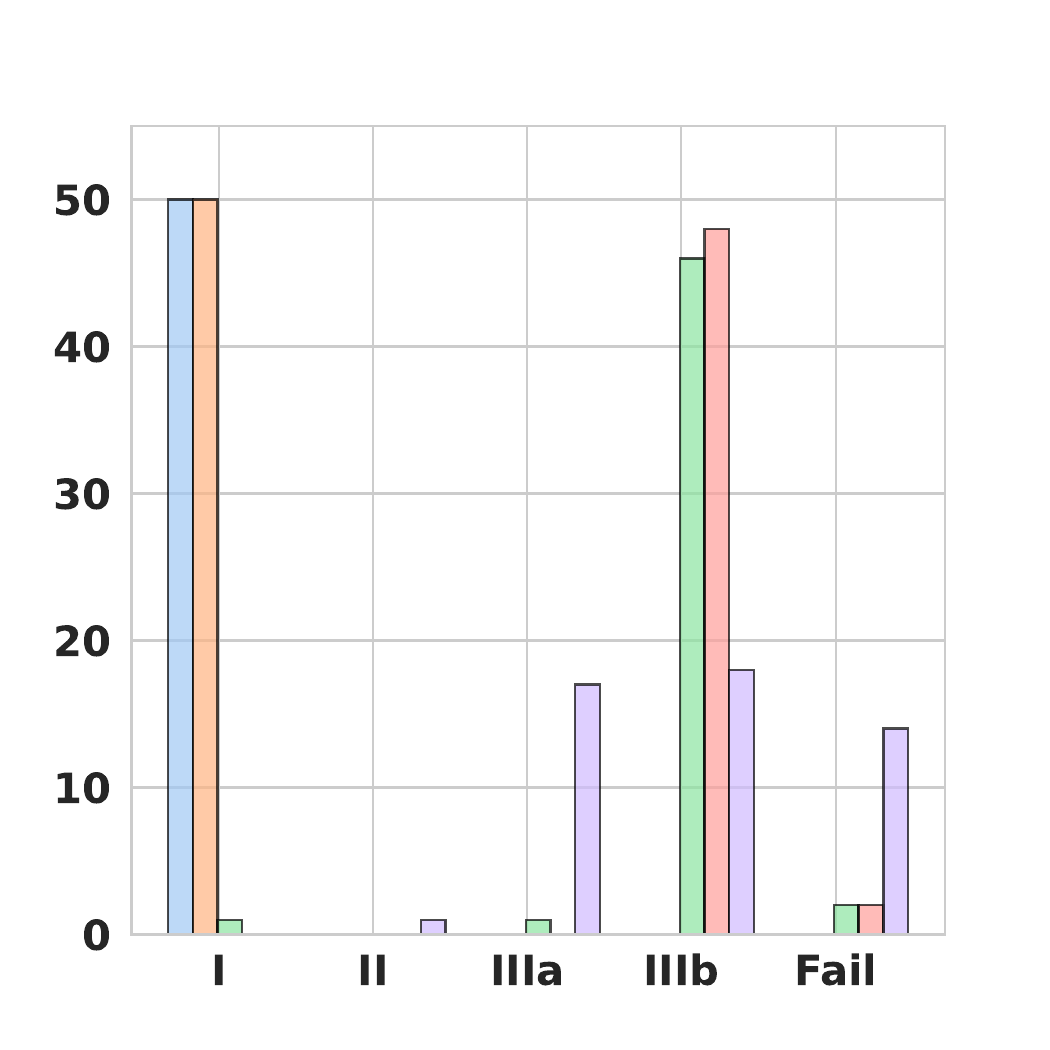}
        \subcaption{}
        \label{fig:classes_Bir}
    \end{subfigure}
    \caption{Classification of coil quality by RROC for 50 coils with different packing densities (PD) and at different catheter positions (\subref{fig:classes_Small}) Small aneurysm geometry, (\subref{fig:classes_Gamm}) narrow neck aneurysm geometry, (\subref{fig:classes_Bir}) Bifurcation aneurysm geometry}
    \label{fig:classes_Inlet}
\end{figure*}

\subsection{RROC}
In this section, the RROC motivated occlusion classification from Section \ref{insilicioclass} is used to classify coil placements. We consider placements in two regions of the aneurysm. In a first test, the placements are generated according to the catheter positions as shown in Fig. \ref{fig:Aneurisms}. Secondly, we simulate placements in the center of mass of the aneurysms. Placements can vary by small perturbations, as it is the case in  practice. This is implemented by adding a perturbation vector to the position of the catheter endpoint. The perturbations are sampled uniformly in a ball of radius \SI{1}{\milli \meter} around the original catheter endpoint. In each aneurysm geometry and for each position, we simulate 50 of such placements and apply our modified occlusion classification. Fig. \ref{fig:classes_Inlet} shows how placements are classified. We consider the PDs \SI{20}{\percent} and \SI{30}{\percent} while using filling coils for both PDs and a framing coil for \SI{20}{\percent} PD. Framing and filling coils distinguish in size and stiffness. For framing / filling coils we set $D_3$ to \SI{5}{\milli\meter} / \SI{2}{\milli\meter} for the small aneurysm, \SI{7}{\milli\meter} / \SI{4}{\milli\meter} for the narrow neck aneurysm and \SI{8}{\milli\meter} / \SI{4}{\milli\meter} for the bifurcation aneurysm. The length $D_1$ is set to \SI{50}{\micro\meter} for framing coils and \SI{40}{\micro\meter} for filling coils, while $D_2$ is in all cases set to \SI{305}{\micro\meter}.

\textbf{Small Aneurysm} Fig. \ref{fig:classes_Small}:
For the small aneurysm, packing densities of \SI{30}{\percent} lead in all cases to Class I. Class II is not assigned except for one of the framing coils. Class IIIa is roughly assigned to one third of the framing coils and Class IIIb to all the filling coils at \SI{20}{\percent} PD such as to one third of all the framing coils. The Class Fail is only assigned to the remaining framing coils.

We interpret the statistics as follows. Using filling coils with sufficient (\SI{30}{\percent}) PD will result in good occlusion while this is not the case when lowering the PD of the filling coils to \SI{20}{\percent}. Framing coils on the other hand can in some cases produce at \SI{20}{\percent} the slightly better occlusion Class IIIa but might also produce failed cases where coil is migrating into the parent vessel.

\textbf{Narrow Neck Aneurysm } Fig. \ref{fig:classes_Gamm}:
In case of the narrow neck aneurysm, all coils at PD \SI{30}{\percent} that were inserted close to the inlet resulted in Class I. For coils inserted at the center with the same PD a few are not in Class I. We also note that there is a small chance for coils of lower packing density to produce a Class I occlusion. Class II is mainly dominated by framing coils, and coils that are inserted at the inlet having a  lower PD. Note that also some high PD coils are assigned to this class. The majority of the framing coils is assigned to Class IIIa while the majority of the filling coils of \SI{20}{\percent} PD is assigned to Class IIIb. For the narrow neck aneurysm the Fail Class was never assigned.

We therefore interpret that the narrow neck aneurysm occludes better when compared to the small aneurysm since less cases are assigned to Class IIIb and the Class Fail. On the other hand, the narrow neck is more challenging to fill which we see in the increased amount of coils assigned to Class II.

\textbf{Bifurcation Aneurysm } Fig. \ref{fig:classes_Bir}:
For the bifurcation aneurysm, all cases with \SI{30}{\percent} PD are assigned to Class I. Class II is only assigned to one of the framing coils. Class IIIa mostly constitutes framing coils. The Class IIIb contains the majority of the lower PD filling coils and a third of the framing coils. Finally, the  Class Fail includes another third of the framing coils and a small fraction of the low PD filling coils.
The bifurcation aneurysm therefore behaves similar to the small aneurysm in terms of RROC.

\begin{figure}[H]
    \centering
    \begin{subfigure}[b]{0.2\textwidth}
    \centering
    \includegraphics[width=\linewidth]{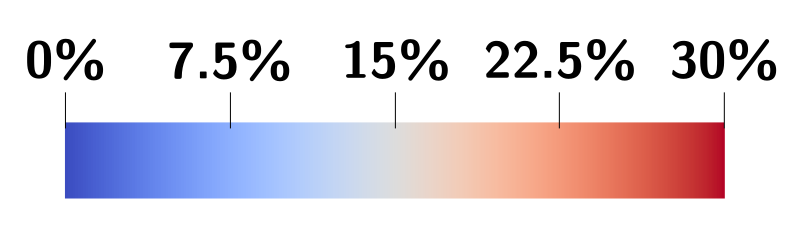}
        \includegraphics[width=\linewidth, trim=10cm 16cm 29cm 8cm, clip]{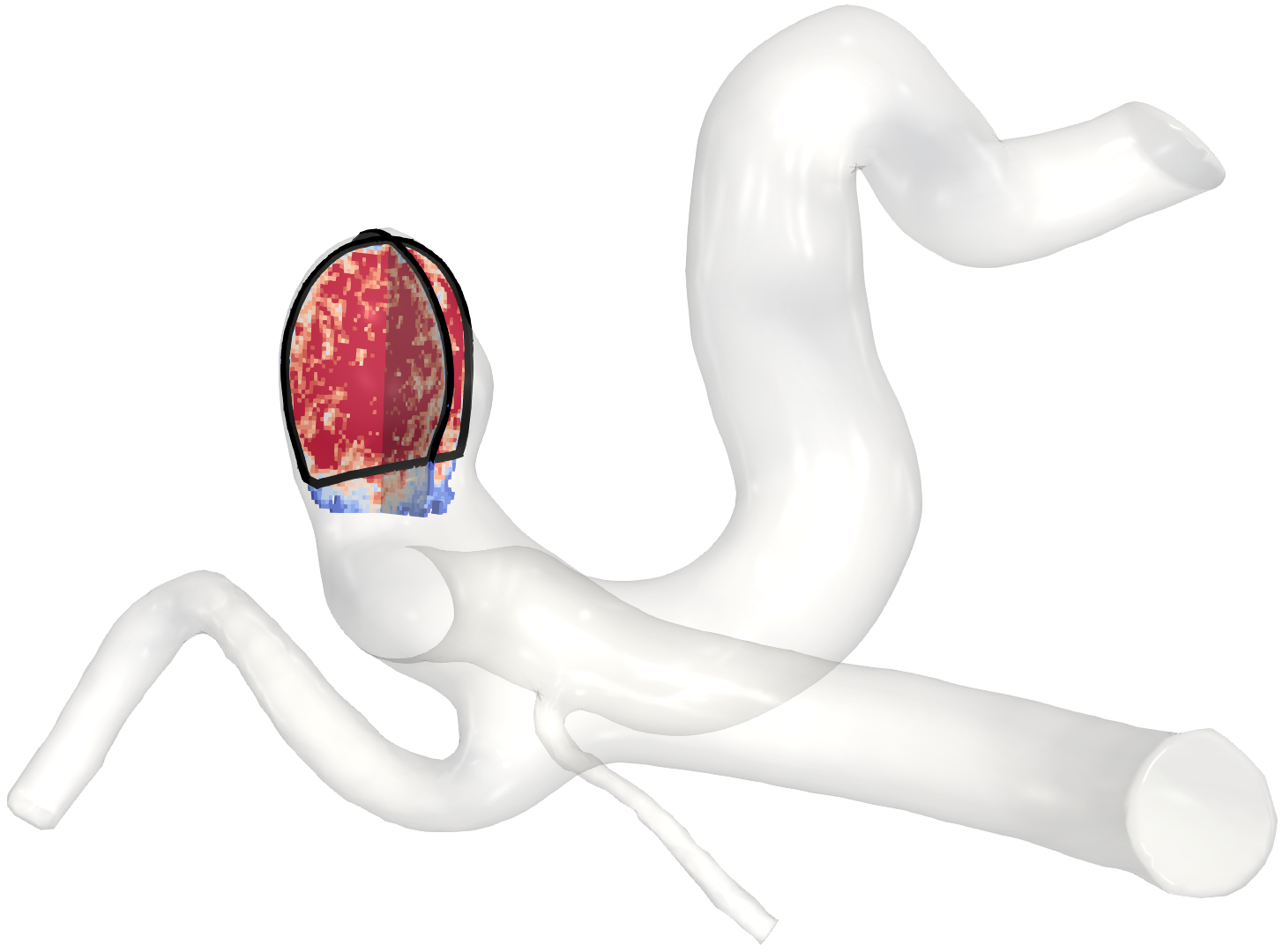}
        \subcaption{}
        \label{fig:Small_Inlet_30PD_ClassI_scaled}
    \end{subfigure}
    \begin{subfigure}[b]{0.2\textwidth}
    \centering
        \includegraphics[width=\linewidth, trim=23cm 9cm 9cm 0cm, clip]{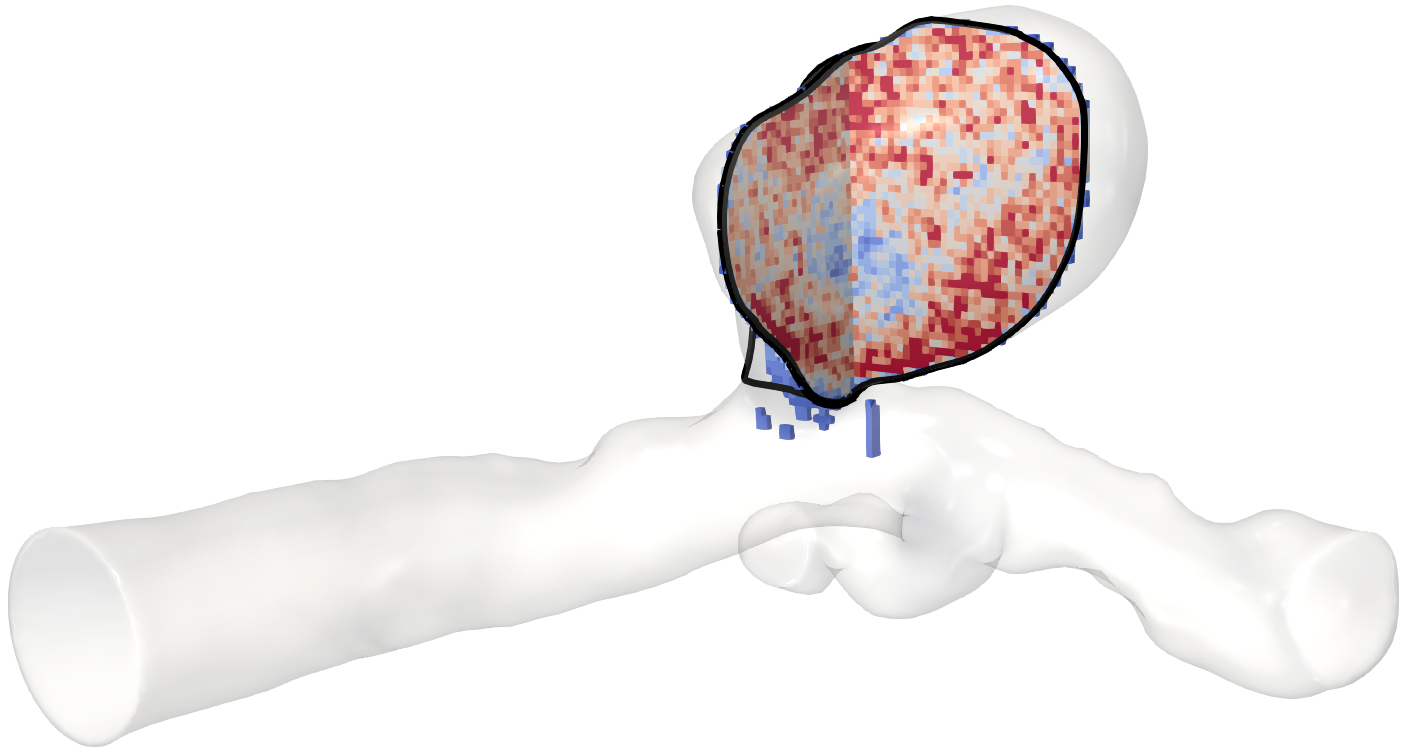}
        \subcaption{}
        \label{fig:Gamm_Inlet_Frame_20PD_ClassIIIaII_scaled}
    \end{subfigure}
    \begin{subfigure}[t]{0.2\textwidth}
    \centering
        \includegraphics[width=\linewidth, trim=10cm 16cm 29cm 8cm, clip]{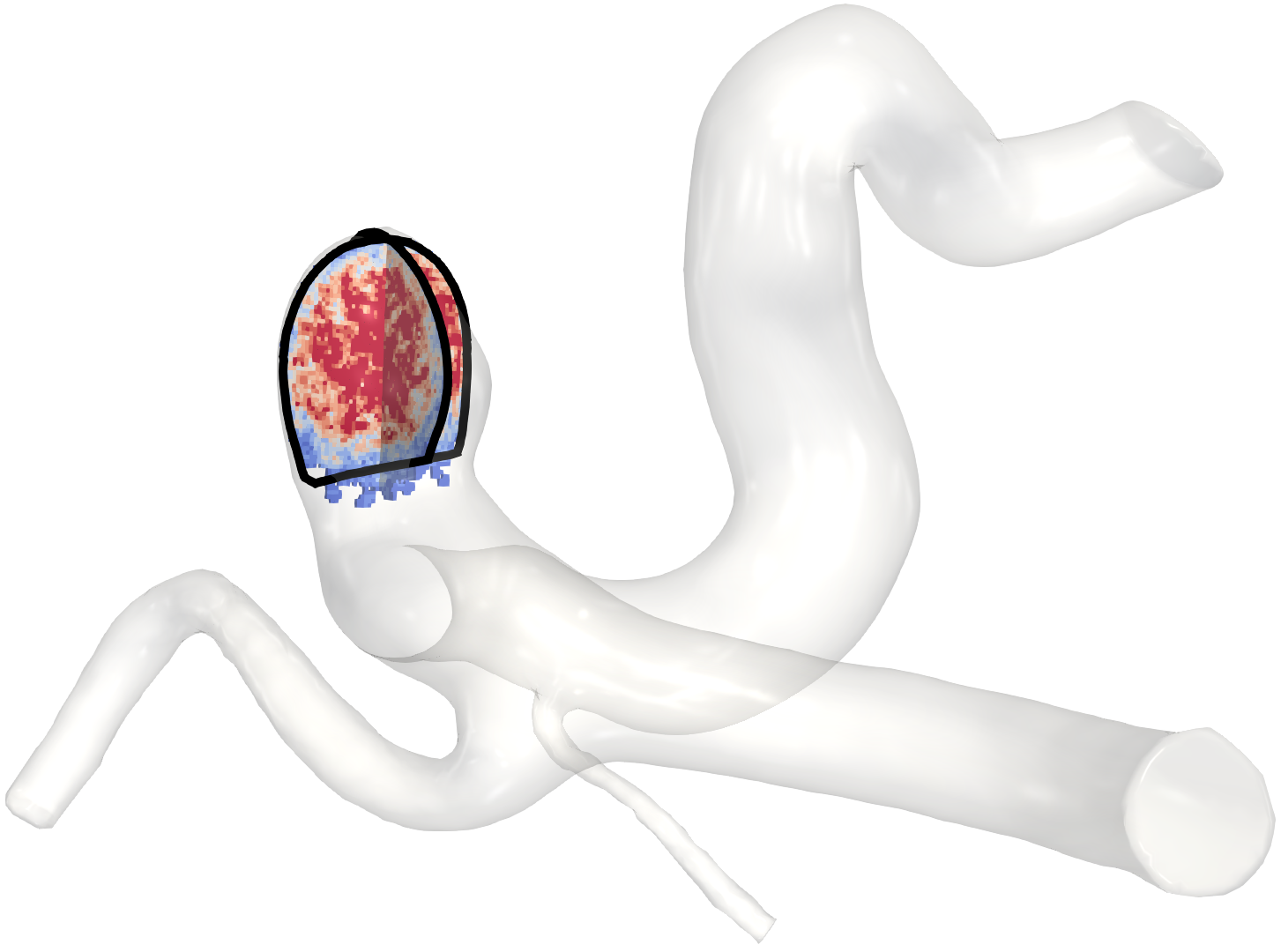}
        \subcaption{}
        \label{fig:Small_Center_20PD_ClassIIIb_scaled}
    \end{subfigure}
    \begin{subfigure}[t]{0.2\textwidth}
    \centering
        \includegraphics[width=\linewidth, trim=29cm 11cm 0cm 0cm, clip]{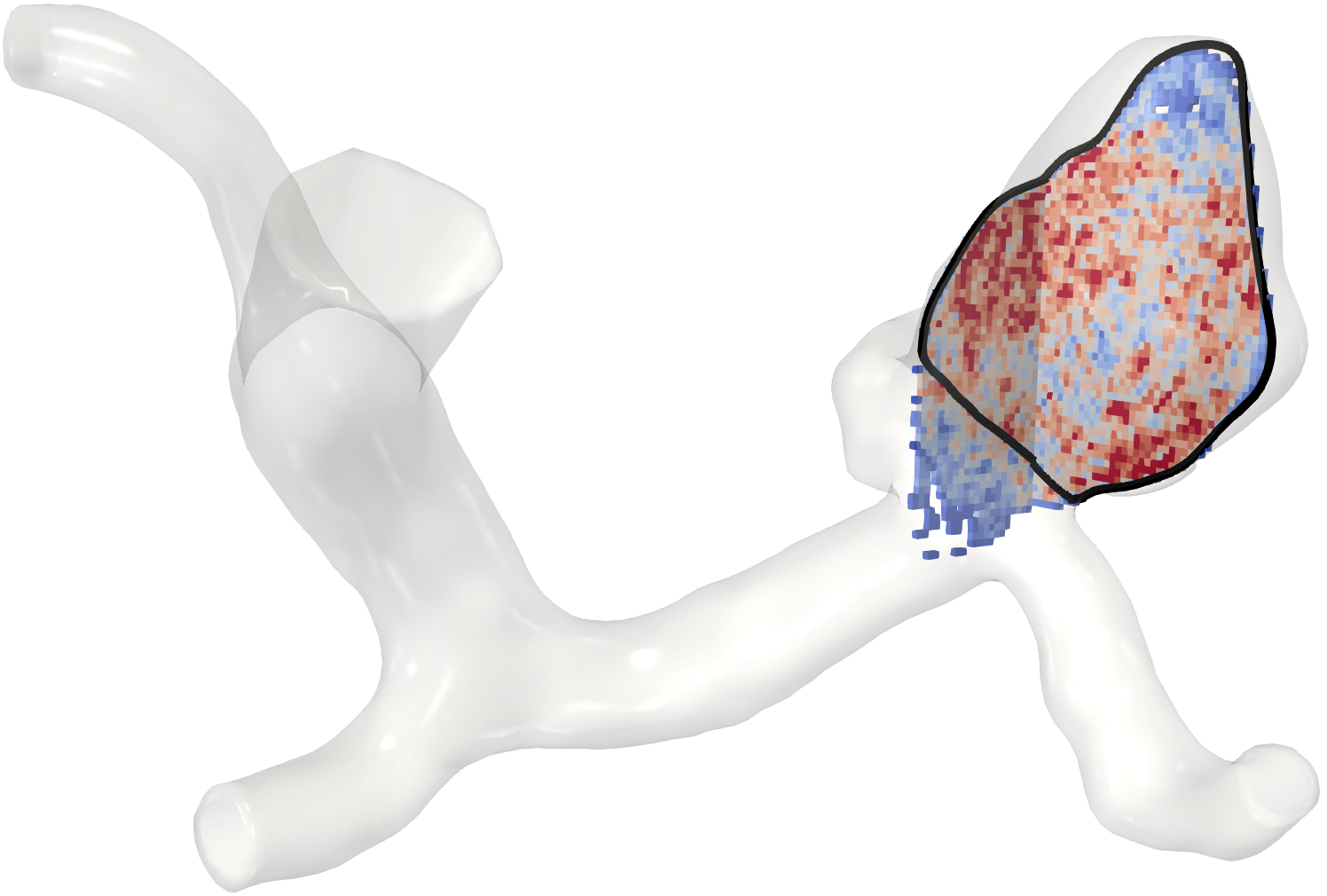}
        \subcaption{}
        \label{fig:Bir_Inlet_Frame_20PD_ClassF_scaled}
    \end{subfigure}
    \caption{Examples of classes assessed by our classifier in Fig. \ref{fig:classes_Inlet} depicted as cross-sections of averaged coil distributions with a sample-size \SI{50}{}. The color-scale corresponds to the PD in the voxels with values over \SI{30}{\percent} being red. (\subref{fig:Small_Inlet_30PD_ClassI_scaled}) Class I from \SI{30}{\percent} PD Inlet \ref{fig:classes_Small}, (\subref{fig:Gamm_Inlet_Frame_20PD_ClassIIIaII_scaled}) lies between Class II / IIIa from \SI{20}{\percent} PD Frame for the narrow neck aneurysm in Fig. \ref{fig:classes_Gamm}, (\subref{fig:Small_Center_20PD_ClassIIIb_scaled}) ClassIIIb from \SI{20}{\percent} PD Frame in \ref{fig:classes_Small} and (\subref{fig:Bir_Inlet_Frame_20PD_ClassF_scaled}) Class Fail from \SI{20}{\percent} PD Inlet in \ref{fig:classes_Bir}.}
    \label{fig:classes_mean_fields}
\end{figure}

In Fig. \ref{fig:classes_mean_fields}, we show the averaged coil distributions of the 50 coils considered in each case calculated by (\ref{EQ:porosity_stats_local}). Fig. \ref{fig:Small_Inlet_30PD_ClassI_scaled} corresponds to Class I. Here the coil distribution  is relatively uniform and sufficiently enough coil covers the neck without extensively wandering into the parent vessel. 
Fig. \ref{fig:Gamm_Inlet_Frame_20PD_ClassIIIaII_scaled} is a representative of Class II and IIIa. Some samples of this average get assigned to Class IIIa due to the low value of the coil distribution in the core. Another large fraction of samples gets assigned Class II since the coil distribution in the neck is not sufficient.
Fig. \ref{fig:Small_Center_20PD_ClassIIIb_scaled} is a Class IIIb example.  As in the RROC, the classification follows from the fact that the wall region at the neck of the aneurysm is poorly filled. 
Finally, in Fig. \ref{fig:Bir_Inlet_Frame_20PD_ClassF_scaled},  a relatively large amount of coil is located below the neck resulting in  Case Fail classification.


\section{Conclusion}
\label{sec:conclusion}
We have presented a mathematical model of endovascular coiling that is based on the DER method. Our model accurately takes into account important properties of the coil design such as its natural shape, bending and torsion response and can be placed efficiently into realistic aneurysm geometries. The broad applicability of the model was shown by testing it for the coiling of three different aneurysms showing that it is applicable in a wide range of cases. We have qualitatively validated the correctness of our model by comparing it to real coils extruded into free space. 

The second part of our study focused on a statistical evaluation of our model. The main tool to achieve this was the voxelization of the coil geometry, which allowed us to define the notion of averaged coils. These were then used to calculate the volume fractions of coils in the core, boundary and neck region of an aneurysm and to study the parameter sensitivity of our coils.  Further we have developed a Raymond Roy Occlusion type Classifier that allowed us to grade embolized coils by their occlusion properties. Finally, the 
relevance of the assigned classes were illustrated by analyzing mean fields of embolized coils for which a majority was graded with a specific class.

In our study, we could reproduce a wide range of phenomena observed in real coiling situations, such as the behavior of filling coils that are used to fill the internal volume of aneurysms and the framing coils used for stabilization. The DER model contains only the minimum degrees of freedom (3 translational and 1 rotational) to represent the natural shape of a coil. This makes it an excellent candidate for an a priori investigation of a coiling procedure outcome in a clinical setup, without the overhead of a fully resolved mechanical model. 
Our RROC motivated classification provides an extension to the traditional RROC and is based on the geometric distribution of the coil, therefore it can be used to speed up classification-analyses of coil occlusion when correctly calibrated.

\backmatter

\bmhead{Supplementary information}

Not applicable

\bmhead{Acknowledgments}

F. Holzberger M. Muhr and B. Wohlmuth gratefully acknowledge the financial support partially provided by the Deutsche Forschungsgemeinschaft under the grant number WO 671/11-1 as well as under project number 465242983 within the priority programme ``SPP 2311: Robust coupling of continuum-biomechanical in silico models to establish active biological system models for later use in clinical applications - Co-design of modeling, numerics and usability'' (WO 671/20-1).

We would like to express our gratitude to Prof. Jan Kirschke and Dr. Julian Schwarting of the Hospital Rechts der Isar in Munich for generously providing us two images of the coiling procedure for an aneurysm shown in Fig. \ref{fig:main}. Their expertise as neuroradiologists and their invaluable contributions have greatly enriched this work.

\section*{Declarations}
Not applicable

\begin{appendices}

\section{Derivation of the twisting momentum}
\label{app:twist_mom}
The material frame is parametrized by the Bishop frame as:
\begin{align*}
    \boldsymbol{D}_1^j &= \cos(\phi^j)\boldsymbol{U}^j + \sin(\phi^j)\boldsymbol{V}^j\\
    \boldsymbol{D}_2^j &= -\sin(\phi^j)\boldsymbol{U}^j + \cos(\phi^j)\boldsymbol{V}^j.
\end{align*}
From that it follows
\begin{align*}
    \frac{\partial \boldsymbol{D}_1^j}{\partial \phi^j} = \boldsymbol{D}_2^j,\quad 
    \frac{\partial \boldsymbol{D}_2^j}{\partial \phi^j} = -\boldsymbol{D}^j_1.
\end{align*}
Now, we recall the form of the bending energy
\begin{align*}
    E_{b1}&=\sum\limits_{i=1}^{n-2} \frac{b}{2\overline{\ell}}(\kappa_{i1}-\overline{\kappa_{i1}})^2,\\ 
    &\text{ with } \kappa_{i1} = \frac{1}{2} \big(\boldsymbol{D}_2^{i-1} + \boldsymbol{D}_2^{i}\big) \cdot \underbrace{\frac{2 \boldsymbol{t}^{i-1}\times \boldsymbol{t}^{i}}{1+\boldsymbol{t}^{i-1}\cdot \boldsymbol{t}^{i}}}_{(\kappa \boldsymbol{b})_i},\\
    &E_{b2}=\sum\limits_{i=1}^{n-2} \frac{b}{2\overline{\ell}}(\kappa_{i2}-\overline{\kappa_{i2}})^2, \\
    &\text{ with } \kappa_{i2} = -\frac{1}{2} \big(\boldsymbol{D}_1^{i-1} + \boldsymbol{D}_1^{i}\big) \cdot \underbrace{\frac{2 \boldsymbol{t}^{i-1}\times \boldsymbol{t}^{i}}{1+\boldsymbol{t}^{i-1}\cdot \boldsymbol{t}^{i}}}_{(\kappa \boldsymbol{b})_i}.
\end{align*}
Let $\alpha:=b/(2\overline{\ell})$ and recall $\boldsymbol{t}^{i}=\frac{\boldsymbol{x}^{i+1}-\boldsymbol{x}^{i}}{\| \boldsymbol{x}^{i+1}-\boldsymbol{x}^{i}\|}$. We proceed by calculating the gradient. First note that:
\begin{align*}
    \frac{\partial \kappa_{i1}}{\partial \phi ^ i} &= \frac{1}{2} \frac{\partial \boldsymbol{D}_2^i}{\partial \phi ^ i}\cdot(\kappa\boldsymbol{b})_i =-\frac{1}{2} \boldsymbol{D}_1^i\cdot(\kappa\boldsymbol{b})_i 
    \text{ and } \\
    \frac{\partial \kappa_{i+1,1}}{\partial \phi ^ i} &= \frac{1}{2} \frac{\partial \boldsymbol{D}_2^i}{\partial \phi ^ i}\cdot(\kappa\boldsymbol{b})_{i+1} =-\frac{1}{2} \boldsymbol{D}_1^i\cdot(\kappa\boldsymbol{b})_{i+1},\\
    \frac{\partial \kappa_{i2}}{\partial \phi ^ i} &= -\frac{1}{2} \frac{\partial \boldsymbol{D}_1^i}{\partial \phi ^ i}\cdot(\kappa\boldsymbol{b})_i =-\frac{1}{2} \boldsymbol{D}_2^i\cdot(\kappa\boldsymbol{b})_i 
    \text{ and } \\
    \frac{\partial \kappa_{i+1,2}}{\partial \phi ^ i} &= -\frac{1}{2} \frac{\partial \boldsymbol{D}_1^i}{\partial \phi ^ i}\cdot(\kappa\boldsymbol{b})_{i+1} =-\frac{1}{2} \boldsymbol{D}_2^i\cdot(\kappa\boldsymbol{b})_{i+1}.
\end{align*}
Using this for the gradient, it yields:
\begin{align*}
    \frac{\partial E_{b1}}{\partial \phi ^ i} =& \alpha  \frac{\partial }{\partial \phi ^ i} \big[ (\kappa_{i1}-\overline{\kappa_{i1}})^2 + (\kappa_{i+1,1}-\overline{\kappa_{i+1,1}})^2\big] \\
    =&-\alpha  \big[ (\kappa_{i1}-\overline{\kappa_{i1}}) \boldsymbol{D}_1^i\cdot(\kappa\boldsymbol{b})_i  \\
    &+ (\kappa_{i+1,1}-\overline{\kappa_{i+1,1}}) \boldsymbol{D}_1^i\cdot(\kappa\boldsymbol{b})_{i+1}  \big],\\
    \frac{\partial E_{b2}}{\partial \phi ^ i} =& \alpha  \frac{\partial }{\partial \phi ^ i} \big[ (\kappa_{i2}-\overline{\kappa_{i2}})^2 + (\kappa_{i+1,2}-\overline{\kappa_{i+1,2}})^2\big] \\
    =&-\alpha  \big[ (\kappa_{i2}-\overline{\kappa_{i2}}) \boldsymbol{D}_2^i\cdot(\kappa\boldsymbol{b})_i  \\
    &+ (\kappa_{i+1,2}-\overline{\kappa_{i+1,2}}) \boldsymbol{D}_2^i\cdot(\kappa\boldsymbol{b})_{i+1}  \big].
\end{align*}
For the derivatives with respect to $\phi^0$ and $\phi^{n-2}$, we get
\begin{align*}
    \frac{\partial E_{b1}}{\partial \phi ^ 0} &=  \alpha  \frac{\partial }{\partial \phi ^ 0}  (\kappa_{1,1}-\overline{\kappa_{1,1}})^2 \\
    &= -\alpha  (\kappa_{1,1}-\overline{\kappa_{1,1}}) \boldsymbol{D}_1^0\cdot(\kappa\boldsymbol{b})_{1},\\
    \frac{\partial E_{b2}}{\partial \phi ^ 0} &=  \alpha  \frac{\partial }{\partial \phi ^ 0}  (\kappa_{1,2}-\overline{\kappa_{1,2}})^2 \\
    &= -\alpha  (\kappa_{1,2}-\overline{\kappa_{1,2}}) \boldsymbol{D}_2^0\cdot(\kappa\boldsymbol{b})_{1},\\
    \frac{\partial E_{b1}}{\partial \phi ^ {n-2}} &=  \alpha  \frac{\partial }{\partial \phi ^ {n-2}}  (\kappa_{n-2,1}-\overline{\kappa_{n-2,1}})^2 \\
    &= -\alpha  (\kappa_{n-2,1}-\overline{\kappa_{n-2,1}}) \boldsymbol{D}_1^{n-2}\cdot(\kappa\boldsymbol{b})_{n-2},\\
    \frac{\partial E_{b2}}{\partial \phi ^ {n-2}} &=  \alpha  \frac{\partial }{\partial \phi ^ {n-2}}  (\kappa_{n-2,2}-\overline{\kappa_{n-2,2}})^2 \\
    &= -\alpha  (\kappa_{n-2,2}-\overline{\kappa_{n-2,2}}) \boldsymbol{D}_2^{n-2}\cdot(\kappa\boldsymbol{b})_{n-2}.
\end{align*}

\section{Coil simulation parameters}
\textbf{General Parameters}
\label{App:Sim_Params}
We first state in table \ref{tab:matparams} the material parameters that were used in each of the conducted simulations if not stated otherwise.
\begin{table}[htbp]
    \renewcommand{\arraystretch}{1.5}
    \centering
    \begin{tabular}{|c|c|}
        \hline
        Material Parameter & Value, Unit \\
        \hline
         $E_w$ & \SI{230}{\giga\pascal} \\
         \hline
         $\mu_w$ & \SI{0.4}{} \\
         \hline
         $\rho$ & \SI{21e3}{\kilo\gram\per\cubic\meter}\\
         \hline
         Simulation Parameter & Value, Unit\\
         \hline
         $\|\boldsymbol{v}_{push}\|$ & \SI{3}{\centi\meter\per\second}
         \\
         \hline
         $k_w$ & \SI{4e2}{\newton\per\meter}\\
         \hline
         $\gamma_w$ & \SI{0.01}{\newton\second\per\meter}\\
         \hline
         $v_{\epsilon}$ & \SI{1e-8}{\meter\per\second}\\
         \hline 
         $\overline{\ell}$ & $D_2$\\
         \hline 
         $\alpha$ & \SI{1e-1}{\joule\per\meter}\\
         \hline 
         $\eta_{\boldsymbol{X}}$ & \SI{1e-2}{\newton \second}\\
         \hline 
         $\eta_{\boldsymbol{\Phi}}$ & \SI{1e-9}{\newton \meter \second }\\
         \hline 
         $\eta_{\mu_{slip,CW}}$ & \SI{0.6}{}\\
         \hline 
         $\eta_{\mu_{stick,CW}}$ & \SI{0.9}{}\\
         \hline 
         $\eta_{\mu_{slip,CC}}$ & \SI{0.6}{}\\
         \hline 
    \end{tabular}
    \caption{Summary of simulation and material parameters.}
    \label{tab:matparams}
\end{table}

\section{Axial strain energy}
\label{app:ax_energy}
The axial strain in the continuous case is given by the difference of the axial strain in the current and reference configuration:
\begin{align}
    \epsilon =  \boldsymbol{x}'(s) -  \overline{\boldsymbol{x}}'(s).
\end{align}
In the discrete setting, we assume that the axial strain is constant on an edge $\boldsymbol{e}^j$ resulting in
\begin{align}
    \epsilon \approx \frac{\|\boldsymbol{e}^j\|}{\|\overline{\boldsymbol{e}}^j\|} - \frac{\|\overline{\boldsymbol{e}}^j\|}{\|\overline{\boldsymbol{e}}^j\|}.
\end{align}
This allows to write the axial strain energy for the discrete setting as
\begin{align}
    \int_0^L \alpha\epsilon^2 ds &= \sum\limits_{j=0}^{n-2}\int_{\boldsymbol{x}_j}^{\boldsymbol{x}_{j+1}}\alpha \epsilon^2 ds \\
    &\approx  \sum\limits_{j=0}^{n-2} \frac{1}{2}\alpha \big(\frac{\|\boldsymbol{e}^j\|}{\|\overline{\boldsymbol{e}}^j\|} - 1\big)^2 \|\overline{\boldsymbol{e}}^j\|,
\end{align}
which is what we have stated above.

\end{appendices}


\setcitestyle{authoryear}
\bibliography{lit}

\end{document}